\documentclass[11pt]{article}

\usepackage[T1]{fontenc}
\usepackage[utf8]{inputenc}
\usepackage{lmodern}
\usepackage{microtype}
\usepackage[margin=1in]{geometry}
\usepackage{graphicx}
\usepackage{booktabs}
\usepackage{array}
\usepackage{amsmath}
\usepackage{amssymb}
\usepackage{authblk}
\usepackage[font=small,labelfont=bf]{caption}
\usepackage{url}
\usepackage{flafter}
\usepackage{float}
\usepackage[section]{placeins}
\usepackage{tikz}
\usetikzlibrary{positioning,fit,calc}
\usepackage[hidelinks]{hyperref}
\hypersetup{
  pdftitle={Homogeneity and heterogeneity of functional element composition of
            genome units across human and seven model organisms},
  pdfauthor={Yanying Wu},
  pdfsubject={Comparative genomics},
  pdfkeywords={comparative genomics; functional elements; transposable elements;
               chromosome composition; compositional homogeneity}}

\graphicspath{{figs_v8/}{figs_v6/}}

\newcommand{\Uadj}{U_{\mathrm{adj}}}
\newcommand{\sm}{\mathrm{sim}}
\newcommand{\Ce}{\textit{C.~elegans}}
\newcommand{\At}{\textit{Arabidopsis}}

\title{\bfseries Homogeneity and heterogeneity of functional element composition\\
of genome units across human and seven model organisms}

\author[ ]{Yanying Wu}
\affil[ ]{Centre for Neural Circuits \& Behaviour, University of Oxford, Oxford, United Kingdom}
\affil[ ]{\normalfont\small\href{mailto:yanying.wu@cncb.ox.ac.uk}{\texttt{yanying.wu@cncb.ox.ac.uk}}}
\date{\number\day~\ifcase\month\or January\or February\or March\or April\or May\or June\or July\or August\or September\or October\or November\or December\fi~\number\year}

\begin{document}
\maketitle

\begin{abstract}
\noindent
Is a chromosome a miniature of its genome --- does it carry the same composition of
functional elements as the genome as a whole? We ask this of eight annotated genomes:
human, mouse, zebrafish, the nematode \textit{Caenorhabditis elegans}, the fruit fly
\textit{Drosophila melanogaster}, budding yeast \textit{Saccharomyces cerevisiae}, and two
plants, \textit{Arabidopsis thaliana} and maize. We measure how similar the composition of
functional elements in a unit of the genome, such as a chromosome, is to that of the whole
genome. We classify the functional elements in a four-level hierarchy that runs from coarse,
general categories to specific ones. The first level sorts every element into five major
classes: protein-coding genes, non-coding RNA (ncRNA) genes, transposable elements (TEs),
pseudogenes, and immunoglobulin and T-cell receptor (IG/TR) gene segments; levels~2 to~4
divide these into progressively finer classes. Besides taking individual chromosomes as
units, we define units in other ways: cytogenetic bands and sub-bands where a band map
exists, equal cuts of each chromosome into an increasing number of parts, and random
genome segments of matching size. We compare the composition of each unit both with the whole
genome and with the other units, and call compositions similar, or homogeneous, when they
differ by less than 5\%, setting each comparison beside what a random sample of the same
size would reach by chance.

The first finding is a \emph{surprising homogeneity}. The level-1 element compositions of the
major chromosomes are similar to each other and to the genome in maize, zebrafish, yeast and
\Ce; in fly, this holds down to the major chromosome arms. Finer classes can be homogeneous
too: in \At{} the same holds for protein function, and in maize for the Wicker TE orders
(the mix of LTR, LINE, SINE, DNA and Helitron elements). In
particular, the reaction classes of enzymes are interchangeable between chromosomes in all
eight genomes.

The second finding is a \emph{structured heterogeneity}. Where a unit departs from its
genome, it does so in one of three recurring ways: clusters of named elements, such as
ribosomal-RNA genes or the immune receptor loci; megabase-scale compartments, such as the
TE-rich heterochromatin around centromeres; and lineage-specific gene families, which
become visible only at the finest levels of classification.

Once chance is accounted for, the genomes fall into groups. For protein function and gene
families, composition along the genome is more uniform in plants and fly than in
vertebrates, and least uniform in \Ce. For the five major classes, and for the TE orders
within them, uniformity instead follows where each genome places its repeats --- spread
along the chromosome arms (maize, zebrafish) or packed around the centromeres
(\At, fly) --- not its lineage.
\end{abstract}

\section{Introduction}

A chromosome is the unit in which a genome is inherited, recombined, packaged in the
nucleus and rearranged during evolution. Whether it is also a unit of \emph{content} ---
whether each chromosome carries a characteristic mix of functional elements, or simply a
sample of the genome-wide mix --- has rarely been asked directly. A reference annotation
makes the question answerable without any new experiment: it assigns every functional
element a class and every class an address, so the class composition of any piece of a
genome can be set against that of the whole. If chromosomes were random draws from one
genome-wide pool, every large enough piece would look like the whole. Where they are not,
the pieces that fail --- and the scale at which they fail --- say something about how the
genome is built.

Element order is known not to be random. Vertebrate genomes are mosaics of isochores\footnote{Isochores: long stretches of DNA, typically hundreds of kilobases to megabases, with fairly uniform guanine--cytosine (GC) content inside each stretch \cite{bernardi,eyrewalker}.} that
differ in GC content, gene density and repeat type
\cite{bernardi,eyrewalker}. Co-expressed and housekeeping genes cluster into chromosomal
domains in yeast, fly and human \cite{cohen,caron,spellman,boutanaev,lercher,versteeg}, a
literature reviewed in \cite{hurst}. Chromatin state and contact maps divide chromosomes
into megabase compartments\footnote{Compartments: alternating megabase blocks of open, gene-active and condensed, silent chromatin, each contacting mostly blocks of its own kind in the nucleus \cite{liebermanaiden}.} \cite{liebermanaiden,filion,kharchenko} and lamina-associated
domains\footnote{Lamina-associated domains: gene-poor, largely silent stretches of chromosome held against the nuclear envelope \cite{guelen}.} \cite{guelen}; centromeres are flanked by megabases of repeat-rich heterochromatin
\cite{copenhaver,hoskins,altemose}; and recombination rate varies systematically along arms
\cite{rockman,comeron}. Many gene families have expanded by tandem duplication in a single
lineage and still sit together where they arose \cite{lespinet}. At the largest scale,
whole chromosome arms are conserved as linkage units across \textit{Drosophila} species, the
Muller elements\footnote{Muller elements A--F: six ancestral \textit{Drosophila} chromosome arms whose gene content has stayed largely together across species, even where arms have fused or been rearranged \cite{muller,bhutkar}.} \cite{muller,bhutkar}. Most of this work asks about expression, chromatin,
sequence or synteny\footnote{Synteny: genes lying on the same chromosome; conserved synteny means the same genes still share a chromosome in different species \cite{passarge}.}. We ask the simpler, annotation-level question of
whether the \emph{class composition} of functional elements --- what kinds of genes,
repeats and gene segments a piece of genome carries --- is local or genome-wide, and we
ask it the same way in eight genomes.

This paper organises the answer around two words. A unit is \emph{homogeneous} with its
genome when its composition matches the genome's to within five percent, so that at least
95\% of the unit's elements (or base pairs) could be paired one-for-one with the genome's
composition. A unit is \emph{heterogeneous} when it does not, \emph{and} the shortfall
exceeds what a random sample of the same size would produce. The second clause matters. A
30-element piece of a nine-class composition cannot score 0.95 even when it is a perfect
random sample, so raw similarity alone confuses small with unusual. We therefore report
every similarity beside its sampling floor. The 5\% tolerance is a convention fixed before
any result was seen, and every conclusion is also checked with two statistics that use no
threshold. All four statistics are defined in \S\ref{sec:scores}.

We lay the evidence out along two axes, both running coarse to fine.

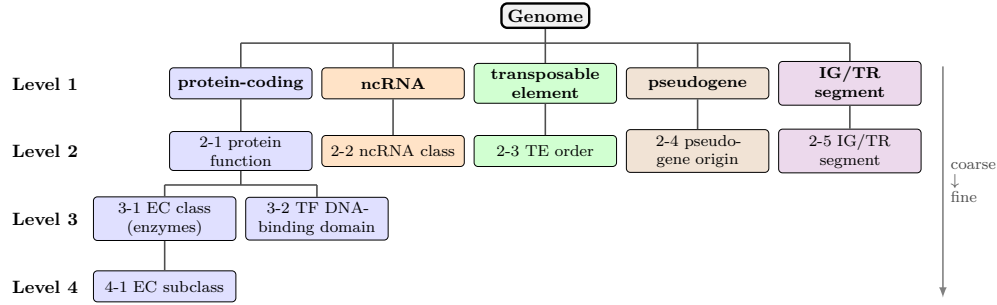
\begin{figure}[H]
\centering
\resizebox{0.8\textwidth}{!}{%
\begin{tikzpicture}[
    font=\scriptsize, >=latex,
    n/.style={draw, rounded corners=2pt, align=center, inner sep=2.5pt,
              minimum height=0.55cm, text width=2.35cm},
    lvl/.style={font=\footnotesize\bfseries, anchor=east},
    edge/.style={thick, gray!60!black}]
  \colorlet{pc}{blue!12}\colorlet{nc}{orange!22}\colorlet{te}{green!18}
  \colorlet{ps}{brown!22}\colorlet{ig}{violet!15}
  \node[draw, very thick, rounded corners=3pt, fill=gray!10, font=\footnotesize\bfseries,
        inner sep=3pt] (root) at (5.4,0) {Genome};
  \foreach \x/\nm/\c/\t in {0/pcg/pc/protein-coding, 2.7/ncg/nc/ncRNA,
      5.4/te/te/transposable element, 8.1/psg/ps/pseudogene, 10.8/igt/ig/{IG/TR segment}}
     \node[n, fill=\c, font=\scriptsize\bfseries] (\nm) at (\x,-1.2) {\t};
  \foreach \x/\nm/\c/\t in {0/fun/pc/2-1 protein function, 2.7/ncc/nc/2-2 ncRNA class,
      5.4/tec/te/2-3 TE order, 8.1/psc/ps/2-4 pseudogene origin, 10.8/igc/ig/{2-5 IG/TR segment}}
     \node[n, fill=\c] (\nm) at (\x,-2.4) {\t};
  \node[n, fill=pc] (ec) at (-1.35,-3.6) {3-1 EC class\\(enzymes)};
  \node[n, fill=pc] (tf) at (1.35,-3.6) {3-2 TF DNA-binding domain};
  \node[n, fill=pc] (ecs) at (-1.35,-4.8) {4-1 EC subclass};
  \draw[edge] (root.south) -- ++(0,-0.25) coordinate (b0);
  \draw[edge] (0,0|-b0) -- (10.8,0|-b0);
  \foreach \nm in {pcg,ncg,te,psg,igt} \draw[edge] (\nm.north |- b0) -- (\nm.north);
  \foreach \a/\b in {pcg/fun,ncg/ncc,te/tec,psg/psc,igt/igc} \draw[edge] (\a) -- (\b);
  \draw[edge] (fun.south) -- ++(0,-0.25) coordinate (b2);
  \draw[edge] (-1.35,0|-b2) -- (1.35,0|-b2);
  \draw[edge] (ec.north |- b2) -- (ec.north);
  \draw[edge] (tf.north |- b2) -- (tf.north);
  \draw[edge] (ec) -- (ecs);
  \foreach \y/\t in {-1.2/Level 1, -2.4/Level 2, -3.6/Level 3, -4.8/Level 4}
     \node[lvl] at (-2.75,\y) {\t};
  \draw[->, thick, gray] (12.45,-0.9) -- (12.45,-5.0)
     node[midway, right, font=\scriptsize, text=gray!60!black, align=left]
     {coarse\\[-1pt]$\downarrow$\\[-1pt]fine};
\end{tikzpicture}}
\caption{\textbf{Class resolution}: the four-level classification hierarchy in brief.
Level~1 splits every annotated functional element into five classes; each deeper level
subdivides one class of the level above. The classes inside each node are listed in
\S\ref{sec:hierarchy}.}
\label{fig:hierarchy-intro}
\end{figure}

\paragraph{Class resolution.} Elements are classified in a four-level hierarchy
(Fig.~\ref{fig:hierarchy-intro}). Level~1 is the element class: protein-coding, ncRNA, TE,
pseudogene and IG/TR; it is weighted by base pairs (bp), with element counts as a check.
Level~2 subdivides each of the five (levels 2-1 to 2-5): protein-coding genes by protein
function, ncRNA genes by RNA class, TEs by order, pseudogenes by origin, and IG/TR segments
by locus and segment type. Level~3 subdivides two functional classes: enzymes by Enzyme
Commission (EC)\footnote{Enzyme Commission (EC) number: a four-part code, such as EC~2.7.11.1, assigned by the Nomenclature Committee of the International Union of Biochemistry and Molecular Biology (IUBMB) to classify an enzyme by the reaction it catalyses. The first digit gives one of seven reaction classes and the second the subclass \cite{webb,ec}.} reaction class (level~3-1) and transcription factors (TFs) by DNA-binding
domain (DBD; level~3-2). Level~4 is the EC subclass (level~4-1). The full hierarchy, with
every class, is given in \S\ref{sec:hierarchy}.

\begin{figure}[H]
\centering
\includegraphics[width=\textwidth]{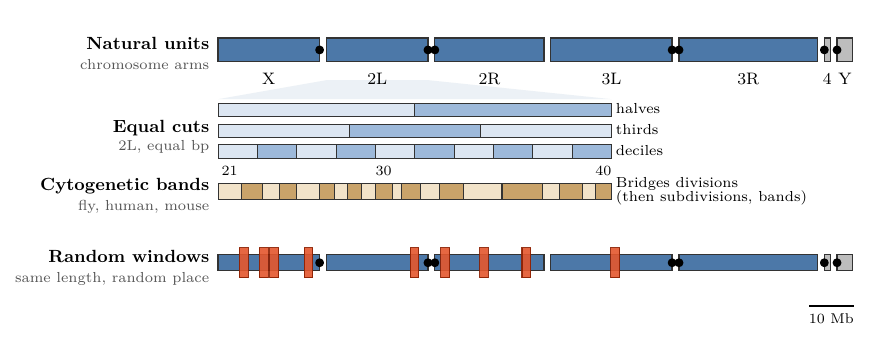}
\caption{\textbf{Unit resolution}, with the fly genome as the example. Natural units are
the chromosome arms (blue; the dot chromosome \texttt{4} and \texttt{Y}, grey, are special
units; black dots mark centromeres). Each natural unit is then cut into equal parts (arm
\texttt{2L} shown as halves, thirds and deciles), into cytogenetic bands where a band map
exists (the 20 Bridges divisions of \texttt{2L} shown), and compared with random windows of
the same length placed anywhere in the genome. Arms are drawn to scale.}
\label{fig:units-intro}
\end{figure}

\paragraph{Unit resolution.} The genome is divided into units in several ways
(Fig.~\ref{fig:units-intro}): the natural unit (a chromosome arm, or a whole chromosome
where arms are not defined) $\to$ halves, thirds, quarters, fifths $\to$ deciles, all equal
in base pairs; where a cytogenetic map exists (fly, human, mouse), its regions or
divisions, bands and sub-bands; and random windows of a given length. Each definition is
described in full in \S\ref{sec:methods}.

Part~I (\S\ref{sec:homog}) collects the homogeneous compositions in that order. Part~II
(\S\ref{sec:heterog}) collects the heterogeneous ones in the same order.
\S\ref{sec:synthesis} tests one hypothesis against both: \emph{functional element
composition along the genome is more uniform in plants and flies than in mammals.} The
hypothesis follows from the vertebrate literature above. If isochores and cis-clustered
paralogue families\footnote{Cis-clustered paralogue families: sets of gene copies made by duplication within one genome (paralogues) that still lie side by side on the same chromosome (in cis) \cite{koonin,lespinet}.} are what make composition vary along a genome
\cite{bernardi,eyrewalker,lespinet}, then genomes without isochores, such as those of
\textit{Drosophila} and \textit{Arabidopsis}, should be more uniform than mammalian ones.
The eight genomes were chosen to span that contrast, with yeast, \Ce{} and zebrafish as
intermediates that can show whether genome size, lineage or annotation is doing the work.

The eight genomes span annotation styles as much as phylogeny. There are two mammals with
annotations rich in long non-coding RNA (lncRNA), pseudogene and IG/TR entries, and with
genomes nearly half TE by length; a teleost\footnote{Teleost: a ray-finned bony fish, here the zebrafish \cite{howe}.}; a holocentric\footnote{Holocentric: the chromosome attaches to the spindle along its whole length rather than at a single centromere \cite{maddox}.} nematode whose ncRNA annotation
is dominated by PIWI-interacting RNA (piRNA) loci; a fly with FlyBase's curated gene groups
and TE insertions; a budding yeast with almost no non-coding annotation; and two plants
whose non-coding classes come from NCBI RefSeq, one of them (maize) 88\% TE by length.
Where a difference between genomes traces to who annotated them rather than how they are
built, we say so. \S\ref{sec:caveats} lists where that happens.

In brief, the findings are as follows. By base pairs, every major chromosome of maize,
zebrafish, yeast and \Ce{} and every major arm of fly matches its genome's level-1
composition to within 5\%; in \At{} the same holds for protein function, and in maize for the
Wicker TE orders (the mix of LTR, LINE, SINE, DNA and Helitron elements); and EC reaction classes are interchangeable between chromosomes in all eight
genomes. Where a unit departs from its genome, the cause is a named element cluster (the
ribosomal-RNA genes of the nucleolus organisers\footnote{Nucleolus organisers: chromosomal sites carrying tandem arrays of ribosomal-RNA genes, around which the nucleolus forms \cite{mcstay}.}, the zinc-finger array on human
chromosome~19, the IG/TR loci), a megabase-scale compartment (the TE- and pseudogene-rich
heterochromatin flanking centromeres), or a lineage-specific gene family that appears only at
the finest levels. Once chance is accounted for, the chromosomes of the plants and fly depart
from their genome's protein-function composition by less than one percentage point, those of
zebrafish, human and mouse by 5--8.5 points, and those of \Ce{} by 13; TF domains give the
same order. The plant-and-fly advantage does not extend to level~1, where the two mammals are
the least uniform genomes by base pairs but \At{} collapses once its chromosomes are cut, nor
to TEs, whose uniformity follows where each genome places its repeats rather than its
lineage.

\section{Data and methods}
\label{sec:methods}

\subsection{Genomes, annotation and units}

Eight assemblies were used, each in its own coordinates with no liftover:
\textit{D.~melanogaster} dm6 (Berkeley Drosophila Genome Project release 6, BDGP6;
FlyBase FB2026\_01) \cite{adams,hoskins,flybase}, human GRCh38 \cite{lander,schneider},
mouse GRCm39 \cite{mouse}, zebrafish GRCz11 \cite{howe}, \textit{S.~cerevisiae} R64
\cite{goffeau}, \Ce{} WBcel235 \cite{celegans,wormbase}, \At{} TAIR10.1 \cite{agi,tair} and
maize Zm-B73-NAM-5.0 \cite{schnable,hufford}. Chromosome names follow FlyBase/Ensembl style
without a \texttt{chr} prefix. Each element is placed by its midpoint, with GFF3
(General Feature Format version~3) coordinates 1-based inclusive and BED (Browser
Extensible Data) and Ensembl tables converted at load.

\paragraph{Genes.} Protein-coding, ncRNA, pseudogene and IG/TR classes come from Ensembl
biotypes \cite{ensembl,frankish}, except in \At{} and maize. For those two we use the
RefSeq annotation of the National Center for Biotechnology Information (NCBI) on the same
assembly \cite{refseq}, because the Ensembl Plants route \cite{ensemblgenomes} returns no
non-coding classes for them. Fly uses FlyBase gene classes \cite{flybase}. Function, enzyme
and TF assignments come from UniProt \cite{uniprot} and InterPro \cite{interpro}, and in
fly from FlyBase curated gene groups.

\paragraph{Transposable elements.} TEs enter at level~1, from one repeat annotation
per genome: FlyBase transposable-element insertions (FBti) in fly \cite{kaminker}; the
RepeatMasker track of the University of California, Santa Cruz (UCSC) Genome Browser in
human (hg38), mouse (mm39), zebrafish (danRer11) and \Ce{} (ce11) \cite{ucsc,repeatmasker};
the chromosomal-feature file of the \textit{Saccharomyces} Genome Database (SGD) in yeast
\cite{goffeau}; the TAIR10 transposable-element table in \At{} \cite{tair}; and the
structural TE annotation of the maize genome database (MaizeGDB) in maize \cite{hufford}.
Gene models that an annotation itself labels as TEs are counted as TEs, not genes. In
yeast, the open reading frames (ORFs) of the Ty retrotransposons are listed as
protein-coding genes; we drop them, so each Ty element is counted once, as a TE.

\paragraph{Units.} A \emph{unit} is a chromosome arm where a centromere position is
available (fly, human) and a whole chromosome otherwise. Mouse, zebrafish, yeast, \Ce,
\At{} and maize are therefore scored on whole chromosomes. \Ce{} chromosomes are holocentric
\cite{maddox} and have no arms or bands. Their known internal structure is a set of
recombination domains --- gene-dense, low-recombination centres flanked by
high-recombination arms \cite{rockman} --- which Part~II returns to.

\begin{figure}[!htbp]
\centering
\includegraphics[width=\textwidth,height=0.7\textheight,keepaspectratio]{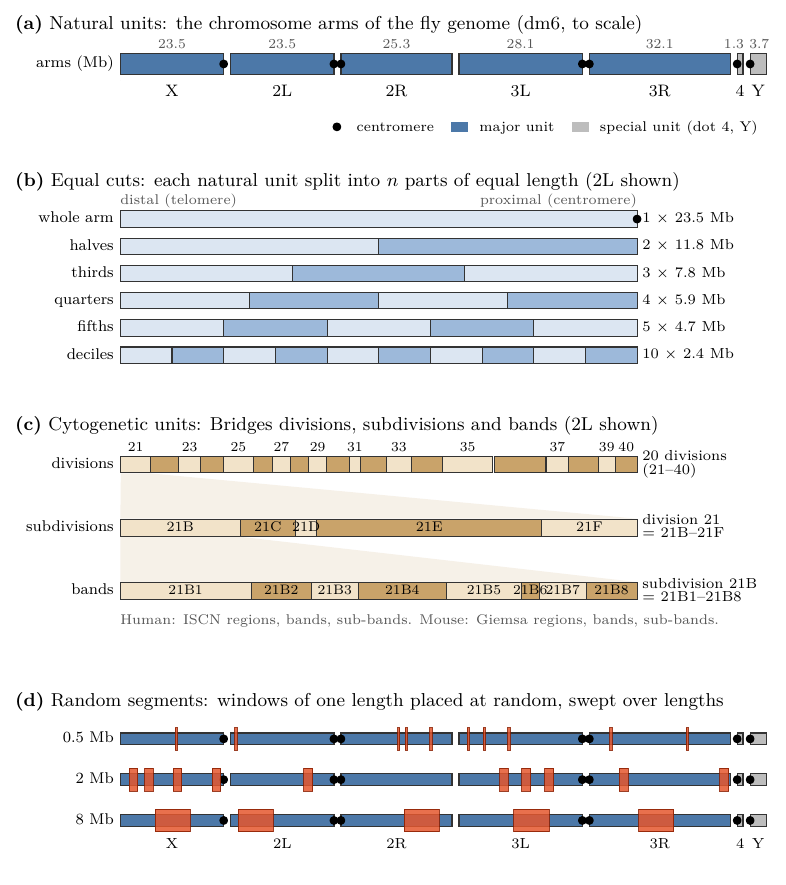}
\caption{\textbf{How units are defined}, with the fly genome (dm6) as the example.
(a)~Natural units: the chromosome arms, drawn to scale with their lengths. Black dots mark
centromeres. The five large arms are major units; the dot chromosome \texttt{4} and
\texttt{Y} are special units, reported separately. In genomes without a centromere annotation
the natural unit is the whole chromosome. (b)~Equal cuts: each natural unit is split into
2, 3, 4, 5 or 10 parts of equal base-pair length, from the distal (telomeric) to the
proximal (centromeric) end; the part lengths shown are for \texttt{2L}. (c)~Cytogenetic
units at three nested resolutions: the 20 Bridges divisions of \texttt{2L} (21--40), the
lettered subdivisions of one division (21B--21F) and the numbered bands of one subdivision
(21B1--21B8). Band boundaries are from the Bridges map distributed with
\texttt{ideogram.js}, drawn proportionally on the dm6 arm. (d)~Random segments: windows of
one fixed length placed at random positions in the major units, here at 0.5, 2 and 8~Mb;
the length is swept to find how long a window must be before it resembles the whole
genome. Every part, of every kind, is scored only if it holds at least 20 elements.}
\label{fig:units}
\end{figure}

Units are cut two ways. \emph{Equal cuts} split each natural unit into 2, 3, 4, 5 or 10
parts of equal base-pair length. \emph{Band cuts} use a cytogenetic map where one exists,
at three resolutions: Bridges polytene divisions, subdivisions and bands in fly
\cite{bridges}; International System for Human Cytogenomic Nomenclature (ISCN) regions,
bands and sub-bands in human \cite{iscn,furey}; and Giemsa regions, bands and sub-bands in
mouse. Both kinds of part are scored at every level, and a part is scored only if it
holds at least 20 elements. A third kind of unit ignores unit
boundaries altogether: random segments of a fixed length placed at random positions in the
genome, with the length swept from well below to well above a typical unit
(\S\ref{sec:scores}). Figure~\ref{fig:units} shows all three on the fly genome.

\paragraph{Special units.} A few units are set aside from every ``major unit'' summary and
discussed on their own: every \texttt{Y}, fly's dot chromosome \texttt{4}, and the human
acrocentric short arms \texttt{13p}, \texttt{14p}, \texttt{15p}, \texttt{21p} and
\texttt{22p}. These are the units a cytogeneticist would exclude in advance ---
heterochromatic, gene-poor, bearing ribosomal DNA (rDNA) or sex-limited
\cite{riddle,carvalho,mcstay,nurk} --- and we name them before looking at any result.

\subsection{The classification hierarchy}
\label{sec:hierarchy}

\begin{figure}[!htbp]
\centering
\resizebox{\textwidth}{!}{%
\begin{tikzpicture}[
    font=\scriptsize, >=latex,
    box/.style={draw, rounded corners=2pt, align=flush left, inner sep=3pt,
                anchor=north, text width=3.55cm},
    l1/.style={draw, rounded corners=2pt, align=center, inner sep=3pt,
               anchor=north, text width=3.55cm, minimum height=0.85cm},
    lvl/.style={font=\footnotesize\bfseries, anchor=west, align=left},
    edge/.style={thick, gray!60!black}]
  \colorlet{pc}{blue!12}\colorlet{nc}{orange!22}\colorlet{te}{green!18}
  \colorlet{ps}{brown!22}\colorlet{ig}{violet!15}
  \foreach \y/\t in {-1.95/Level 1, -4.9/Level 2, -8.5/Level 3, -11.1/Level 4}
     \node[lvl] at (-3.9,\y) {\t};
  \foreach \y in {-2.85,-7.15,-10.2} \draw[gray!40, dashed] (-3.9,\y) -- (18.4,\y);
  \node[draw, very thick, rounded corners=3pt, fill=gray!10, font=\footnotesize\bfseries,
        align=center, anchor=north, inner sep=4pt] (root) at (8.2,0.3)
       {Genome\\[-1pt]{\normalfont\scriptsize(all annotated functional elements on the assembly)}};
  \node[l1, fill=pc] (pcg) at (0,-1.5)   {\textbf{protein-coding}};
  \node[l1, fill=nc] (ncg) at (4.1,-1.5) {\textbf{ncRNA}};
  \node[l1, fill=te] (te)  at (8.2,-1.5) {\textbf{transposable element}};
  \node[l1, fill=ps] (psg) at (12.3,-1.5){\textbf{pseudogene}};
  \node[l1, fill=ig] (igt) at (16.4,-1.5){\textbf{IG/TR segment}};
  \draw[edge] (root.south) -- ++(0,-0.35) coordinate (b0);
  \draw[edge] (0,0|-b0) -- (16.4,0|-b0);
  \foreach \n in {pcg,ncg,te,psg,igt} \draw[edge] (\n.north |- b0) -- (\n.north);
  \node[box, fill=pc] (fun) at (0,-3.05)
       {\textbf{2-1 protein function}\\(11 classes)\\
        enzyme $\cdot$ TF $\cdot$ receptor $\cdot$ transporter/channel $\cdot$
        signalling $\cdot$ proteostasis $\cdot$ RNA processing/translation $\cdot$
        trafficking $\cdot$ structural/extracellular $\cdot$ chromatin $\cdot$
        immunity/defence};
  \node[box, fill=nc] (ncc) at (4.1,-3.05)
       {\textbf{2-2 ncRNA class}\\(8 classes)\\
        lncRNA $\cdot$ miRNA $\cdot$ tRNA $\cdot$ rRNA $\cdot$ snRNA $\cdot$
        snoRNA/scaRNA $\cdot$ piRNA $\cdot$ other ncRNA};
  \node[box, fill=te] (tec) at (8.2,-3.05)
       {\textbf{2-3 TE order}\\(5 classes; by bp)\\
        LTR $\cdot$ LINE $\cdot$ SINE $\cdot$ DNA transposon (TIR) $\cdot$ Helitron};
  \node[box, fill=ps] (psc) at (12.3,-3.05)
       {\textbf{2-4 pseudogene origin}\\(3 classes)\\
        processed $\cdot$ unprocessed $\cdot$ unitary};
  \node[box, fill=ig] (igc) at (16.4,-3.05)
       {\textbf{2-5 IG/TR segment}\\(8 classes)\\
        IG C, D, J, V $\cdot$ TR C, D, J, V};
  \foreach \a/\b in {pcg/fun,ncg/ncc,te/tec,psg/psc,igt/igc} \draw[edge] (\a.south) -- (\b.north);
  \node[box, fill=pc] (ec) at (0,-7.65)
       {\textbf{3-1 EC class}\\(7 classes, enzymes)\\
        oxidoreductase $\cdot$ transferase $\cdot$ hydrolase $\cdot$ lyase $\cdot$
        isomerase $\cdot$ ligase $\cdot$ translocase};
  \node[box, fill=pc] (tf) at (4.1,-7.65)
       {\textbf{3-2 TF DNA-binding domain}\\(13 InterPro families, TFs)\\
        homeodomain $\cdot$ C2H2 zinc finger $\cdot$ bHLH $\cdot$ bZIP $\cdot$
        nuclear receptor $\cdot$ forkhead $\cdot$ \ldots};
  \draw[edge] (fun.south) -- ++(0,-0.3) coordinate (b2);
  \draw[edge] (0,0|-b2) -- (4.1,0|-b2);
  \draw[edge] (ec.north |- b2) -- (ec.north);
  \draw[edge] (tf.north |- b2) -- (tf.north);
  \node[box, fill=pc] (ecs) at (0,-10.5)
       {\textbf{4-1 EC subclass}\\(top 12 + other)\\ e.g.\ EC 2.7, phosphotransferases};
  \draw[edge] (ec.south) -- (ecs.north);
  \node[font=\scriptsize, text=gray!30!black, anchor=north west, align=flush left]
       at (9.6,-7.65) {weights and denominators:\\[1pt]
        level 1 --- base pairs of all elements (count as a check)\\
        2-1 --- protein-coding genes with a named function\\
        2-2, 2-4, 2-5 --- genes of that level-1 class\\
        2-3 --- TE base pairs\\
        3-1, 4-1 --- enzymes; 3-2 --- classified TFs};
\end{tikzpicture}}
\caption{\textbf{The classification hierarchy}, drawn as a tree rooted at the genome.
Level~1 splits every annotated functional element into five classes. Each is subdivided at
level~2 (2-1 to 2-5). Protein function is subdivided again at level~3, enzymes by reaction
class (3-1) and transcription factors by DNA-binding domain (3-2), and EC classes once more
at level~4 (4-1). Buckets that record annotation state rather than class
(``other/unclassified'' elements, ``other annotated family'' and ``no family annotation''
proteins, unclassified TEs and pseudogenes) are omitted here and dropped before any score.
Abbreviations are defined in the text.}
\label{fig:hierarchy}
\end{figure}
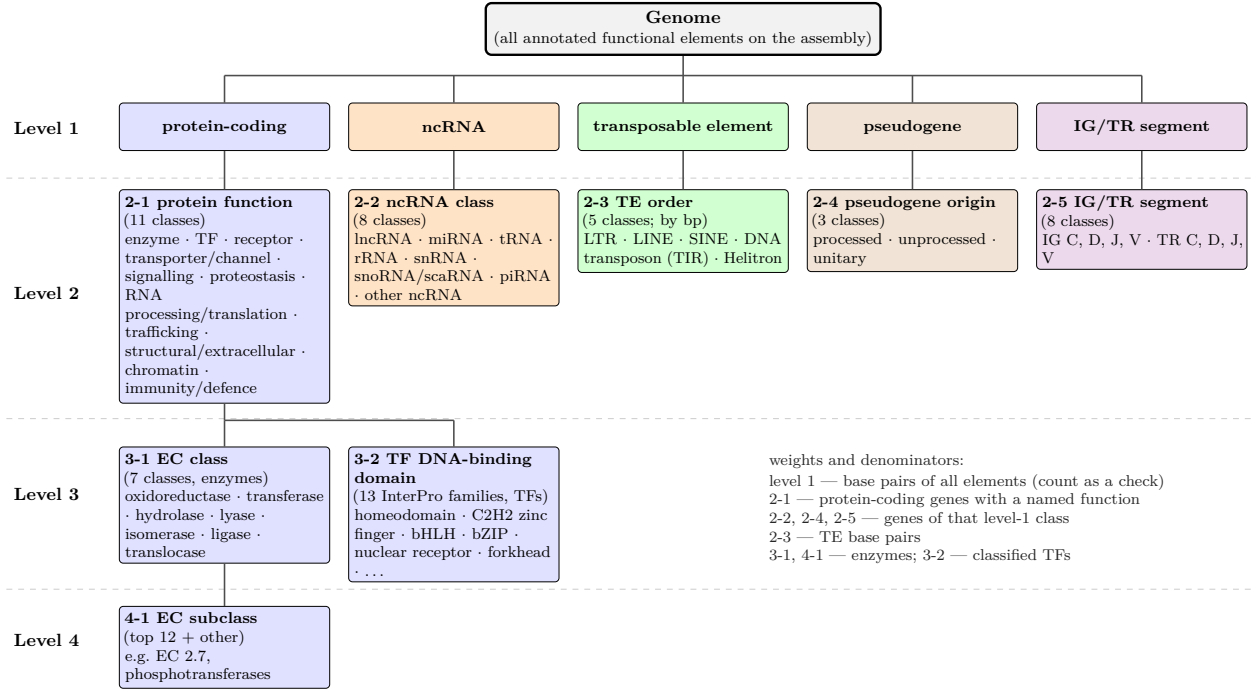

A composition can look homogeneous at one resolution and heterogeneous at another. A
coarse class pools many families, so a cluster of one family is diluted by the rest. A
fine class isolates the family, but leaves each unit with few elements, so sampling noise
grows. We therefore classify every element at several nested levels and ask at which
resolution, if any, a unit stops resembling its genome (Fig.~\ref{fig:hierarchy}). The
hierarchy follows three principles.
\begin{enumerate}
\item \textbf{Established vocabularies.} Each level uses an externally curated
classification rather than one devised for this study: Ensembl/GENCODE biotypes
\cite{frankish}, InterPro families \cite{interpro}, the Wicker classification of
transposable elements \cite{wicker}, the EC list of the IUBMB \cite{ec}, and the classification of
transcription factors by their DNA-binding domain \cite{wingender,lambert}. This keeps
classes reproducible and comparable across the eight genomes.
\item \textbf{Nesting.} Each deeper level subdivides one class of the level above, as the
ranks of the Gene Ontology \cite{go}, the EC list and TF classifications do. Every step
down therefore scores fewer elements over more classes, which is why the sampling floor
(\S\ref{sec:scores}) matters most at the deepest levels.
\item \textbf{From what an element is to how its product works.} Level~1 records what kind
of element a feature is; level~2 what its product does or where it came from; levels~3
and~4 how a protein does it (the reaction an enzyme catalyses, the domain a TF uses to bind
DNA). The coarse levels cover the whole annotated genome; the fine levels cover only part of
the proteome, but they are fine enough to name individual lineage-specific gene families.
\end{enumerate}

\paragraph{Why TEs are a level-1 class, and why level~1 is weighted by base pairs.}
Pooling TE copies with genes swamps every count: TE copies are 98\% of all level-1 elements in human, mouse and
zebrafish, 96\% in maize, 52\% in \Ce, 45\% in \At, 24\% in fly and 6\% in yeast
(RepeatMasker also splits one element into many fragments). Counted, level~1 would then
mostly measure TE density and repeat-annotation practice. Yet TEs are the largest class of
functional element by length in five of the eight genomes, and leaving them out would hide
most of the sequence of the large genomes. They therefore join level~1, and level~1 is
\emph{weighted by base pairs}: each element contributes its
annotated length (gene span, or TE copy length), so that a 200~bp Alu and a 50~kb gene are
weighed by the sequence they occupy. Element counts are reported beside every level-1
base-pair result as a check. Overlaps are not resolved (an intronic TE is counted in both
the gene and the TE), so ``element base pairs'' can exceed assembly length. Levels below
level~1 are counted per element, except TE order (2-3), which keeps its base-pair weight.

\paragraph{The levels.}
\begin{itemize}
\item \textbf{Level 1 --- element class.} Five scored classes:
  \begin{itemize}
  \item \emph{protein-coding} genes, whose transcripts are translated into protein;
  \item \emph{non-coding RNA (ncRNA) genes}, whose functional product is the RNA itself;
  \item \emph{transposable elements (TEs)}, sequences that can move or copy themselves to
  new genomic sites, and their remnants \cite{wicker};
  \item \emph{pseudogenes}, gene copies that have lost coding capacity \cite{pei};
  \item \emph{immunoglobulin and T-cell receptor (IG/TR) segments}, the gene segments that
  are recombined in lymphocytes to make antibodies and T-cell receptors; vertebrates only
  \cite{lefranc}.
  \end{itemize}
A sixth bucket, \emph{other/unclassified} (provisional or artefactual gene models), is
drawn in figures but dropped before any score.
\item \textbf{Level 2-1 --- protein function} (protein-coding genes). Eleven named
functions: enzyme, transcription factor, receptor, transporter/channel, signalling,
proteostasis, RNA processing/translation, structural/extracellular, trafficking,
chromatin/chromosome and immunity/defence. Enzymes are genes with a curated EC number; the
other ten classes come from InterPro family and domain names \cite{interpro}. Two
annotation-state buckets (``other annotated family'', ``no family annotation'') are drawn
in composition figures but \emph{dropped before any score}.
\item \textbf{Level 2-2 --- ncRNA class} (ncRNA genes). Long non-coding RNA (lncRNA),
transcripts longer than 200 nucleotides (nt) that are not translated \cite{mattick};
microRNA (miRNA), $\sim$22-nt RNAs that repress target messenger RNAs (mRNAs)
\cite{bartel}; transfer RNA (tRNA), the adaptor that carries amino acids to the ribosome
\cite{gtrnadb}; ribosomal RNA (rRNA), the structural and catalytic RNA of the ribosome,
encoded in tandem rDNA arrays at the nucleolus organiser regions (NORs) \cite{mcstay};
small nuclear RNA (snRNA), the RNA core of the spliceosome \cite{will}; small nucleolar RNA
(snoRNA) and small Cajal-body-specific RNA (scaRNA), which guide chemical modification of
rRNA and snRNA \cite{kiss,darzacq}; PIWI-interacting RNA (piRNA), 21--35-nt RNAs that
silence TEs in the germline, annotated as genes only in \Ce, where they are called 21U-RNAs
\cite{ozata,ruby}; and \emph{other ncRNA}, the remaining small RNA biotypes.
\item \textbf{Level 2-3 --- TE order} (TEs, by base pairs). The Wicker orders
\cite{wicker}: long terminal repeat (LTR) retrotransposons, long interspersed nuclear
elements (LINEs), short interspersed nuclear elements (SINEs), DNA transposons with
terminal inverted repeats (TIRs), and Helitrons (rolling-circle transposons). Weighted by
base pairs, because mean copy length differs three- to five-fold between repeat
annotations. Yeast has a single order (LTR, the Ty elements) and is not scored.
\item \textbf{Level 2-4 --- pseudogene origin} (pseudogenes). \emph{Processed}
pseudogenes arise by retrotransposition of an mRNA; \emph{unprocessed} by duplication of a
gene with its introns; \emph{unitary} by loss of function of a gene that has no functional
copy left \cite{pei}. Only the human and mouse annotations make this distinction for
enough pseudogenes to score; zebrafish's 324 pseudogenes are reported but not scored.
\item \textbf{Level 2-5 --- IG/TR segment} (IG/TR segments). The constant (C), diversity
(D), joining (J) and variable (V) segments of the immunoglobulin (IG) and T-cell receptor
(TR) loci \cite{lefranc}. Scored in human and mouse.
\item \textbf{Level 3-1 --- EC class} (enzymes). The seven IUBMB reaction classes
\cite{ec}: oxidoreductases, transferases, hydrolases, lyases, isomerases, ligases and
translocases. EC class names the chemistry an enzyme performs, independently of its
sequence family.
\item \textbf{Level 3-2 --- TF DNA-binding domain} (TFs). A protein-coding gene is a TF if
it carries an InterPro DNA-binding domain, and is classed by that domain
\cite{interpro,wingender,lambert}: the most frequent families across species, for example
homeodomain, C2H2 zinc finger, basic helix--loop--helix (bHLH) and basic leucine zipper
(bZIP). Fly's FlyBase superclass vocabulary is not comparable, so fly is left out of this
level.
\item \textbf{Level 4-1 --- EC subclass} (enzymes). The second digit of the EC number (for
example EC~2.7, transferases of phosphorus-containing groups, which include the kinases).
The shared top-12 subclasses are kept, with the tail folded into one bucket.
\end{itemize}

\subsection{Scores}
\label{sec:scores}

Every comparison in the paper is summarised by one of four statistics, each answering a
different question (Table~\ref{tab:terms}). Similarity asks how close one part is to its
genome; the match share asks how much of a genome is a near-copy of the whole; excess
dissimilarity asks whether the departures are larger than sampling allows; and $\Uadj$ asks,
without any threshold or cut, how much an element's unit tells us about its class.

\begin{figure}[!htbp]
\centering
\includegraphics[width=\textwidth]{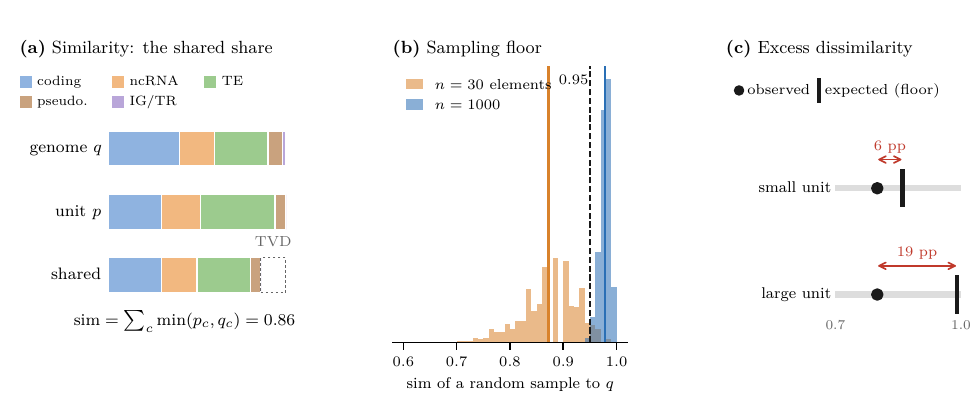}
\caption{\textbf{How similarity is scored} (illustration with toy compositions).
(a)~A unit's composition $p$ and its genome's composition $q$ share, class by class,
$\min(p_c,q_c)$; the sum is the similarity, and the remainder is the total variation
distance (TVD). (b)~Random samples drawn from the genome itself fall short of
$\sm = 1$, and the shortfall grows as the sample shrinks: a 30-element sample of this
composition rarely reaches 0.95, a 1{,}000-element one almost always does. The mean of
such draws is the sampling floor. (c)~The same observed similarity can mean little or a
lot. Excess dissimilarity is the distance from the observed value down from its floor:
small for a small unit that sits near its floor, large for a large unit that does not.}
\label{fig:scores}
\end{figure}

\paragraph{Similarity and its sampling floor.}
For a part with composition $p$ and a genome with composition $q$, $\sm = \sum_c
\min(p_c,q_c) = 1 - \mathrm{TVD}$, where $\mathrm{TVD} = \tfrac12\sum_c |p_c - q_c|$ is the
total variation distance \cite{levin}: the share of the part that would have to change
class to match the genome (Fig.~\ref{fig:scores}a). Similarity has a ceiling set by size.
A part with few elements is a noisy sample even when it is drawn at random from the genome,
so its similarity falls short of 1 for no biological reason (Fig.~\ref{fig:scores}b). We
estimate that \emph{sampling floor}, the \emph{expected} similarity, as the mean $\sm$ of
200 multinomial draws of the part's element count from $q$: the most a part of that size
could score by sampling alone. The \emph{excess dissimilarity} of a set of parts is the
median of $(\mathrm{expected} - \sm)$ in percentage points (pp): 0 means the parts look like
random draws from the genome, and larger means regional structure that sampling cannot
produce (Fig.~\ref{fig:scores}c). For the base-pair-weighted levels (level~1 and TE order)
the draws are of elements while $\sm$ is in base pairs, so the floor is optimistic and the
excess is an upper bound.

\paragraph{Match share.} For each species, level and cut, the \emph{match share} is the
fraction of major parts with $\sm \geq 0.95$. It answers the plain question ``how much of
this genome is a near-copy of the whole?'' and is the headline number of Part~I.

\paragraph{Across-unit uniformity.} The bias-corrected mutual information $\Uadj =
(I_{\mathrm{obs}} - \bar I_{\mathrm{null}})/H(\mathrm{class})$ is the share of an element's
class entropy $H$ that knowing its unit predicts \cite{cover}, above a size-matched null. It
is computed on element counts at every level, and its $p$-value comes from 4{,}000
permutations of unit labels, as $(b+1)/(m+1)$ for $b$ of $m$ permutations at least as
extreme, so that no $p$-value is reported as zero \cite{phipson}.

\paragraph{Within-unit tests and nulls.} These are a permutation $\chi^2$ test on parts
$\times$ classes, with bias-corrected Cramér's $V$ as effect size \cite{bergsma}; a
width-matched re-lay null for band partitions; a random-segment length sweep against
within- and between-unit label shuffles; and a fixed tandem-array rule (densest 200~kb
window holding $\geq 10$ genes and $\geq 15\%$ of the class on the unit, at $\geq 3\times$
enrichment). These tests need independent draws, so they are run on element counts. Counting TE
copies, however, mostly measures how each repeat annotation fragments its elements:
RepeatMasker reports one element as many hits, whereas FlyBase, TAIR and MaizeGDB list
whole insertions, so copy counts are comparable neither between genomes nor with genes.
The equal-cut partition tests therefore count \emph{gene classes only} --- protein-coding,
ncRNA, pseudogene and IG/TR --- in all eight genomes, and where TEs sit along a unit is read
from the base-pair-weighted analyses (level~1 by base pairs and TE order). The band tests
likewise count genes only in human and mouse; in fly they keep the FlyBase TE insertions,
which are whole, canonical copies.

All scans are implemented in \texttt{fly\_genome.homogeneity} and collected in notebook
52. Per-species figures come from notebooks 22--53 (\S\ref{sec:data}).

\begin{table}[!htbp]
\centering
\caption{Key terms used throughout the paper.}
\label{tab:terms}
\small
\begin{tabular}{p{3.1cm}p{11.4cm}}
\toprule
term & meaning \\
\midrule
natural unit & a chromosome arm where a centromere position is available (fly, human), a
whole chromosome otherwise \\
major / special unit & special units (every \texttt{Y}, fly \texttt{4}, human acrocentric
p arms) are named in advance and reported apart; all others are major \\
part & a piece of a natural unit: an equal cut, a band unit or a random segment \\
composition & the share of each class among a part's elements (or base pairs) at one
level of the hierarchy \\
similarity, $\sm$ & $\sum_c \min(p_c,q_c) = 1 - \mathrm{TVD}$ between a part and its whole
genome; 1 is identical, 0 shares nothing \\
homogeneous & $\sm \geq 0.95$: the part matches its genome to within 5\% \\
sampling floor & the expected $\sm$ of a random sample of the part's size drawn from the
genome \\
excess dissimilarity & median (floor $-$ observed $\sm$) over parts, in percentage points;
0 means no structure beyond sampling \\
match share & the percentage of major parts with $\sm \geq 0.95$ \\
$\Uadj$ & bias-corrected mutual information between unit and class, as a share of class
entropy; 0 means the unit predicts nothing about class \\
\bottomrule
\end{tabular}
\end{table}

\section{Overview: one scan, eight genomes, nine levels, six cuts}
\label{sec:overview}

Before turning to individual levels and genomes, we summarise the whole scan. Every genome
was scored at every level of the hierarchy and at every cut, from natural units down to
deciles. Four figures summarise the result, and five patterns in them, running from the
coarse to the fine end of the hierarchy, organise the rest of the paper; the numbers behind
each pattern are given in Parts~I and~II.

\begin{figure}[!htbp]
\centering
\includegraphics[width=\textwidth]{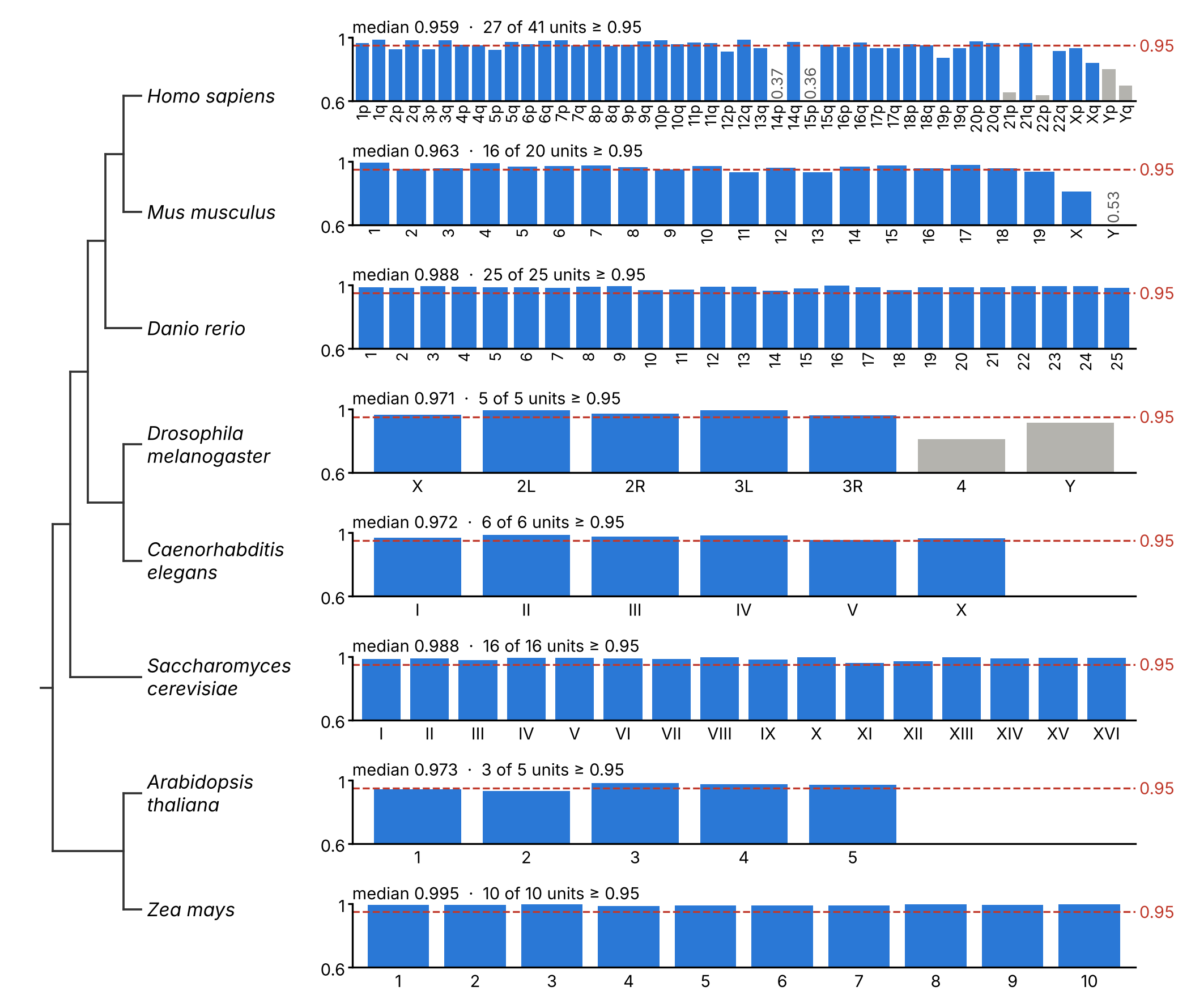}
\caption{\textbf{The eight genomes on their species tree, beside each natural unit's
level-1 similarity to its own genome, by base pairs.} Left: topology after TimeTree
\cite{timetree}, with fly and \Ce{} grouped as Ecdysozoa \cite{aguinaldo}. Branch lengths
are not to scale. Right: one bar per natural unit --- an arm where the assembly has arms
(human, fly), a whole chromosome otherwise --- at $\sm = 1 - \mathrm{TVD}$ against the
genome's level-1 composition. The dashed line is 0.95. Grey bars are special units; bars
below 0.6 are cut and carry their value. The panel title gives the median over major units
and how many reach 0.95.}
\label{fig:tree}
\end{figure}

\paragraph{1. Homogeneity is common at the coarse end.} The simplest version of the
question is whether each natural unit carries its genome's level-1 composition, by base
pairs. Figure~\ref{fig:tree} answers it for all eight genomes, placed on their species
tree. In maize, zebrafish, yeast, \Ce{} and fly nearly every major unit clears 0.95 (medians
0.971--0.995); human and mouse spread below it (medians 0.959 and 0.963). Further down the
hierarchy, at EC class, every \At{} and fly unit matches its genome as well
(\S\ref{sec:homog}).

\begin{figure}[!htbp]
\centering
\includegraphics[width=\textwidth]{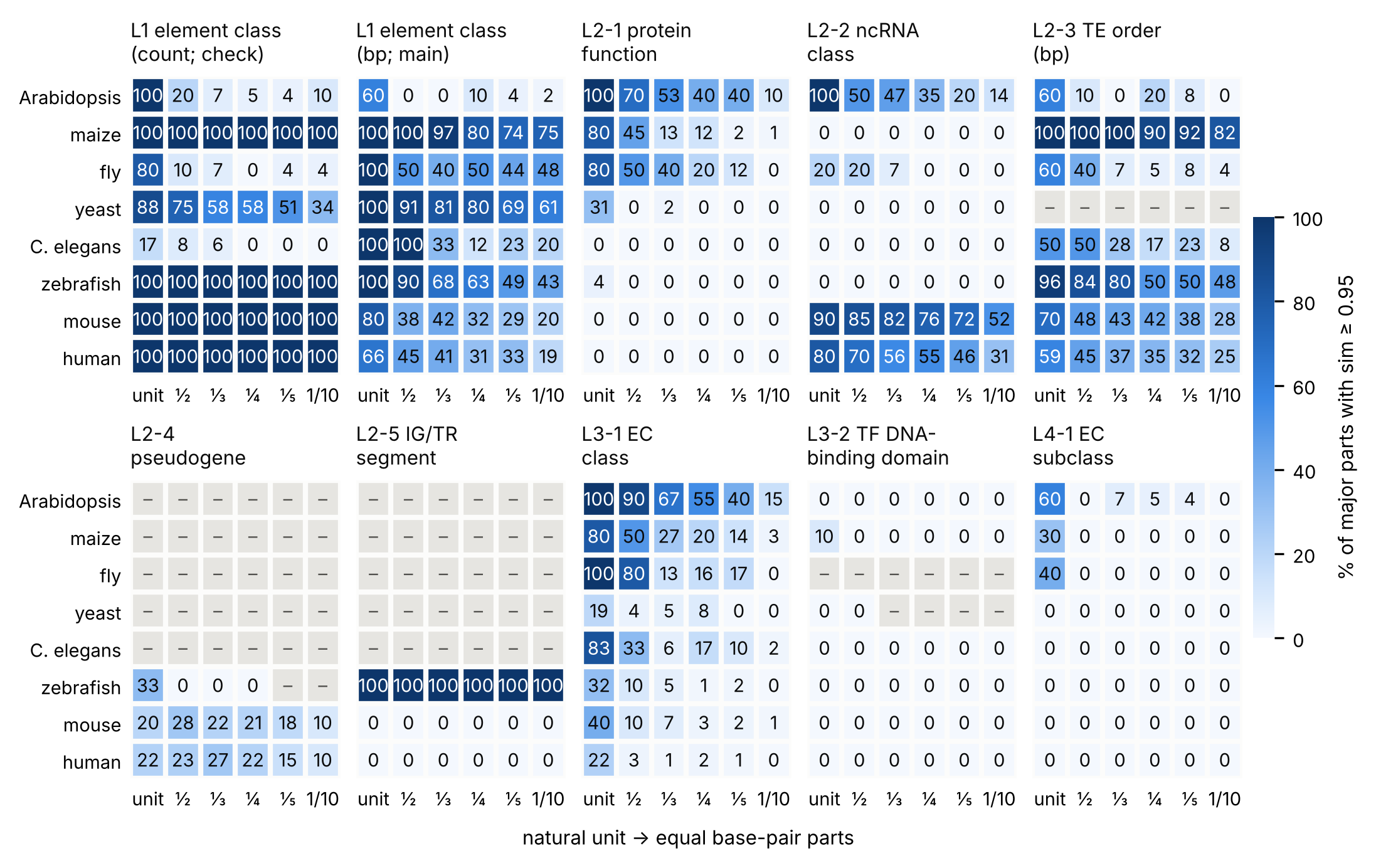}
\caption{\textbf{Match share.} The percentage of major parts whose composition matches the
whole genome at $\sm \geq 0.95$. One panel per level, coarse (top left) to fine (bottom
right); level~1 is shown both by base pairs (the main view) and by count (the check).
Rows are species in the order of the hypothesis, plants to mammals. Columns go from the
natural unit to deciles. Grey cells are levels a species cannot supply. Special units
(all \texttt{Y}, fly \texttt{4}, human acrocentric p arms) are excluded, as are parts with
fewer than 20 elements.}
\label{fig:matchshare}
\end{figure}

\begin{figure}[!htbp]
\centering
\includegraphics[width=\textwidth]{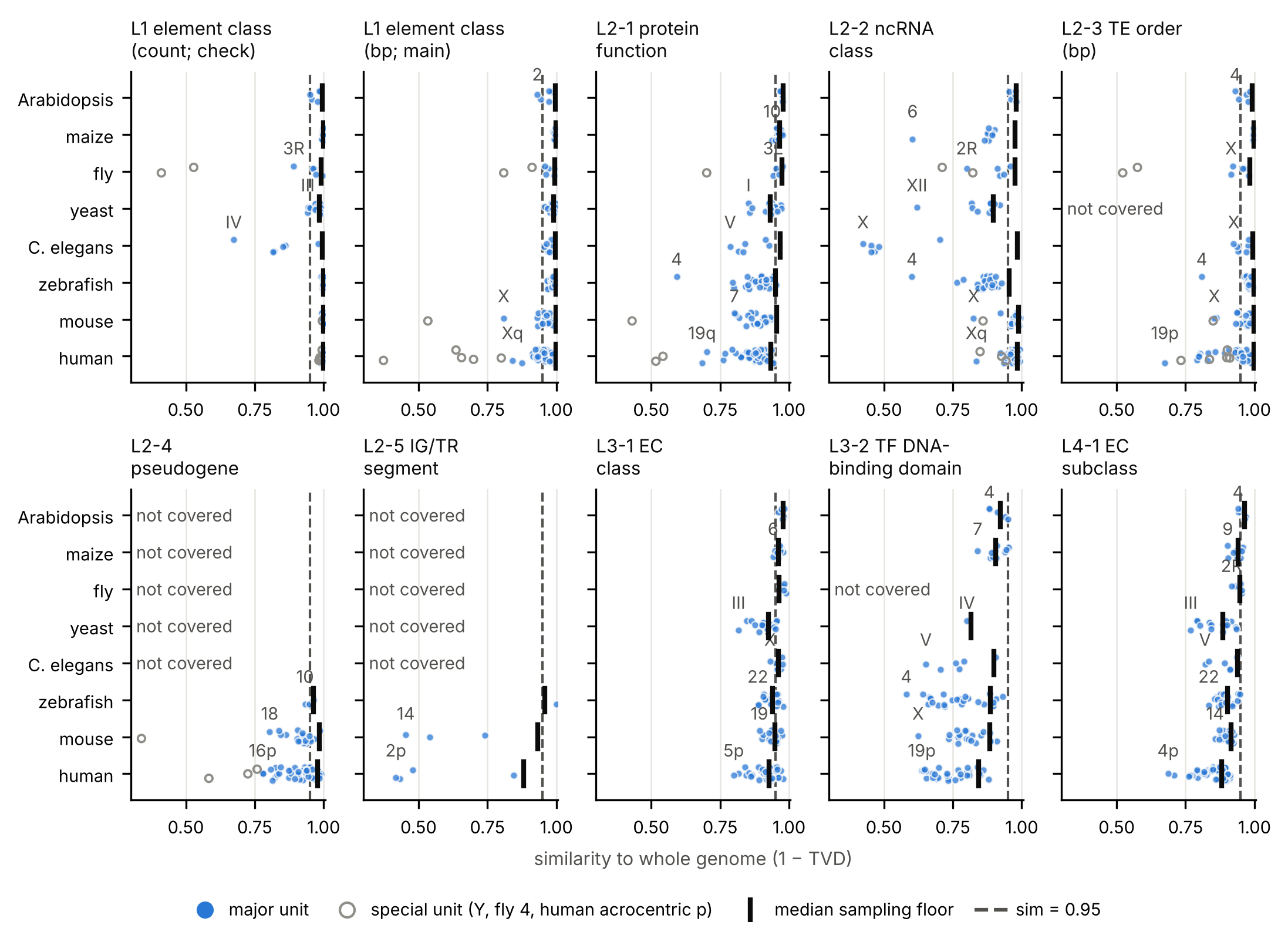}
\caption{\textbf{Every natural unit's similarity to its own genome}, at every level.
Filled dots are major units; hollow grey dots are special units. The black tick on each row
is the median expected similarity of a random draw of the same size --- the sampling floor.
The dashed line is 0.95. The least similar major unit is named where it falls below 0.95.}
\label{fig:unitsim}
\end{figure}

\paragraph{2. Homogeneity shrinks as either axis gets finer.} The match share summarises
the whole scan in one grid of species, levels and cuts (Fig.~\ref{fig:matchshare}).
Cutting units into deciles, or moving from element class to EC subclass, lowers it nearly
everywhere. Much of that fall is the sampling floor: a small part, or a composition spread
over many classes, cannot reach 0.95 even when it is a perfect random sample of its genome.
Figure~\ref{fig:unitsim} makes the floor visible by setting every natural unit's
similarity, at every level, beside the median similarity of a random draw of the same size
(black ticks). Where a unit falls below 0.95 but sits on its tick, it is small, not unusual.

\begin{figure}[!htbp]
\centering
\includegraphics[width=\textwidth]{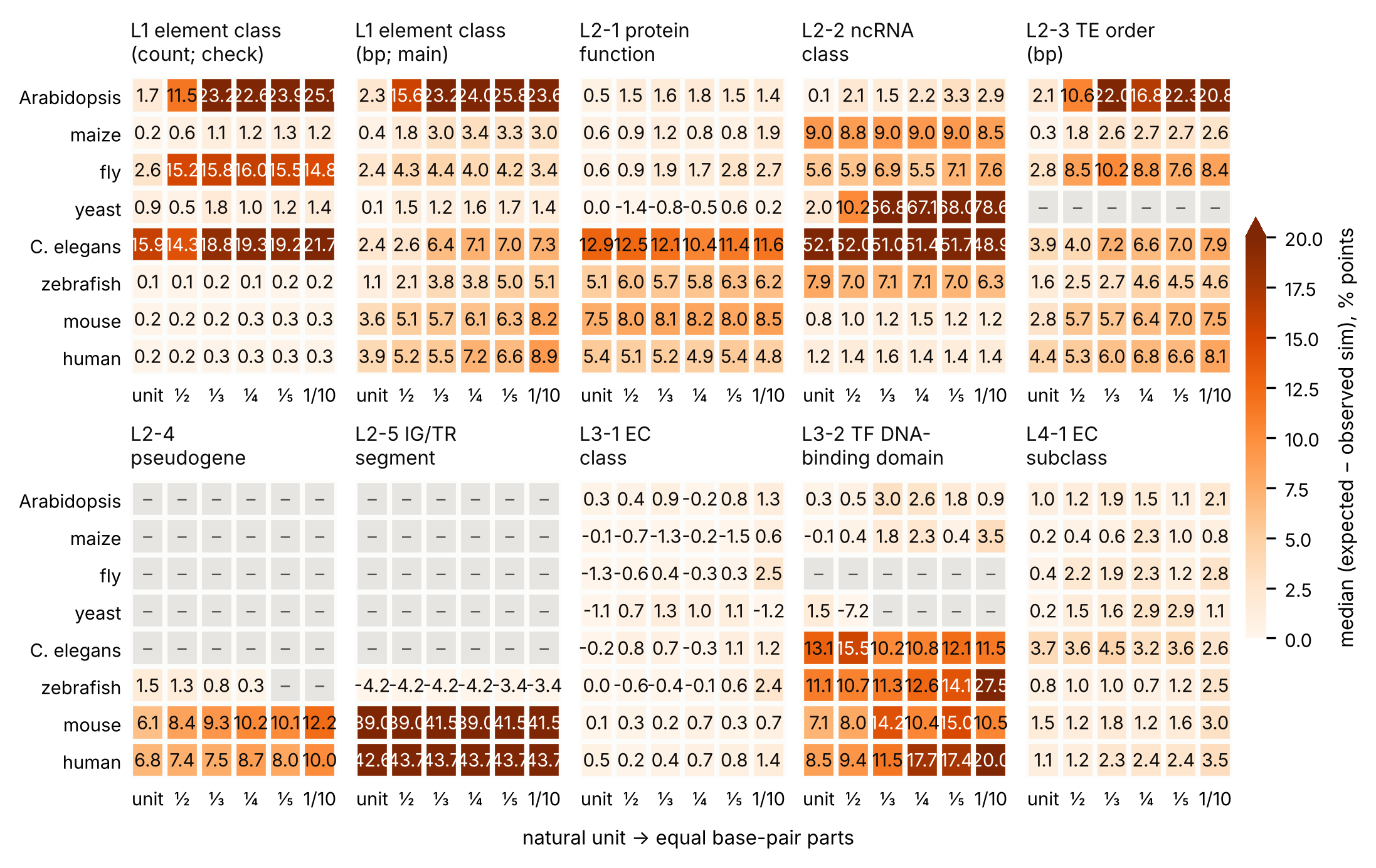}
\caption{\textbf{Excess dissimilarity}: the median of (expected $-$ observed) similarity
over major parts, in percentage points. This is the departure from the genome composition
that sampling cannot explain. It removes the mechanical fall in raw similarity as parts
shrink and classes multiply; 0 means the parts behave like random draws. Colour saturates
at 20 points. For the base-pair-weighted levels the floor is computed on counts and the
value is an upper bound.}
\label{fig:excess}
\end{figure}

\paragraph{3. After the floor is removed, the species separate.} Subtracting each part's
observed similarity from its sampling floor gives the excess dissimilarity
(Fig.~\ref{fig:excess}), the departure from the genome that sampling cannot explain; this is
the figure to read for structure. Protein function separates the species most clearly:
yeast, the plants and fly sit close to zero, the vertebrates several points higher, and
\Ce{} highest of all (\S\ref{sec:het-fn}).

\paragraph{4. Base-pair weighting reorders level~1.} Counted, level~1 is a statement about
TE copy number and is uninformative in the TE-rich genomes: their units score $\approx
0.997$ by count (Fig.~\ref{fig:matchshare}, level-1 count panel). By base pairs, maize and
zebrafish are the most homogeneous genomes, the two mammals the least homogeneous at
whole-unit scale, and \At{} the least homogeneous once its chromosomes are cut
(Fig.~\ref{fig:excess}), because each \At{} chromosome averages a TE- and pseudogene-rich
pericentromere with gene-rich arms (\S\ref{sec:het-cuts}).

\paragraph{5. The new level-2 classes are the most clustered of all.} IG/TR segment,
pseudogene origin and ncRNA class depart furthest from their genomes
(Figs.~\ref{fig:excess} and~\ref{fig:unitsim}), and each does so through a handful of named
loci: the IG/TR loci themselves, unprocessed pseudogenes beside their parent genes, and the
piRNA, rDNA and snRNA arrays (\S\ref{sec:het-nc}, \S\ref{sec:het-psig}).

\section{Part I --- Homogeneity: where units of the genome resemble the whole genome as well as each other}
\label{sec:homog}

\subsection{Level 1, natural units: most major units are near-copies by base pairs}
\label{sec:homog-l1}

The level-1 composition differs enormously \emph{between} genomes
(Table~\ref{tab:level1}). By base pairs, protein-coding genes take 95\% of yeast's element
length and 8\% of maize's, where 90\% is TE. Human and mouse are about one-third
protein-coding, one-third TE and one-fifth to one-quarter ncRNA, the last mostly
long intronic lncRNA spans. By count the picture is different again: TE copies are 98\% of
mammalian and zebrafish elements. \emph{Within} a genome, however, the major units are
close copies of one another and of the whole.

\begin{table}[!htbp]
\centering
\caption{Level-1 composition of eight genomes. Left block: percentage of element base
pairs (the main view). Right block: element counts. IG/TR segments are under 0.1\% of base
pairs everywhere. ``bp'' is the summed element length (overlaps not resolved). \At{} and
maize genes are read from NCBI RefSeq, fly from FlyBase, the other five from Ensembl; TE
sources are listed in \S\ref{sec:methods}.}
\label{tab:level1}
\small
\setlength{\tabcolsep}{4pt}
\begin{tabular}{lrrrrrrrrrr}
\toprule
 & \multicolumn{5}{c}{\% of element bp} & \multicolumn{5}{c}{element count} \\
\cmidrule(lr){2-6}\cmidrule(lr){7-11}
species & coding & ncRNA & TE & pseudo. & bp (Mb) & coding & ncRNA & TE & pseudo. & IG/TR \\
\midrule
yeast       & 94.7 &  0.9 &  4.4 &  0.1 &     9.3 &  6{,}572 &    397 &        434 &    12 &   0 \\
fly         & 87.5 &  4.6 &  7.6 &  0.3 &   111.1 & 13{,}968 & 3{,}497 &    5{,}734 &   330 &   0 \\
\Ce         & 80.1 &  2.5 & 13.4 &  4.0 &    80.5 & 19{,}973 & 24{,}789 &   49{,}813 & 2{,}128 & 0 \\
\At         & 63.2 &  2.6 & 24.1 & 10.1 &   103.6 & 27{,}444 & 5{,}612 &   31{,}189 & 4{,}843 &   0 \\
zebrafish   & 52.2 &  4.1 & 43.6 &  0.1 & 1{,}506.6 & 25{,}094 & 6{,}438 & 2{,}525{,}565 & 324 & 150 \\
mouse       & 39.2 & 20.4 & 39.4 &  1.0 & 2{,}825.9 & 21{,}757 & 38{,}965 & 3{,}758{,}776 & 13{,}818 & 493 \\
human       & 36.4 & 26.7 & 35.8 &  1.1 & 3{,}905.3 & 20{,}094 & 41{,}953 & 4{,}491{,}330 & 15{,}199 & 412 \\
maize       &  8.2 &  0.8 & 89.7 &  1.3 & 2{,}097.4 & 33{,}681 & 8{,}304 & 1{,}267{,}368 & 5{,}034 & 0 \\
\bottomrule
\end{tabular}
\end{table}

\begin{figure}[!htbp]
\centering
\includegraphics[width=\textwidth]{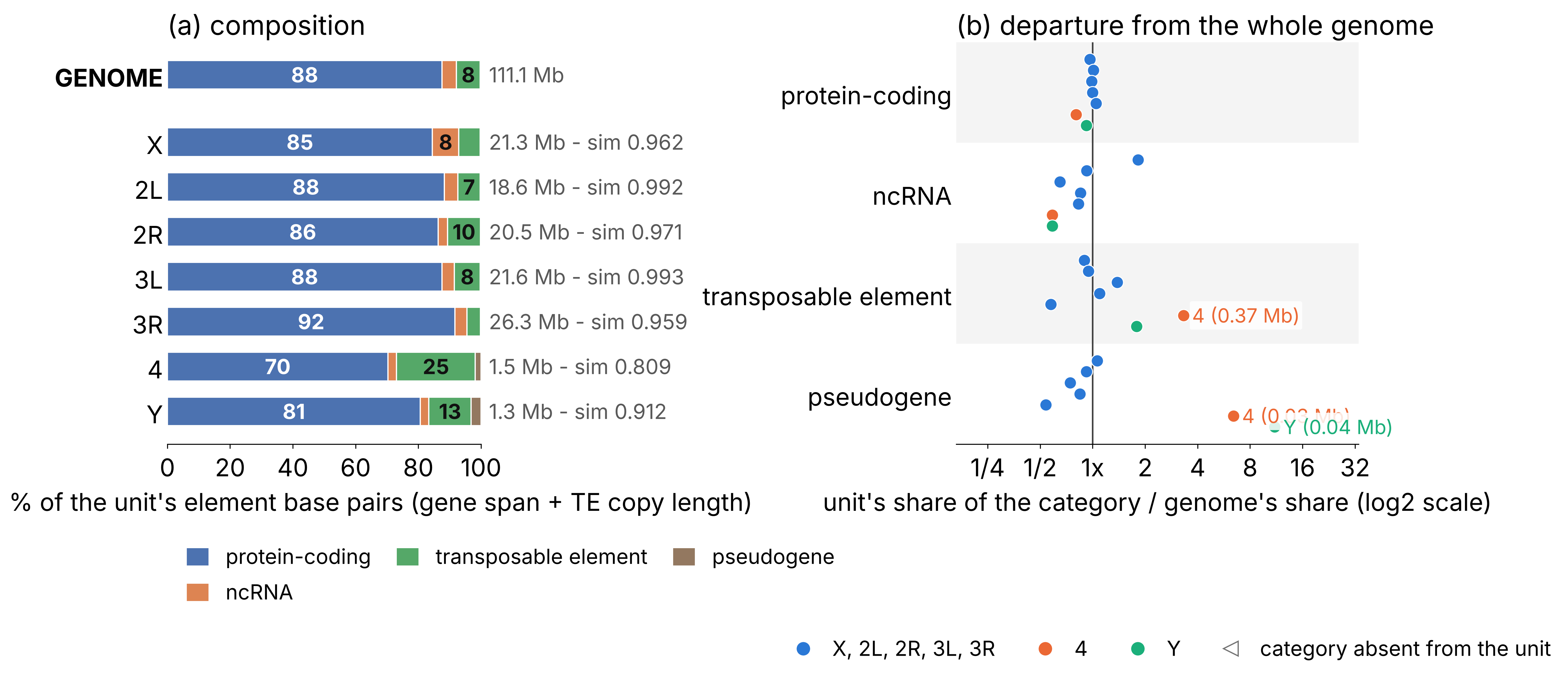}
\caption{Fly level~1 by base pairs, arm by arm, against the whole genome. (a) Composition,
with each arm's similarity to the genome. (b) Each arm's share of each class divided by
the genome's share. The five major arms are near-copies of the genome; \texttt{4} and
\texttt{Y} are not.}
\label{fig:flyl1}
\end{figure}

Fly is the cleanest illustration (Fig.~\ref{fig:flyl1}). Its five large arms hold
85--92\% protein-coding and 7--10\% TE base pairs, and each scores $\sm = 0.959$--$0.993$
against the genome. Only the two special units fall away: dot chromosome \texttt{4}
(0.809, 25\% TE by base pairs) and \texttt{Y} (0.912). The ratio panel shows why the arms
pass: every class sits within a factor of two of the genome. By count the arms fare worse
(match share 80\%), because FBti insertions are 24\% of fly's elements and cluster near
the centromeres; base-pair weighting gives each short insertion its true weight.

\begin{figure}[!htbp]
\centering
\includegraphics[width=\textwidth,height=0.8\textheight,keepaspectratio]{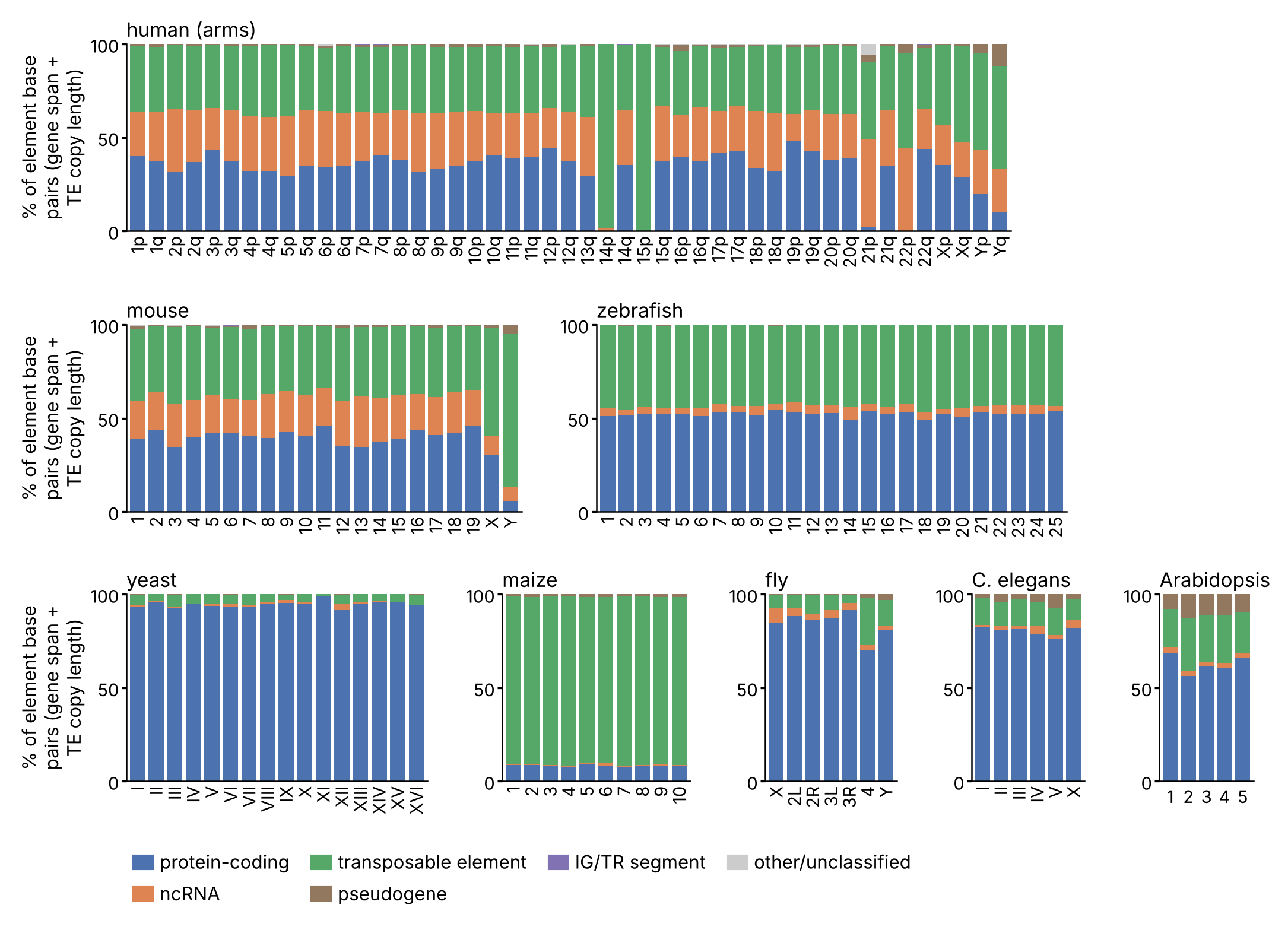}
\caption{Level-1 composition of every unit in all eight genomes by base pairs. The genomes are
packed into three rows, and one bar has the same width in every panel; hatched bars are
units with too few elements for a composition. Colours are fixed per class throughout the paper. Within most genomes the
bars are close copies of each other; the departures are named in Part~II.}
\label{fig:l1unit}
\end{figure}

The same holds across most of the eight genomes (Fig.~\ref{fig:l1unit}, and
Fig.~\ref{fig:tree}). In \Ce, all six chromosomes reach $\sm \geq 0.95$ (median 0.972); the
piRNA cluster on \texttt{IV}, which dominates \Ce{} level~1 by count (\S\ref{sec:het-l1}),
is a few hundred kilobases of 21-nt loci and nearly vanishes by base pairs. In \At, three
of the five chromosomes pass (\texttt{3}, \texttt{4} and \texttt{5} at 0.973--0.982), and
\texttt{1} and \texttt{2} fall just short (0.943, 0.931).

\begin{figure}[!htbp]
\centering
\includegraphics[width=\textwidth]{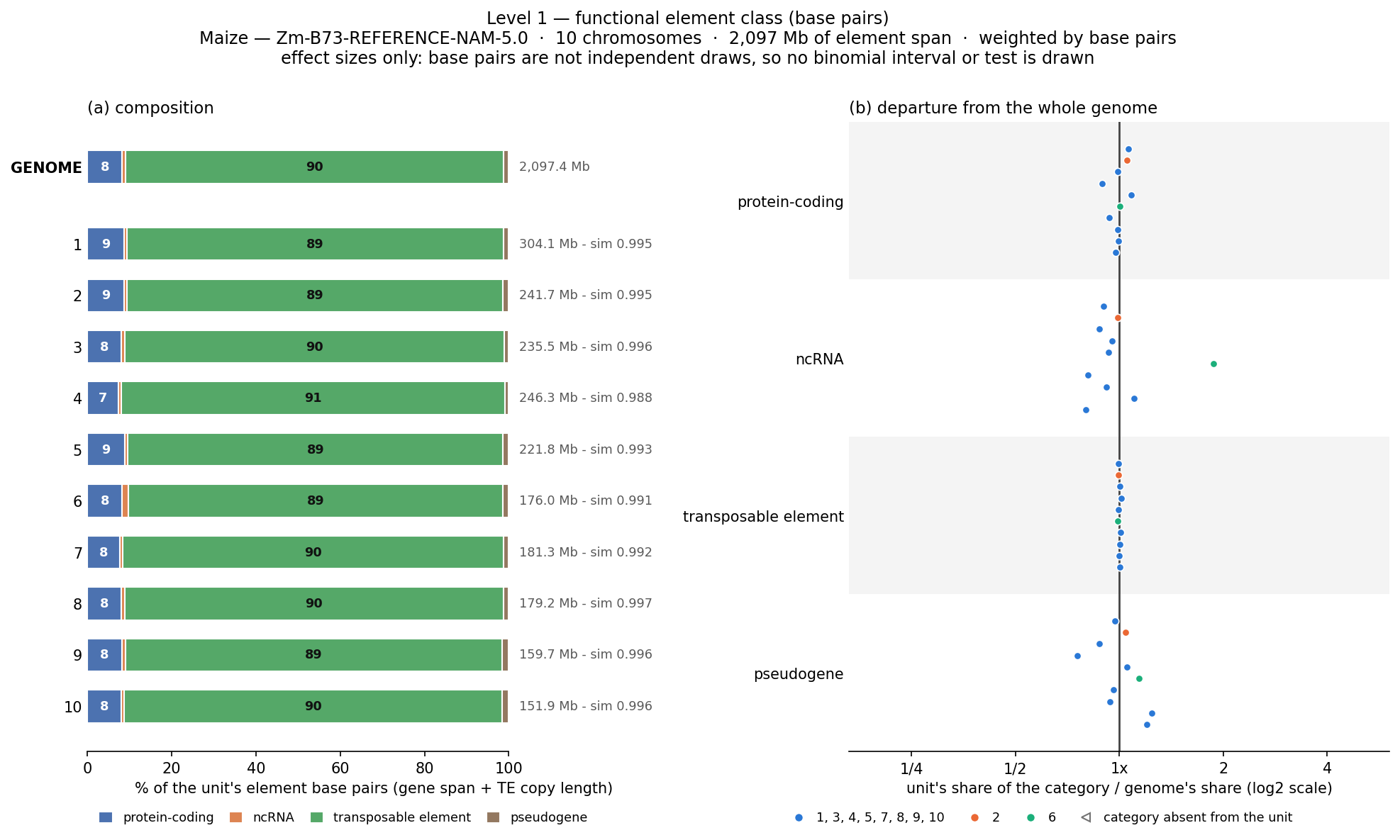}
\caption{Maize level~1 by base pairs, chromosome by chromosome, against the whole genome.
(a) Composition, with each chromosome's similarity to the genome. (b) Each chromosome's
share of each class divided by the genome's share. All ten chromosomes are near-copies of
the genome; the one visible departure, ncRNA on \texttt{6}, is the rRNA array of the
nucleolus organiser.}
\label{fig:mzl1}
\end{figure}

\begin{figure}[!htbp]
\centering
\includegraphics[width=\textwidth,height=0.8\textheight,keepaspectratio]{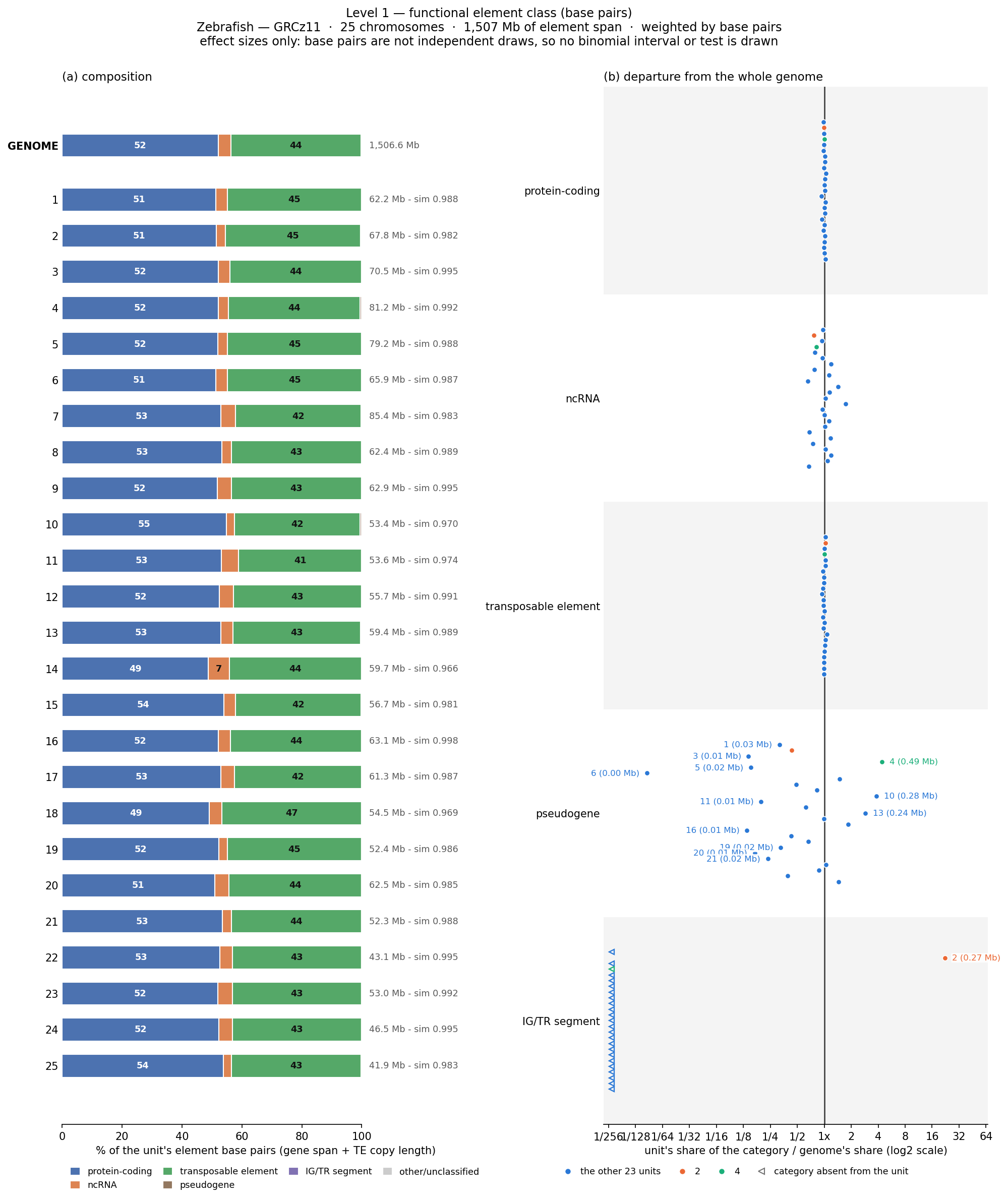}
\caption{Zebrafish level~1 by base pairs, chromosome by chromosome, against the whole
genome. (a) Composition, with each chromosome's similarity to the genome. (b) Each
chromosome's share of each class divided by the genome's share. Protein-coding genes and
TEs sit close to the genome line on all 25 chromosomes; only the very small pseudogene and
IG/TR classes scatter.}
\label{fig:drl1}
\end{figure}

\begin{figure}[!htbp]
\centering
\includegraphics[width=\textwidth]{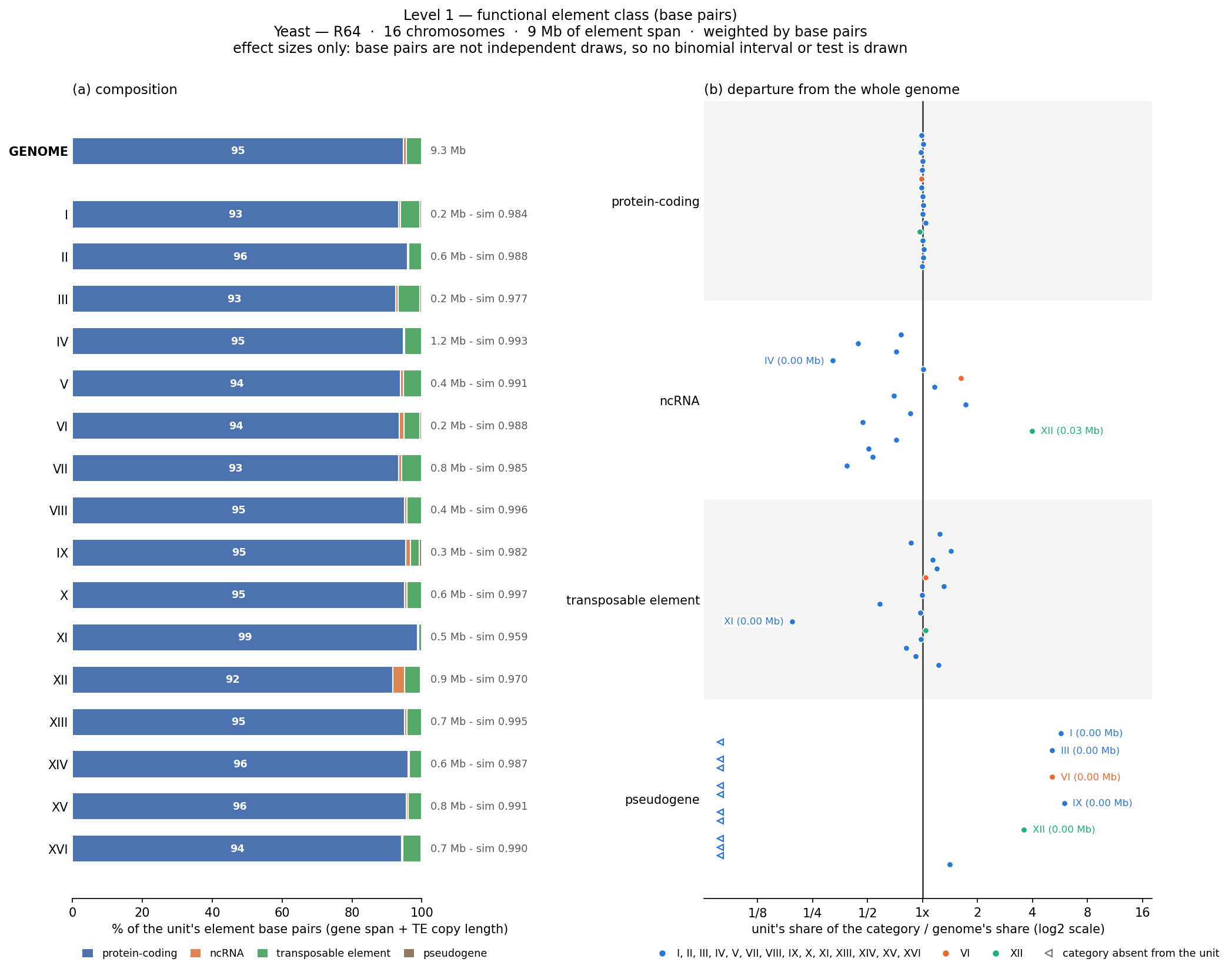}
\caption{Yeast level~1 by base pairs, chromosome by chromosome, against the whole genome.
(a) Composition, with each chromosome's similarity to the genome. (b) Each chromosome's
share of each class divided by the genome's share. Protein-coding genes hold 92--99\% of
every chromosome; the TE, ncRNA and pseudogene classes, each made of few elements, scatter
widely in relative terms but move the composition by only a few points.}
\label{fig:scl1}
\end{figure}

In particular, all chromosomes of maize, zebrafish and yeast show striking similarity with
their whole genome, and therefore with one another
(Figs.~\ref{fig:mzl1}--\ref{fig:scl1}). In each of the three, the classes that make up
most of the element length (TE and protein-coding genes in maize and zebrafish,
protein-coding genes in yeast) stay within 11\% of the genome's share on every chromosome;
only small classes move further, and they are too small to shift the composition by more
than a few points.

\emph{Maize} (Fig.~\ref{fig:mzl1}). Each of the ten chromosomes is 89--91\% TE and
7--9\% protein-coding by base pairs, against 89.7\% and 8.2\% for the genome, and each
scores $\sm = 0.988$--$0.997$, which makes maize the most homogeneous genome at this
level. The TE share of every chromosome is within 1\% of the
genome's (0.99--1.01 times), and the protein-coding share within 11\% (0.89--1.08 times).
The one visible departure is ncRNA on chromosome \texttt{6}, at 1.9 times the genome's
share: the rRNA array of the nucleolus organiser (\S\ref{sec:het-nc}). It is 1.5\% of the
chromosome's element length and leaves \texttt{6} at 0.991.

\emph{Zebrafish} (Fig.~\ref{fig:drl1}). The 25 chromosomes are 49--55\% protein-coding and
41--47\% TE, against 52.2\% and 43.6\% for the genome, at $\sm = 0.966$--$0.998$. Both
large classes sit within 7\% of the genome's share on every chromosome (protein-coding
0.93--1.05 times, TE 0.94--1.07 times). The three lowest scores have three different
causes: \texttt{14} (0.966) carries 7.0\% ncRNA, 1.7 times the genome's share; \texttt{18}
(0.969) is slightly TE-rich (46.6\%); and \texttt{10} (0.970) is ncRNA-poor (2.7\%). The
small classes are concentrated --- the IG/TR segments, 0.02\% of the genome, reach 0.4\% of
\texttt{2}, where the TR-$\alpha/\delta$ locus lies (\S\ref{sec:het-psig}) --- but
together they are under 0.2\% of the genome and barely touch the similarity.

\emph{Yeast} (Fig.~\ref{fig:scl1}). The 16 chromosomes are 92--99\% protein-coding by base
pairs, against 94.7\% for the genome, at $\sm = 0.959$--$0.997$; the protein-coding share
of every chromosome is within 4\% of the genome's (0.97--1.04 times). The small classes vary
more in relative terms because they are built from few elements: the genome has only 434 Ty
copies, so the TE share runs from 0.8\% on \texttt{XI} (0.19 times the genome's) to 6.2\%
on \texttt{III} (1.4 times), and \texttt{XI}'s scarcity of Ty copies is what makes it the
lowest-scoring chromosome. ncRNA reaches 3.4\% on \texttt{XII}, four times the genome's
share, from the rDNA locus and the snoRNA and tRNA genes around it (\S\ref{sec:het-nc}).

\begin{figure}[!htbp]
\centering
\includegraphics[width=\textwidth,height=0.8\textheight,keepaspectratio]{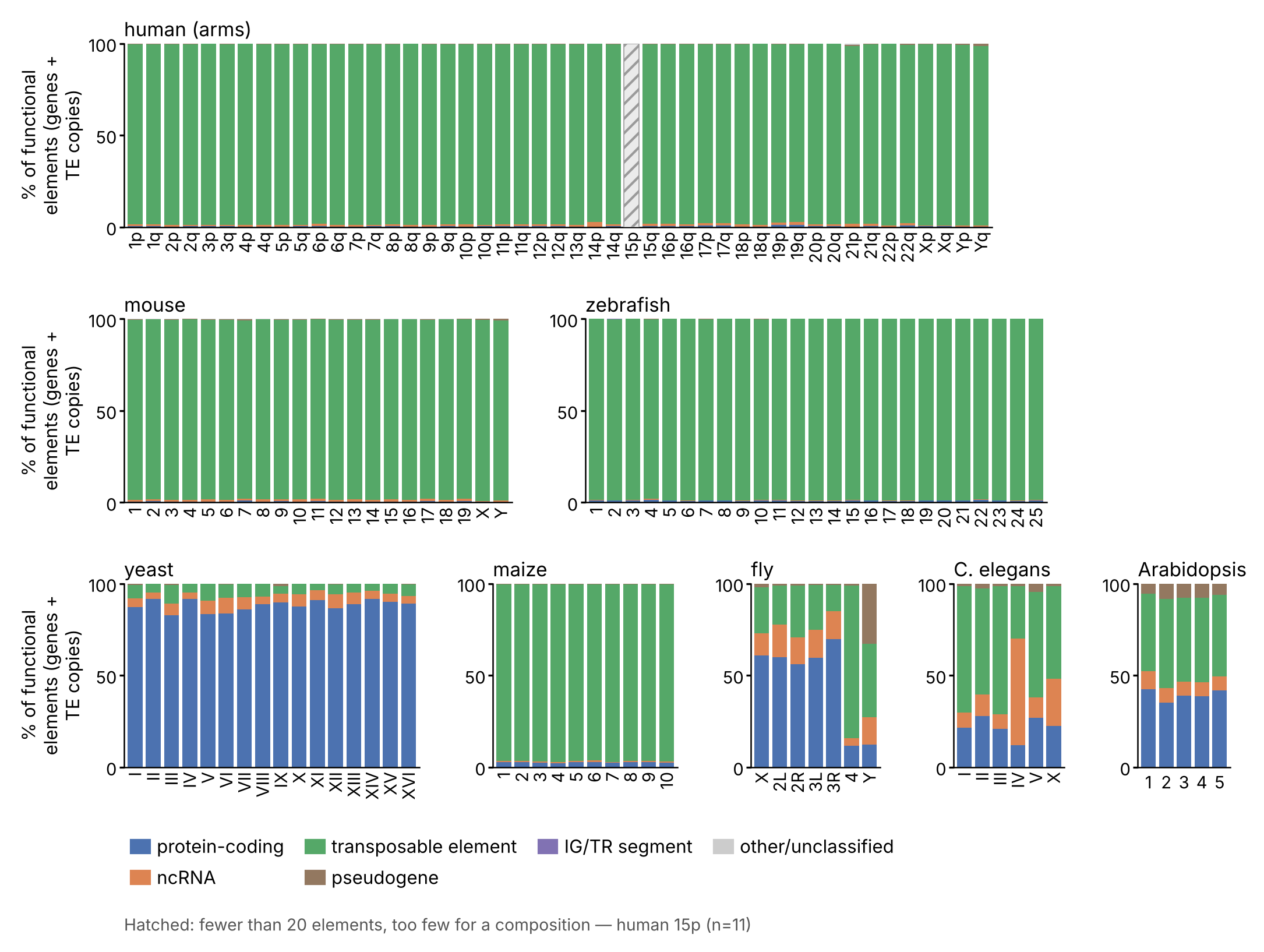}
\caption{The count check: level-1 composition per unit by element count. In human, mouse,
zebrafish and maize every bar is almost entirely TE copies, which is why counts cannot be
the main view at this level.}
\label{fig:l1unitcount}
\end{figure}

Excess dissimilarity at level~1 for whole units is 0.1 points in yeast, 0.4 in maize, 1.1
in zebrafish and 2.3--2.4 in \At, fly and \Ce{} --- small against 3.6 in mouse and 3.9 in
human. Counting elements instead of base pairs gives the check in
Fig.~\ref{fig:l1unitcount}: in human, mouse, zebrafish and maize every bar is then almost
entirely TE copies, which is why counts cannot be the main view at this level.

\subsection{Level 2-1, natural units: protein function is not organised by chromosome}
\label{sec:homog-fn}

\begin{figure}[!htbp]
\centering
\includegraphics[width=\textwidth]{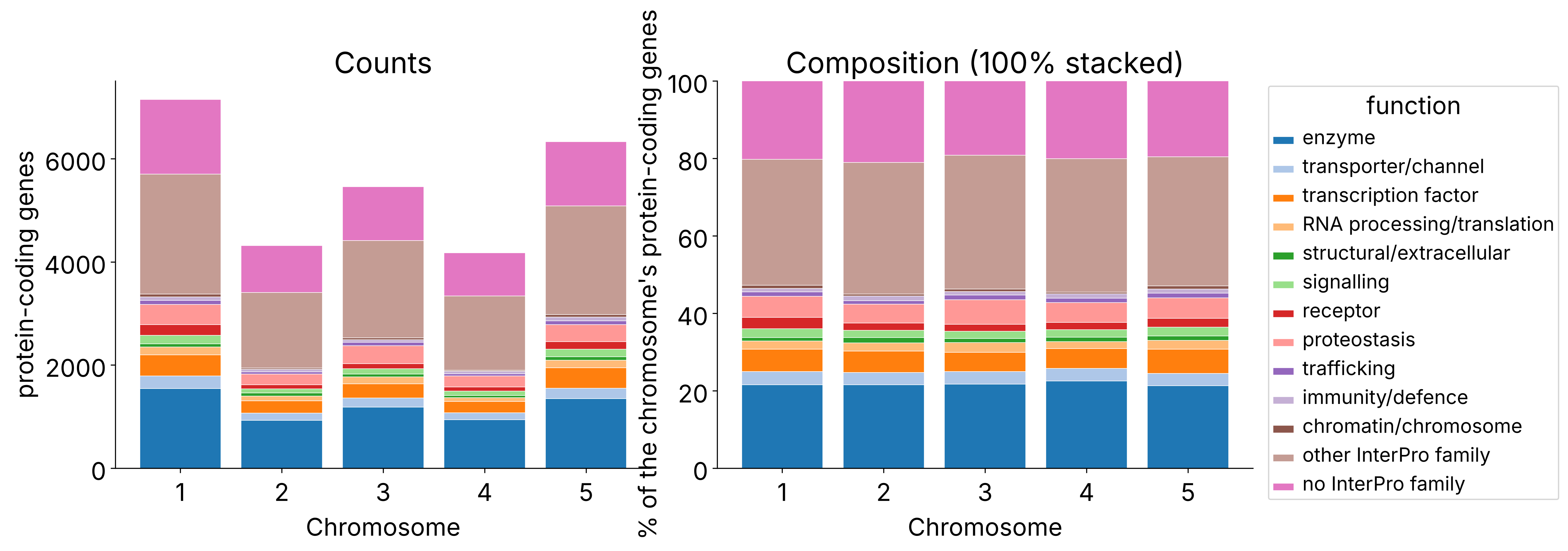}
\caption{\At{} level~2-1: broad protein function of every chromosome, as counts (left) and
as a 100\% stack (right). The five chromosomes differ two-fold in gene number and not at
all, to the eye, in functional composition.}
\label{fig:atfn}
\end{figure}

This is the most striking homogeneity in the study, and the one that separates the species
most cleanly. In \At{}, the named-function composition of the five chromosomes is so
alike that the 100\% stacks are hard to tell apart by eye (Fig.~\ref{fig:atfn}). Scored on
named functions only, the five chromosomes lie at $\sm = 0.967$--$0.977$. The match share
is 100\% for whole chromosomes, 70\% for halves and still 53\% for thirds.

\begin{figure}[!htbp]
\centering
\includegraphics[width=\textwidth,height=0.8\textheight,keepaspectratio]{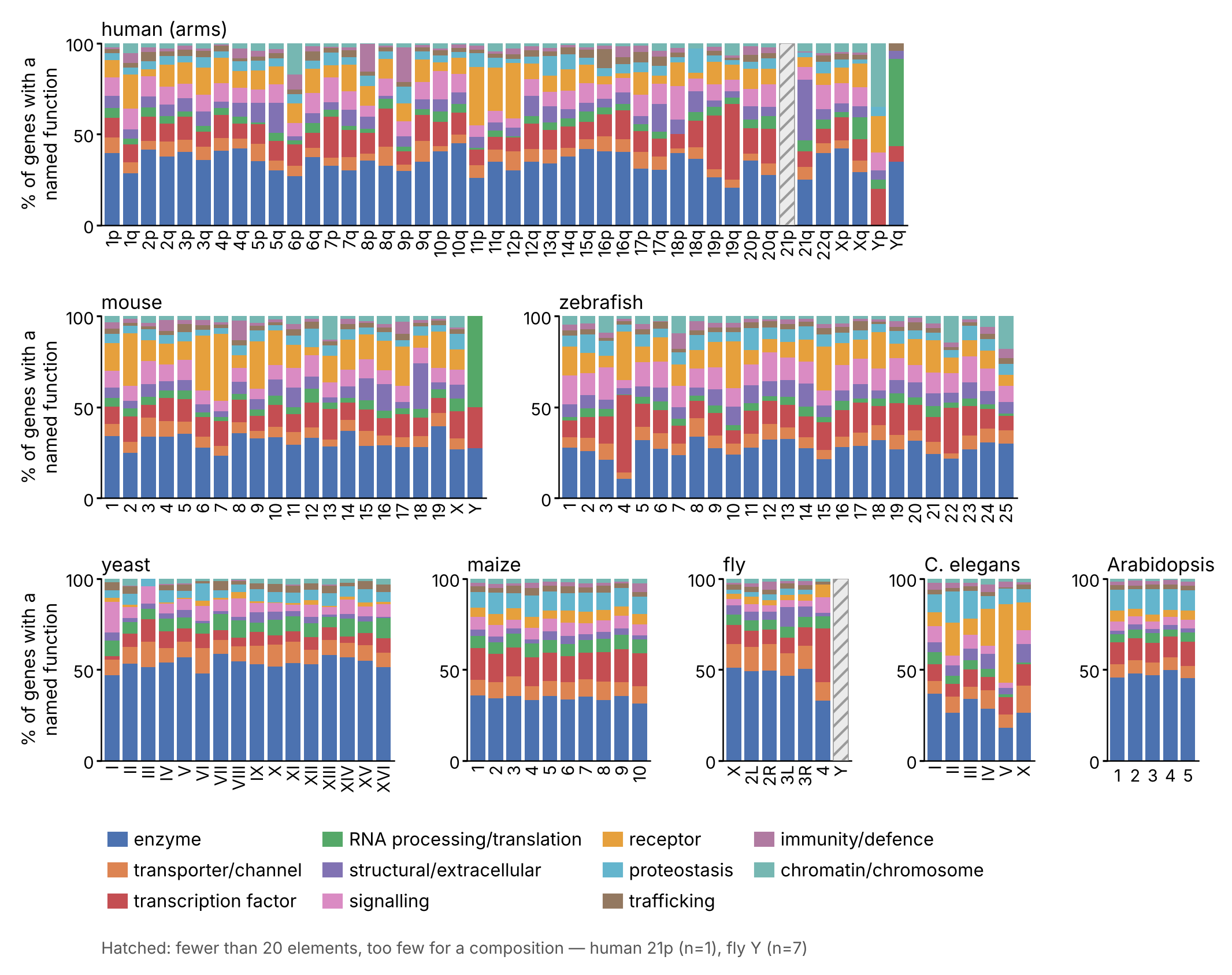}
\caption{Level-2-1 protein function per unit, named functions only, all eight genomes.
Flat within the plants, yeast and fly; the vertebrate and \Ce{} exceptions are in
Part~II.}
\label{fig:fnunit}
\end{figure}

Fly and maize are only slightly behind, each at 80\% match share for whole units, with fly
dot \texttt{4} the only real outlier. Yeast's sixteen chromosomes do not reach 0.95 on raw
similarity (31\%), but only because each holds a few hundred named genes. Yeast's excess
dissimilarity at this level is 0.0 points for whole chromosomes, the lowest of any genome,
and its $\Uadj$ is indistinguishable from zero ($-0.0015$, $p = 0.82$): its chromosomes
are random draws from one functional pool. Table~\ref{tab:fn} gives the genome-wide
function compositions that these chromosomes reproduce, and Fig.~\ref{fig:fnunit} shows
them unit by unit in all eight genomes.

\begin{table}[!htbp]
\centering
\caption{Level-2-1 protein function, as a percentage of protein-coding genes with a
\emph{named} function (the six largest named buckets shown). ``named'' is the
denominator; ``cov.'' is that denominator as a percentage of all protein-coding genes.}
\label{tab:fn}
\small
\begin{tabular}{lrrrrrrrr}
\toprule
species & enzyme & TF & receptor & transp. & signal. & proteost. & named & cov. \\
\midrule
yeast       &  54.3 &  6.8 &  1.4 &  8.9 &  6.2 &  5.8 &  3{,}152 & 48.0 \\
fly         &  49.1 &  9.4 &  3.2 & 13.2 &  3.6 &  2.9 &  7{,}942 & 56.9 \\
\At         &  46.7 & 12.1 &  4.8 &  7.0 &  4.6 & 11.7 & 12{,}712 & 46.4 \\
maize       &  34.4 & 16.1 &  5.6 &  9.0 &  6.4 & 10.1 & 11{,}767 & 30.1 \\
\Ce         &  26.7 &  9.1 & 22.5 &  8.7 &  5.6 & 10.8 &  8{,}739 & 43.8 \\
human       &  34.0 & 13.4 & 11.6 &  6.5 &  9.4 &  4.7 & 12{,}619 & 62.8 \\
mouse       &  30.4 & 11.7 & 18.8 &  6.2 &  8.6 &  4.8 & 13{,}837 & 63.6 \\
zebrafish   &  26.2 & 14.9 & 13.2 &  6.8 & 11.9 &  6.4 & 15{,}255 & 60.8 \\
\bottomrule
\end{tabular}
\end{table}

\subsection{Level 2-2, natural units: ncRNA class in \At{} and the mammals}
\label{sec:homog-nc}

\begin{figure}[!htbp]
\centering
\includegraphics[width=\textwidth,height=0.8\textheight,keepaspectratio]{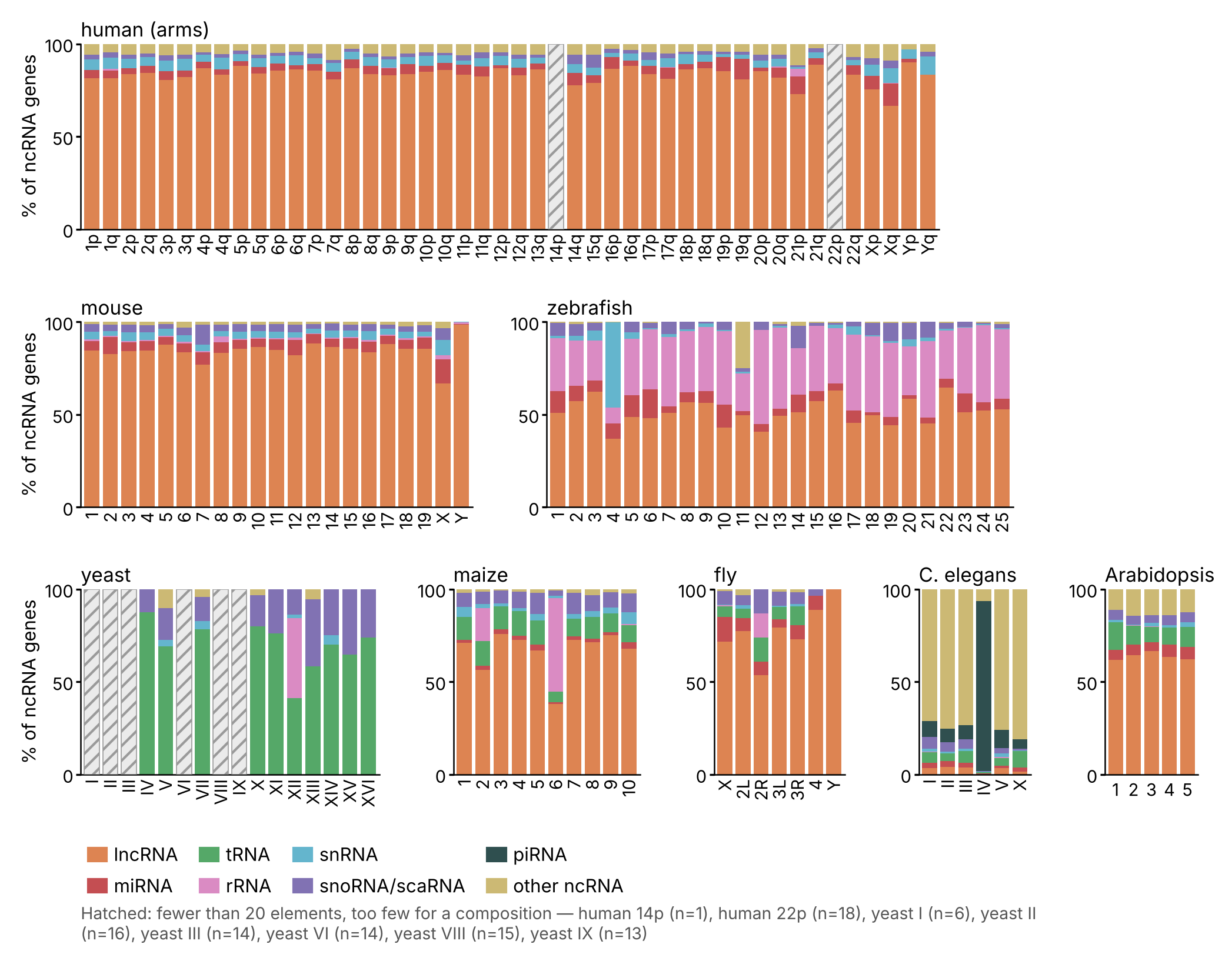}
\caption{Level-2-2 ncRNA class per unit, all eight genomes. Flat within \At, human and
mouse; the exceptions --- \Ce{} \texttt{IV} (piRNA), zebrafish \texttt{4} (snRNA), maize
\texttt{6} (rRNA) and yeast \texttt{XII} (rRNA) --- are named in Part~II.}
\label{fig:ncunit}
\end{figure}

Table~\ref{tab:ncrna} gives the ncRNA classes. The ncRNA class composition is homogeneous
where the ncRNA annotation is dominated by one broadly spread class. \At{} is the clearest
case: every chromosome matches its genome (median $\sm = 0.976$, match share 100\% for
whole chromosomes and still 14\% at deciles) and $\Uadj = 0.0035$, the lowest of any genome.
Human and mouse, whose ncRNA annotation is 83--84\% lncRNA, are next: 81\% of human arms
and 90\% of mouse chromosomes match, and their excess is only 0.8--1.4 points at every cut
--- the lowest of any genome at this level. Their remaining structure is a few arrays
(\S\ref{sec:het-nc}). The ncRNA classes are shown per unit in Fig.~\ref{fig:ncunit}.

\begin{table}[!htbp]
\centering
\caption{Level-2-2 ncRNA class: gene counts. Human, mouse and zebrafish tRNA genes are not
annotated as Ensembl genes; piRNA loci are annotated only in \Ce.}
\label{tab:ncrna}
\small
\setlength{\tabcolsep}{4pt}
\begin{tabular}{lrrrrrrrrr}
\toprule
species & lncRNA & miRNA & tRNA & rRNA & snRNA & sno/scaRNA & piRNA & other & total \\
\midrule
yeast     &      0 &     0 &   275 &    22 &     6 &     77 &       0 &     17 &    397 \\
fly       & 2{,}488 &   260 &   290 &    99 &    30 &    275 &       0 &     55 & 3{,}497 \\
\At       & 3{,}559 &   326 &   625 &     4 &    82 &    287 &       0 &    729 & 5{,}612 \\
maize     & 5{,}321 &   164 &   914 &   884 &   241 &    642 &       0 &    138 & 8{,}304 \\
\Ce       &    294 &   261 &   612 &    20 &   129 &    346 & 15{,}363 & 7{,}764 & 24{,}789 \\
zebrafish & 3{,}247 &   431 &     0 & 1{,}919 &   487 &    253 &       0 &    101 & 6{,}438 \\
mouse     & 32{,}884 & 2{,}201 &   0 &   354 & 1{,}381 & 1{,}558 &       0 &    587 & 38{,}965 \\
human     & 34{,}866 & 1{,}878 &   0 &    47 & 1{,}901 & 1{,}037 &       0 & 2{,}224 & 41{,}953 \\
\bottomrule
\end{tabular}
\end{table}

\subsection{Level 2-3, natural units to deciles: the maize repeat landscape}
\label{sec:homog-te}

\begin{figure}[!htbp]
\centering
\includegraphics[width=\textwidth]{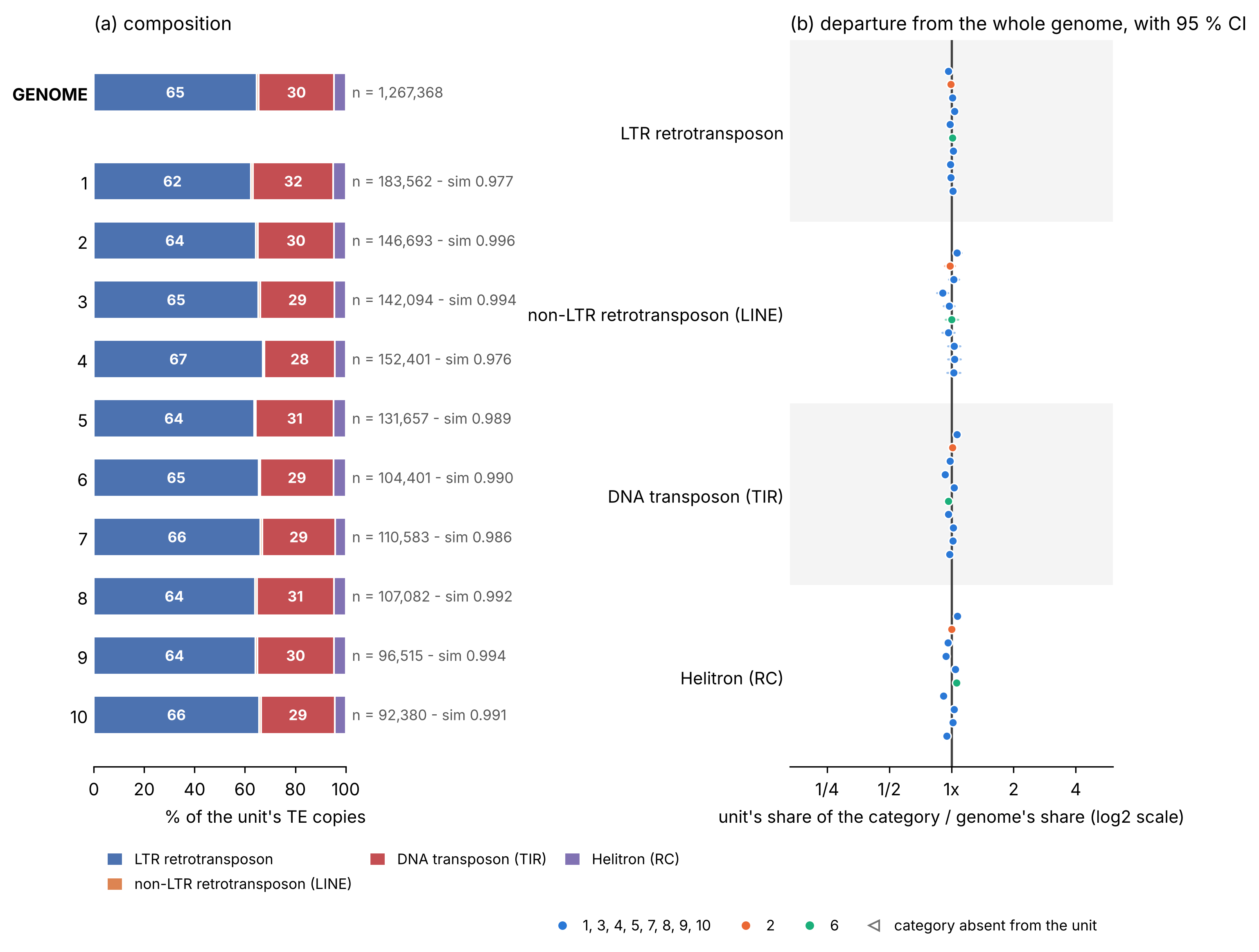}
\caption{Maize level-2-3 TE order, chromosome by chromosome, against the genome, by TE copy count. The
departure panel (b) shows no ratio beyond a few percent. Many cells are nevertheless
``significant'', because maize's copy counts are large enough to make any difference
detectable --- the reason this paper reads similarity beside a sampling floor rather than
$p$-values.}
\label{fig:maizete}
\end{figure}

The TE share of each assembly ranges from 3.4\% (yeast) and 6.1\% (fly) through 10.7\%
(\Ce) and 20.9\% (\At) and 41--49\% (mouse, human, zebrafish) to 88.3\% in maize, where TEs
lie largely as nested LTR retrotransposons between genes \cite{sanmiguel,schnable}. Maize's
TE order composition is also the most homogeneous composition measured anywhere in this
study (Fig.~\ref{fig:maizete}). Maize's
dominant order is LTR (64.7\% of copies), zebrafish's DNA/TIR (77.4\%), \At's Helitron
(41.5\%) and fly's DNA/TIR (55.6\% of copies, but LTR leads by base pairs). Human and mouse
have most copies in SINEs and most base pairs in LINEs. By base pairs every maize
chromosome scores $\sm = 0.996$ at the median against the genome. The homogeneity
survives cutting: every half and third matches at 0.95, as do 90--92\% of quarters and
fifths and 82\% of deciles, each only 15--30~Mb long. Zebrafish is the only other genome
that comes close (96\% of whole chromosomes, 48\% of deciles). The mammals sit at 59--70\%
for whole units and fall to 25--29\% at deciles. $\Uadj$ is 0.0005 in maize, the lowest
value at any level in any genome.

\begin{figure}[!htbp]
\centering
\includegraphics[width=\textwidth,height=0.8\textheight,keepaspectratio]{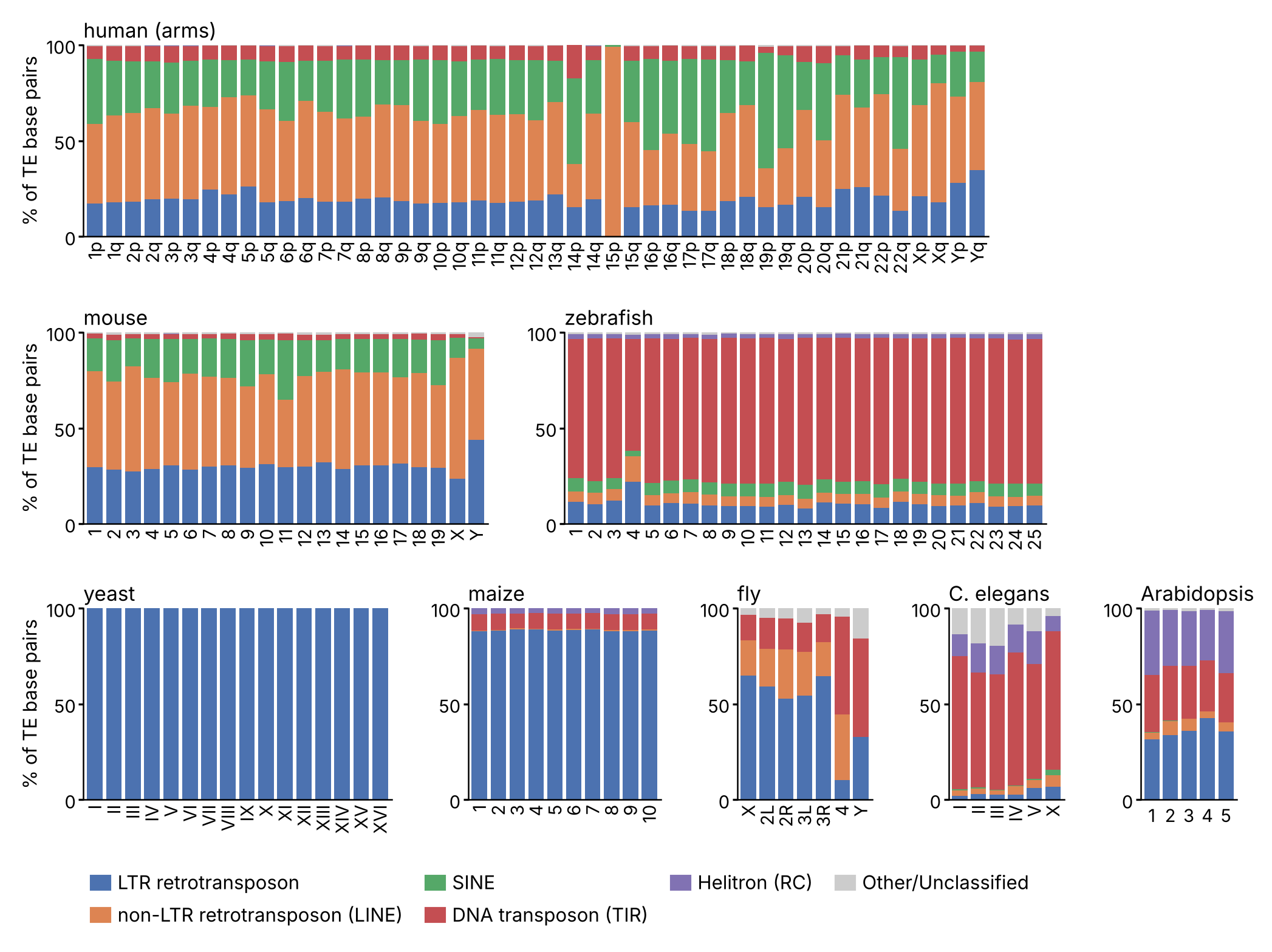}
\caption{TE order composition per unit by base pairs, all genomes. Flat within maize and
zebrafish; there are four distinct repeat landscapes between species.}
\label{fig:teunit}
\end{figure}

Seen unit by unit across all eight genomes (Fig.~\ref{fig:teunit}), TE order is flat within
maize and zebrafish, while the genomes themselves fall into four distinct repeat
landscapes.

\subsection{Level 3-1, natural units: EC class is interchangeable everywhere}
\label{sec:homog-ec}

\begin{figure}[!htbp]
\centering
\includegraphics[width=\textwidth,height=0.8\textheight,keepaspectratio]{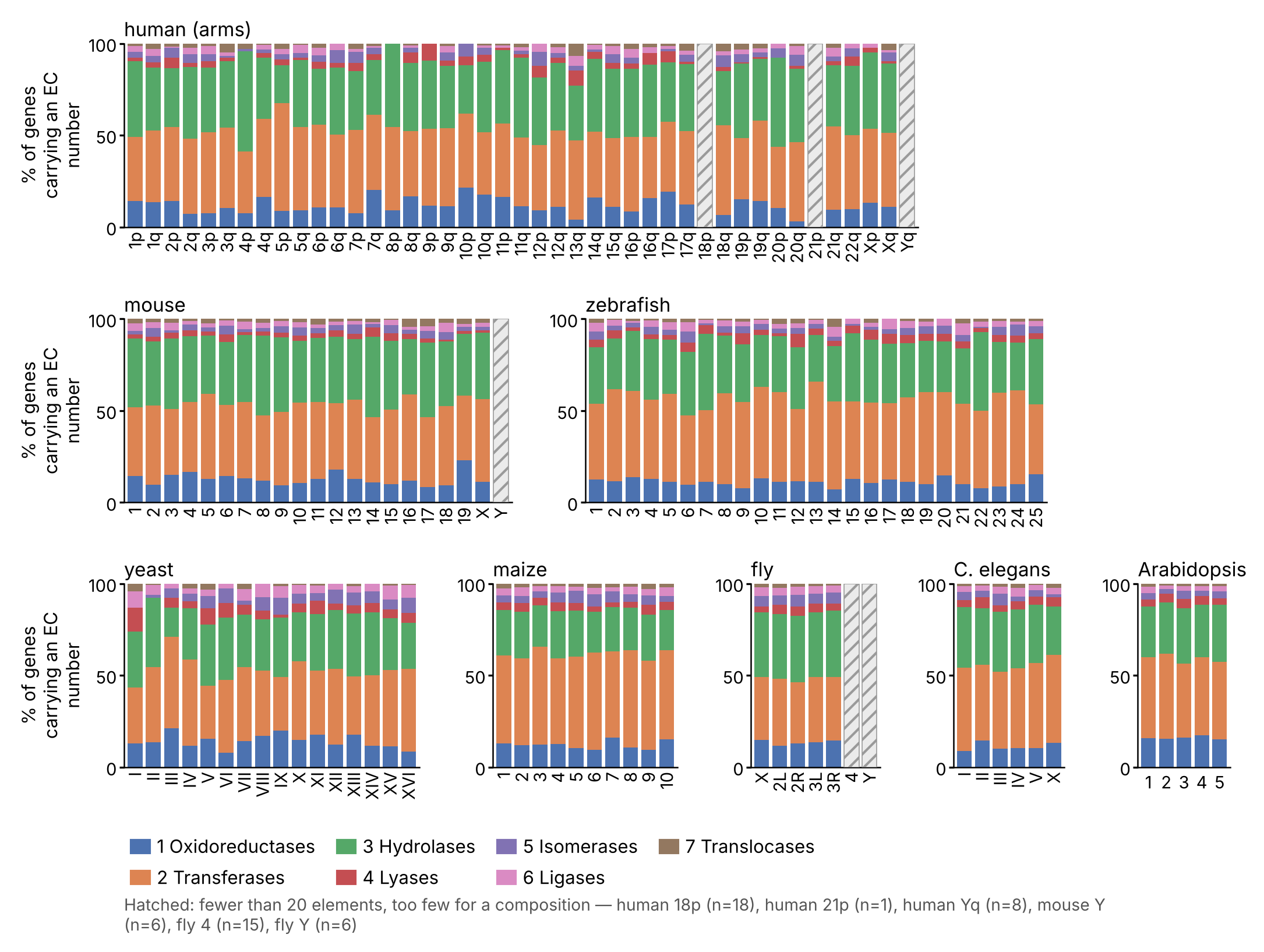}
\caption{Level-3-1 EC class per unit, all genomes. No chromosome-scale organisation in any
species.}
\label{fig:ecunit}
\end{figure}

\begin{figure}[!htbp]
\centering
\includegraphics[width=\textwidth]{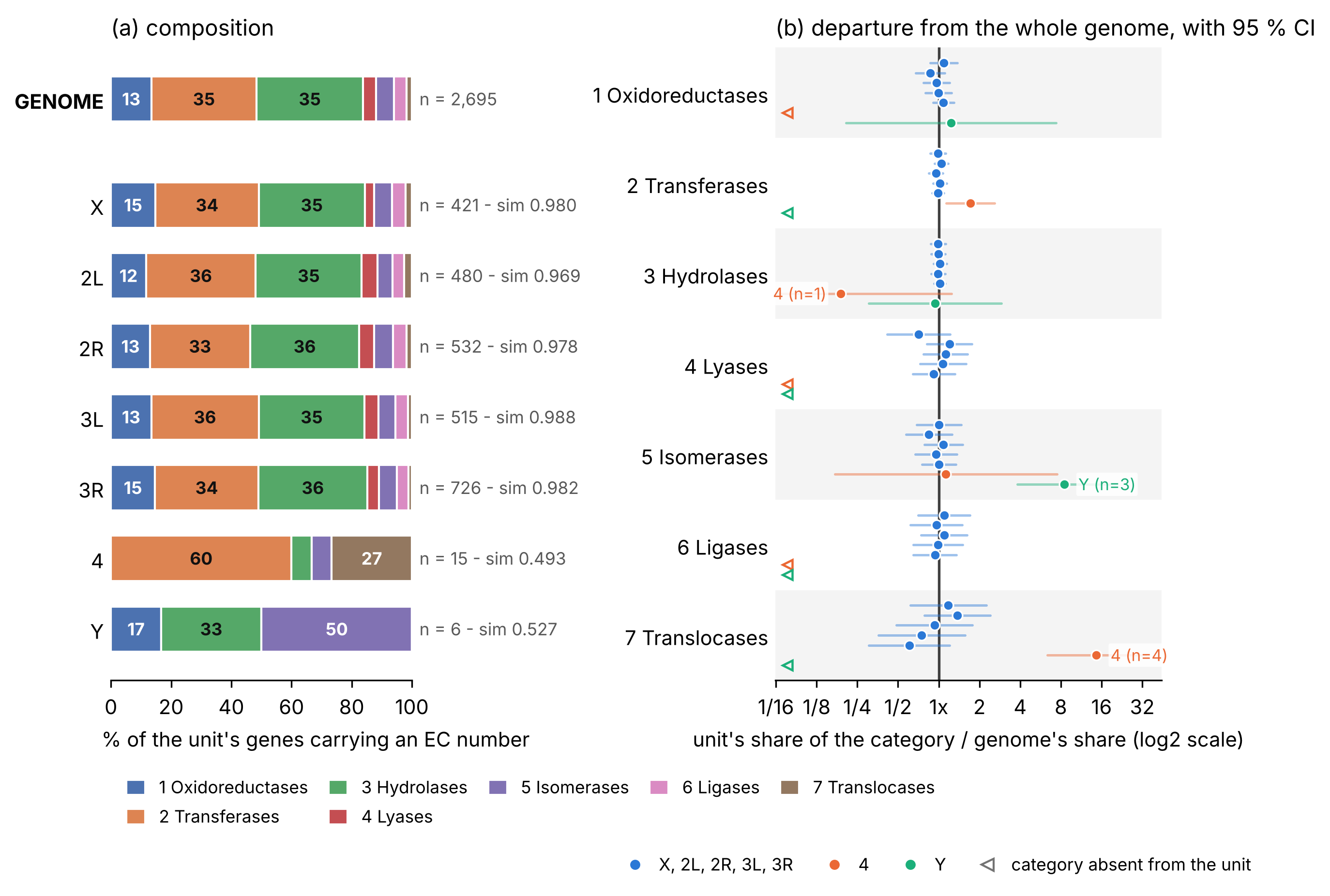}
\caption{Fly EC class, arm by arm, against the genome. Every arm, including chromosome
\texttt{4} where it has enough enzymes to score, sits on the genome line.}
\label{fig:flyec}
\end{figure}

One level deeper, homogeneity becomes universal. The seven IUBMB reaction classes are
distributed across chromosomes as if at random in all eight genomes (Table~\ref{tab:ec},
Fig.~\ref{fig:ecunit}). $\Uadj$ is within noise of zero everywhere ($-0.0015$ to
$0.0030$). No species reaches $p < 0.04$, and none would survive correction for eight
tests. Excess dissimilarity is within $\pm 2.5$ points in every genome at every cut:
whatever fails to reach 0.95 at this level fails because it is small. Fly's arms score
0.969--0.988 and \At's chromosomes 0.961--0.983, and both have a 100\% match share for
whole units (Fig.~\ref{fig:flyec}).

\begin{table}[!htbp]
\centering
\caption{Level-3-1 EC class as a percentage of EC-carrying genes. Transferases plus
hydrolases are 70--77\% of every genome.}
\label{tab:ec}
\small
\begin{tabular}{lrrrrrrrr}
\toprule
species & 1 Ox. & 2 Transf. & 3 Hydrol. & 4 Lyases & 5 Isom. & 6 Lig. & 7 Transloc. & total \\
\midrule
\At        & 15.9 & 43.0 & 29.1 & 4.3 & 3.5 & 2.7 & 1.5 & 6{,}370 \\
fly        & 13.5 & 34.9 & 35.5 & 4.3 & 5.9 & 4.1 & 1.8 & 2{,}695 \\
yeast      & 13.6 & 39.7 & 30.3 & 4.6 & 5.5 & 4.9 & 1.3 & 1{,}790 \\
\Ce        & 11.0 & 44.3 & 31.4 & 4.5 & 3.8 & 3.5 & 1.5 & 2{,}430 \\
mouse      & 12.8 & 40.3 & 36.5 & 2.6 & 3.1 & 2.7 & 2.0 & 4{,}510 \\
human      & 12.1 & 39.9 & 37.1 & 2.9 & 3.3 & 2.7 & 1.9 & 4{,}623 \\
maize      & 12.0 & 49.4 & 24.1 & 4.1 & 4.8 & 3.5 & 2.0 & 4{,}365 \\
zebrafish  & 11.2 & 45.6 & 31.7 & 3.8 & 3.2 & 3.3 & 1.3 & 4{,}293 \\
\bottomrule
\end{tabular}
\end{table}

The homogeneity also holds \emph{between} genomes. The class ranking is identical in all
eight, and the median pairwise distance between two genomes' EC compositions is a third of
that between their function compositions. Reaction chemistry is the most conserved layer of
the hierarchy, both along a genome and across 1.5 billion years of divergence.

\subsection{Levels 3-2 and 4-1, natural units: the plants and fly stay close}
\label{sec:homog-ecsub}

At EC subclass most genomes reject a shared per-chromosome composition (Part~II). The effect
is small, however, and ranks the species in the order of the hypothesis. Whole units match
the genome at 0.95 in 60\% of \At{} chromosomes, 40\% of fly arms and 30\% of maize
chromosomes, and in none of the units of any other genome. Excess dissimilarity for whole
units is at most 3.7 points everywhere, lowest in maize, yeast and fly. TF DNA-binding
domain separates the groups cleanly: the two plants sit at no more than 3--3.5 points of
excess at any cut, the vertebrates and \Ce{} at 7--27.5, and maize is the only genome with
any whole chromosome at 0.95.

\subsection{Finer cuts: homogeneity down to the decile}
\label{sec:homog-cuts}

Chromosome-scale homogeneity can hide internal structure --- two unlike halves that
average to the genome. The equal-partition scans show where it does not.

\begin{figure}[!htbp]
\centering
\includegraphics[width=\textwidth]{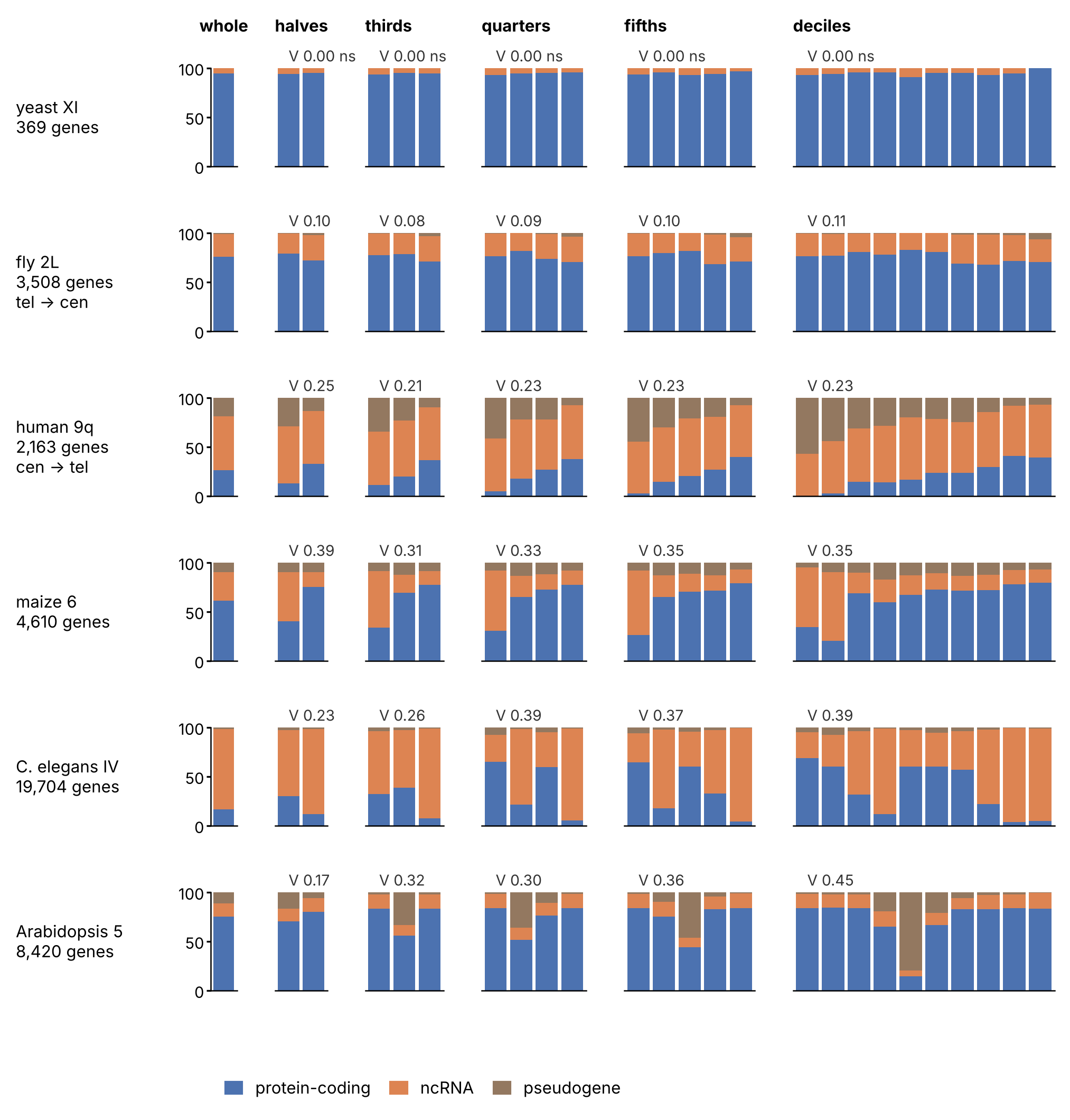}
\caption{\textbf{Equal cuts of six natural units, gene classes only.} Each row is one unit
cut into 1, 2, 3, 4, 5 and 10 parts of equal base-pair length; each bar is one part's
composition by gene count (protein-coding, ncRNA, pseudogene; TE copies are not counted,
\S\ref{sec:scores}). Parts run from the low-coordinate end; for \texttt{2L} that is the
telomere, for \texttt{9q} the centromere. Above each cut is the bias-corrected Cram\'er's
$V$ of the parts~$\times$~classes table (``ns'': permutation $p \geq 0.05$). Yeast
\texttt{XI} does not change at any cut; fly \texttt{2L} changes gently; human \texttt{9q}
runs as a gradient from a pseudogene-rich pericentromere to a gene-rich distal end; maize
\texttt{6} carries the rRNA genes of its nucleolus organiser in the first fifth;
\Ce{} \texttt{IV} carries piRNA loci in its two arm domains; and \At{} \texttt{5} has a
single pseudogene-rich block at its centromere.}
\label{fig:cuts}
\end{figure}

Maize is the clearest case at level~1 by base pairs: its match share is 100\% for halves,
97\% for thirds and still 75\% for deciles, each 15--30~Mb long, with excess at most 3.4
points. Yeast follows (91\% for halves, 61\% for deciles --- pieces of 50--150~kb), then
zebrafish (90\% for halves, 43\% for deciles). The gene-class partition tests show the same from the side of the genes
(Fig.~\ref{fig:cuts}, Table~\ref{tab:partitions}). Yeast \texttt{XI}'s gene composition
does not change at any cut ($V = 0.00$, $p \geq 0.66$ at every cut;
Fig.~\ref{fig:cuts}, top row), and across all sixteen yeast chromosomes only 31\% of
unit~$\times$~cut tests reject a shared composition, against 66--99\% in the other
genomes. Fly arm \texttt{2L} changes only gently along its length ($V = 0.08$--$0.11$):
protein-coding genes are 76--83\% of genes in the six distal deciles and 68--72\% in the
four proximal ones, where ncRNA genes take their place.

A random window gives the same answer without reference to units
(Table~\ref{tab:seglen}). A randomly placed maize segment reaches the genome's level-1
composition at the median once it is 0.9~Mb long, $2.7\times$ the length a label-shuffled
genome needs. Zebrafish and human need $2.5$--$2.6\times$ their shuffle. In these genomes
most of the length requirement is counting noise.

\begin{table}[!htbp]
\centering
\caption{The shortest random segment that resembles the whole genome at level~1 by base
pairs at the median (``lenient'') and at the 5th percentile (``strict''), against the
lenient length of a genome-wide label shuffle. ``none'': not reached at any length up to
the longest unit.}
\label{tab:seglen}
\small
\begin{tabular}{lrrrr}
\toprule
species & lenient & strict & shuffle & lenient / shuffle \\
\midrule
maize       & 0.91~Mb  & 54~Mb   & 0.34~Mb & $2.7\times$ \\
zebrafish   & 10.6~Mb  & 60~Mb   & 4.3~Mb  & $2.5\times$ \\
human       & 38.0~Mb  & none    & 14.5~Mb & $2.6\times$ \\
mouse       & 58.5~Mb  & none    & 11.9~Mb & $4.9\times$ \\
yeast       & 65~kb    & 0.71~Mb & 14~kb   & $4.7\times$ \\
fly         & 7.9~Mb   & 19.9~Mb & 0.52~Mb & $15\times$ \\
\At         & 20.0~Mb  & 21.1~Mb & 0.43~Mb & $47\times$ \\
\Ce         & 9.4~Mb   & 19.3~Mb & 0.16~Mb & $60\times$ \\
\bottomrule
\end{tabular}
\end{table}

\section{Part II --- Heterogeneity: where units of the genome differ from the whole genome as well as from each other}
\label{sec:heterog}

\begin{figure}[!htbp]
\centering
\includegraphics[width=\textwidth]{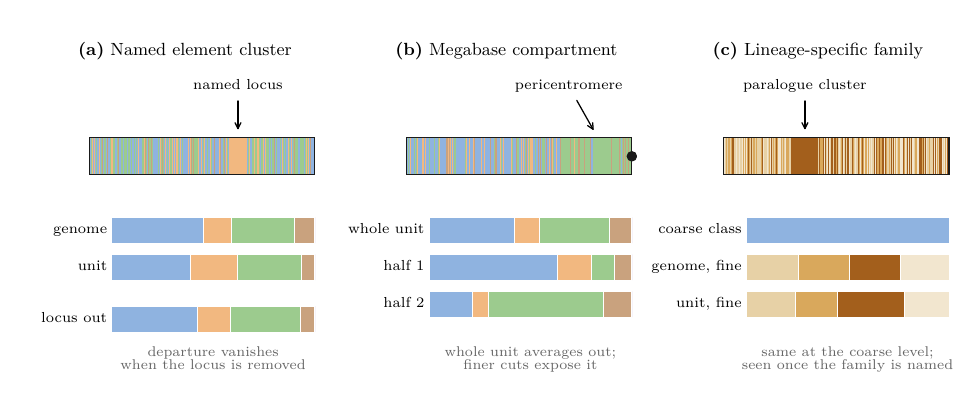}
\caption{\textbf{Three ways a unit departs from its genome} (illustration). Each strip is
one chromosome drawn as a run of elements coloured by class. (a)~A named element cluster,
such as a ribosomal-RNA array, shifts the composition of its unit; removing the locus
restores the match. (b)~A broad compartment, such as TE-rich pericentromeric
heterochromatin (dot: centromere), is averaged away in the whole unit and exposed when the
unit is cut into halves. (c)~A paralogue cluster of one lineage-specific family is
invisible while its genes are pooled into one coarse class and appears once the
classification names the family.}
\label{fig:mechanisms}
\end{figure}

Part~I's homogeneity has three kinds of exception, and each leaves a different
signature (Fig.~\ref{fig:mechanisms}). \emph{Named element clusters} --- tandem arrays,
nucleolus organisers, IG/TR loci --- are local: they make the unit that carries them depart
from the genome, and the departure dissolves when their elements are removed.
\emph{Megabase-scale compartments} --- pericentromeric heterochromatin, isochores and
recombination domains --- are broad: a whole-unit average can hide them, and they appear
only once the unit is cut. \emph{Lineage-specific gene families} are invisible at coarse
levels, where the family is pooled with the rest of its class, and appear only at levels 3
and 4, where the family is named. We take the exceptions in the same order as Part~I,
coarse to fine, and name the mechanism behind each.

\subsection{Level 1, natural units}
\label{sec:het-l1}

\begin{figure}[!htbp]
\centering
\includegraphics[width=\textwidth,height=0.82\textheight,keepaspectratio]{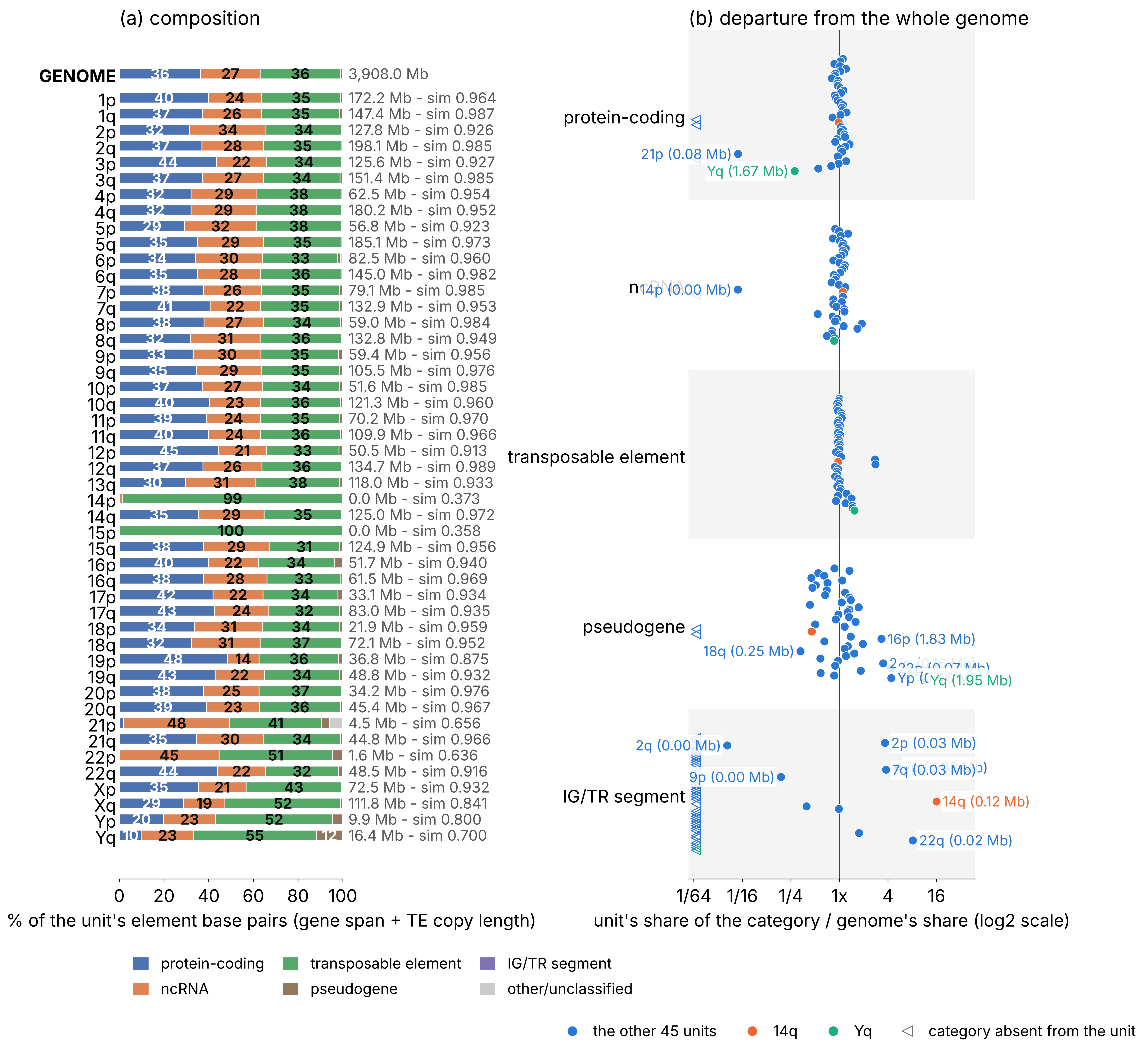}
\caption{Human level~1, arm by arm, against the genome. The arms spread far more widely
around the genome line than fly's (Fig.~\ref{fig:flyl1}).}
\label{fig:hsl1}
\end{figure}

\paragraph{Human and mouse are the least homogeneous genomes by base pairs.} Only 66\% of
human's major arms and 80\% of mouse's chromosomes reach 0.95 at level~1
(Figs.~\ref{fig:tree},~\ref{fig:hsl1}). Excess dissimilarity is 3.6--3.9 points for whole
units, rising to 8--9 at deciles, larger than in every other genome except \At{} once cut.
The least similar major units are mouse \texttt{X} (0.809), human \texttt{Xq} (0.841),
\texttt{19p} (0.875), \texttt{12p} (0.913) and \texttt{22q} (0.916). Both \texttt{X}s are
TE-rich (human \texttt{Xq} 62.7\% TE by base pairs, the highest of any human arm) and
gene-poor, in line with the long-noted enrichment of LINE-1 elements on the mammalian
\texttt{X} \cite{bailey}; \texttt{19p} is the reverse, gene-dense and TE-poor. The special units are the
extremes: human \texttt{14p} 0.373, \texttt{15p} 0.358, \texttt{21p} 0.655, \texttt{22p}
0.635 and \texttt{Yq} 0.699; mouse \texttt{Y} 0.533.

\begin{figure}[!htbp]
\centering
\includegraphics[width=\textwidth]{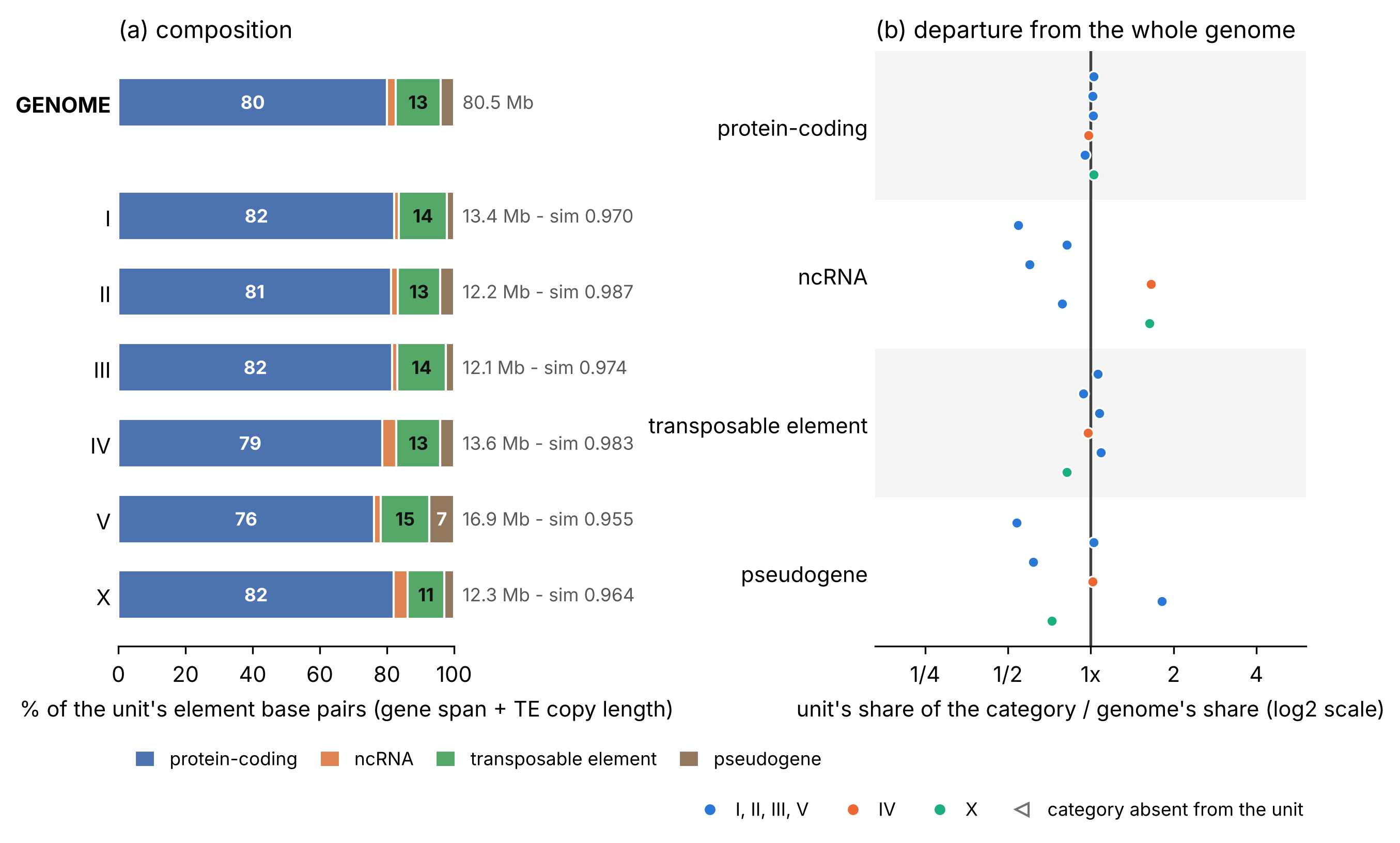}
\caption{\Ce{} level~1, chromosome by chromosome. By count, chromosome \texttt{IV}'s
piRNA excess, and every other chromosome's matching deficit, make each chromosome unlike
the genome; by base pairs the effect nearly vanishes.}
\label{fig:cel1}
\end{figure}

\paragraph{\Ce{}: heterogeneous by count, not by base pairs.} By count, \Ce{} is the most
heterogeneous genome at level~1 by far ($\Uadj = 0.109$, against 0.002--0.041 elsewhere;
Fig.~\ref{fig:cel1}): its chromosomes score only 0.674 (\texttt{IV}) to 0.861
(\texttt{II}), and no part reaches 0.95 at any cut beyond whole chromosomes. The cause is
one class on one chromosome. Chromosome \texttt{IV} holds 96\% of the genome's 15{,}363
piRNA loci \cite{ruby}. Because piRNAs are 16\% of all \Ce{} elements, every other
chromosome is short of them and \texttt{IV} has too many. Dropping the two most divergent
chromosomes takes $\Uadj$ from 0.109 to 0.021. By base pairs the same loci are a few
hundred kilobases and the effect nearly vanishes (\S\ref{sec:homog-l1}). The count and base-pair views
disagree here because the class in question is tiny and numerous. That is the case the
count check exists for.

\paragraph{The remaining level-1 exceptions are special units and named loci.} Fly's
\texttt{4} and \texttt{Y} (\S\ref{sec:homog-l1}) are heterochromatic special units, and
\texttt{4} is the most TE-rich fly unit (27.7\% of its assembly length). \At{} \texttt{1}
and \texttt{2} fall just short because of their large pericentromeres and, on \texttt{2},
the NOR (\S\ref{sec:het-cuts}). Zebrafish \texttt{4}, maize \texttt{6} and yeast
\texttt{XII} pass at level~1 by base pairs, but carry the snRNA cluster, the 45S NOR and
the \texttt{RDN} rDNA locus that dominate level~2-2 (\S\ref{sec:het-nc}).

\subsection{Level 2-1, natural units: zinc fingers and receptors}
\label{sec:het-fn}

\begin{figure}[!htbp]
\centering
\includegraphics[width=\textwidth,height=0.82\textheight,keepaspectratio]{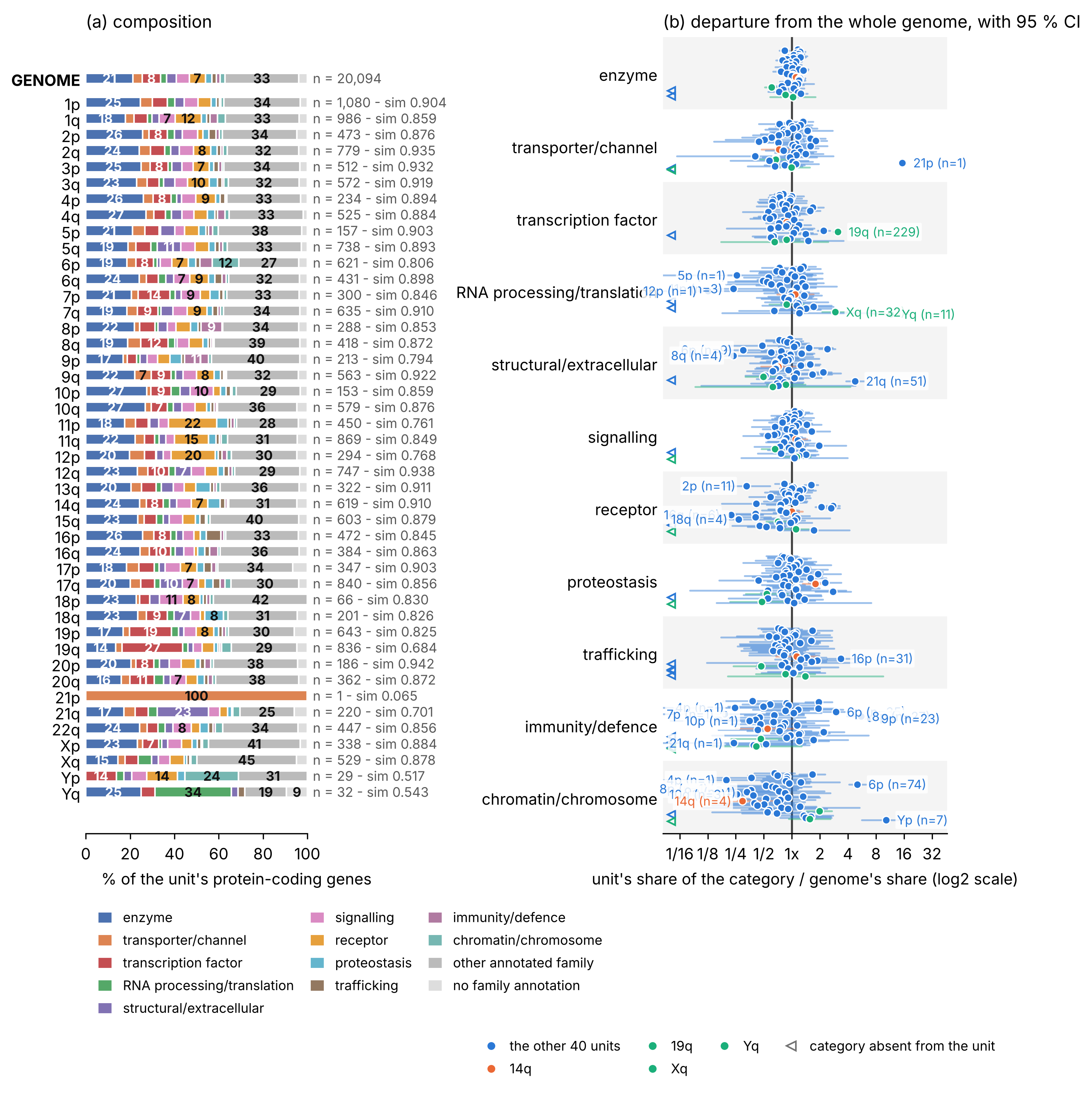}
\caption{Human level-2-1 function, arm by arm. \texttt{19q}'s transcription-factor excess
is the KRAB zinc-finger array; the receptor excesses are olfactory-receptor clusters.}
\label{fig:hsfn}
\end{figure}

In the vertebrates and \Ce, protein function is organised by chromosome, and the
organisation is carried by paralogue clusters. Zebrafish \texttt{4} ($\sm = 0.592$) and
human \texttt{19q} (0.684) and \texttt{21q} (0.701) are the least similar units
(Fig.~\ref{fig:hsfn}). \texttt{19q} is 41.6\% transcription factors against 13.4\%
genome-wide: the Krüppel-associated box (KRAB) zinc-finger array
\cite{huntley,grimwood,imbeault}. Zebrafish \texttt{4} is the same phenomenon in another
lineage, at 42.3\% transcription factors. \Ce{} \texttt{V} (0.786) holds 22\% receptors
against 10\% genome-wide --- the 7-transmembrane chemoreceptor families \cite{robertson}
--- and mouse \texttt{7} carries olfactory-receptor \cite{zhangfirestein} and \textit{Zfp}
clusters. As a result, essentially no vertebrate or \Ce{} unit reaches 0.95 on function at
any cut. Excess dissimilarity is 5.1--6.3 (zebrafish), 4.8--5.4 (human), 7.5--8.5
(mouse) and 11.6--12.9 (\Ce) points, against $\leq 2.8$ in the plants, yeast and fly.

\subsection{Level 2-2, natural units: one class per genome}
\label{sec:het-nc}

ncRNA class is the most heterogeneous level-2 class in five genomes, and in each the
heterogeneity is one class in one place (Fig.~\ref{fig:ncunit}).
\begin{itemize}
\item \textbf{\Ce{} piRNA.} $\Uadj = 0.417$, the largest value in the study. Chromosome
\texttt{IV} holds 64.7\% of the worm's ncRNA genes, and 91.7\% of them are piRNA. Every
chromosome scores 0.42--0.48 against the genome (excess 52 points). Dropping \texttt{IV}
takes $\Uadj$ to 0.022, and dropping the piRNA class takes it to 0.026.
\item \textbf{Maize rRNA.} Chromosome \texttt{6} carries the 45S NOR \cite{mcclintock},
and one of its halves scores 0.381 ($\Uadj = 0.134$).
\item \textbf{Zebrafish snRNA.} Chromosome \texttt{4} holds 394 of the genome's 487 snRNA
genes, on its heterochromatic long arm \cite{howe} ($\Uadj = 0.131$).
\item \textbf{Yeast rRNA.} The \texttt{RDN} rDNA locus on \texttt{XII}: one half of
\texttt{XII} scores 0.329, and yeast's ncRNA excess rises to 79 points at deciles, where
pieces hold only a few genes.
\item \textbf{Fly.} $\Uadj = 0.061$, mostly \texttt{2R}'s 5S rDNA and lncRNA clusters;
dropping the two most divergent arms takes it to 0.011.
\end{itemize}

\subsection{Level 2-3, natural units: repeats follow repeat biology}
\label{sec:het-te}

\begin{figure}[!htbp]
\centering
\includegraphics[width=\textwidth,height=0.82\textheight,keepaspectratio]{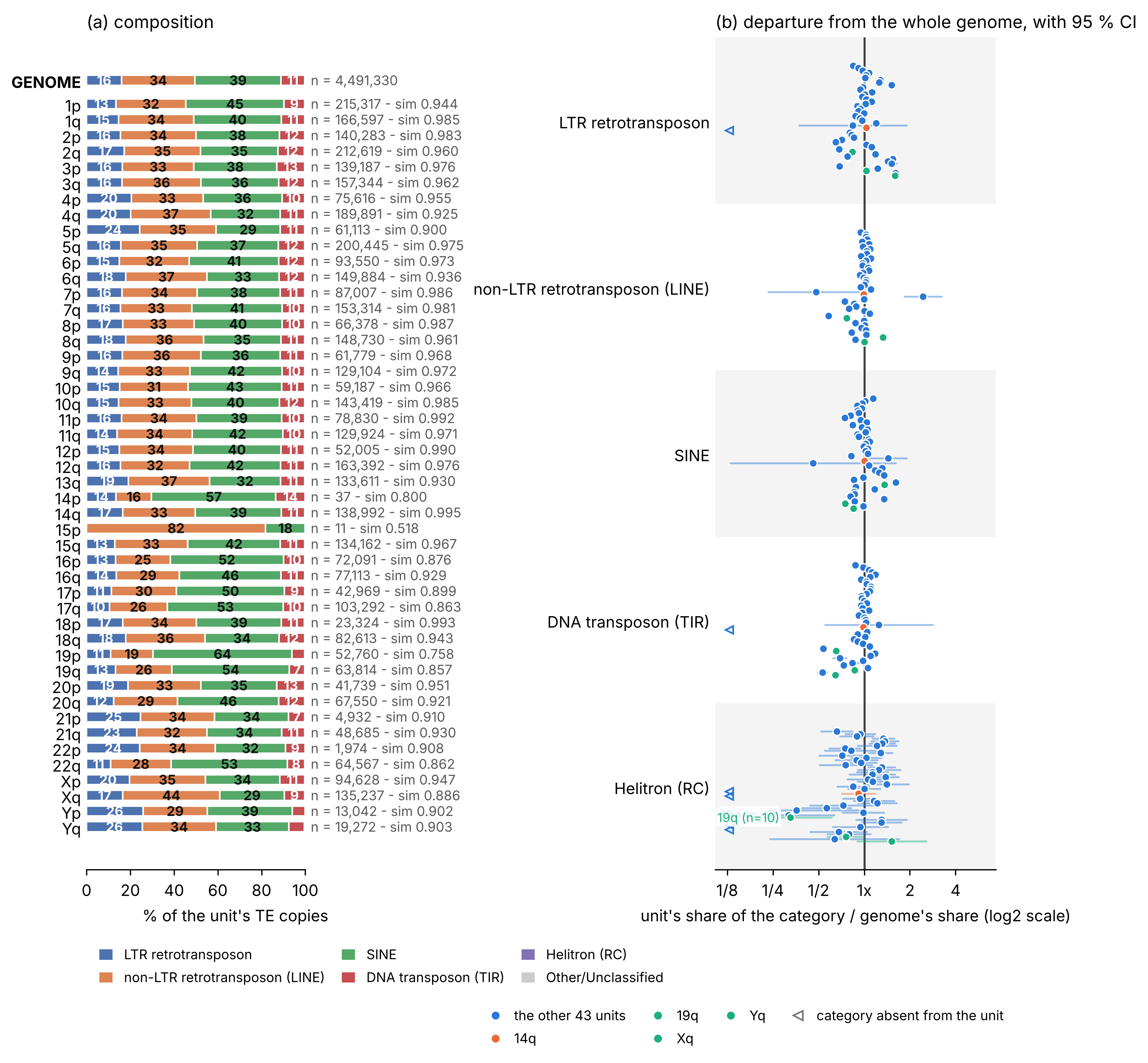}
\caption{Human TE order, arm by arm, by base pairs. \texttt{19p}/\texttt{19q} are
SINE-rich and LINE-poor.}
\label{fig:hste}
\end{figure}

TE order is the level at which the plant/fly-versus-vertebrate order breaks. Whole
chromosomes are fairly homogeneous everywhere except for named cases: human \texttt{19p}
(0.675, SINE-enriched by $\log_2 = 1.11$ and LINE-depleted by $-1.14$ --- the isochore
effect at its extreme \cite{eyrewalker,versteeg}; Fig.~\ref{fig:hste}), \texttt{19q}
(0.791), \texttt{17q} and \texttt{22q} (0.799); fly dot \texttt{4} (DNA transposons
enriched by $\log_2 = 1.59$); and zebrafish \texttt{4} (LTR and LINE both enriched,
$\log_2 = 1.05$ and $1.21$). Cutting the units changes the picture. \At's TE excess rises
from 2.1 points for whole chromosomes to 15--22 for thirds and smaller pieces, and fly's
from 2.8 to 7--10. In both genomes that is the pericentromere \cite{copenhaver,hoskins}: a
retrotransposon-rich, gene-poor block that a whole-chromosome average blends with the arms,
and that a finer cut isolates. Maize, whose TEs are spread along the whole chromosome,
shows nothing of the kind (excess $\leq 2.7$ at any cut).

\subsection{Levels 2-4 and 2-5, natural units: pseudogenes and IG/TR loci}
\label{sec:het-psig}

\begin{figure}[!htbp]
\centering
\includegraphics[width=\textwidth,height=0.45\textheight,keepaspectratio]{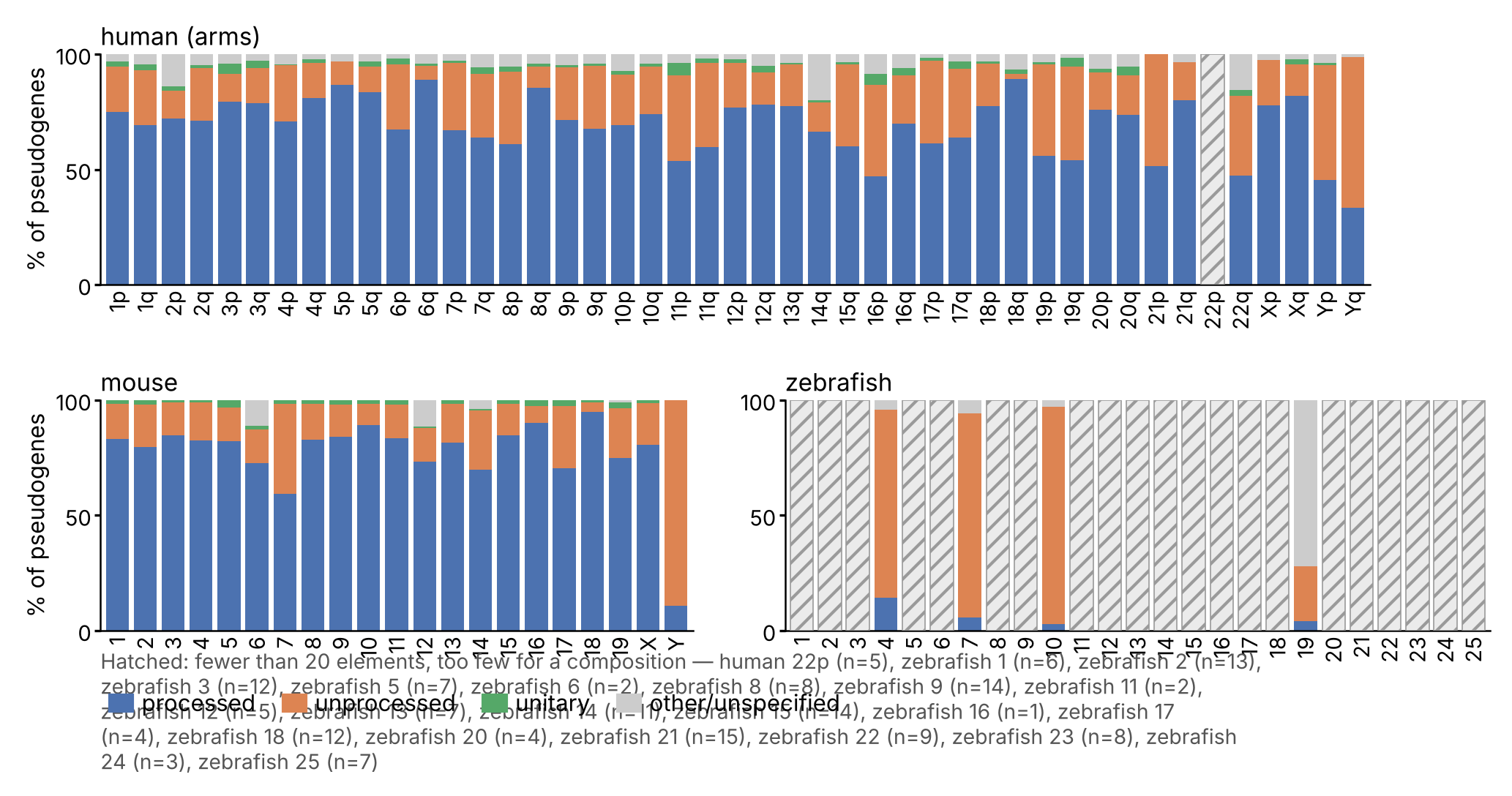}
\caption{Level-2-4 pseudogene origin per unit (human, mouse, zebrafish).}
\label{fig:psunit}
\end{figure}

\paragraph{Pseudogene origin.} Human and mouse annotate 10{,}634 and 10{,}249 processed,
3{,}535 and 3{,}153 unprocessed and 296 and 206 unitary pseudogenes. Processed
pseudogenes are reverse-transcribed mRNAs inserted more or less anywhere; unprocessed
pseudogenes are duplicates that sit beside their parent genes and so inherit their
clustering. The ratio of the two therefore varies between units, and no human or mouse unit
reaches 0.95 at any but the coarsest cut (match share 20--22\% for whole units). The least
similar units are human \texttt{16p} (0.779), \texttt{18q} (0.780), \texttt{6q} (0.808) and
\texttt{19q} (0.813), and mouse \texttt{18} (0.804). Excess is 6--12 points, and $\Uadj$ is
0.060 in human and 0.144 in mouse; dropping mouse's two most divergent chromosomes takes it
to 0.032 (Fig.~\ref{fig:psunit}).

\begin{figure}[!htbp]
\centering
\includegraphics[width=\textwidth,height=0.45\textheight,keepaspectratio]{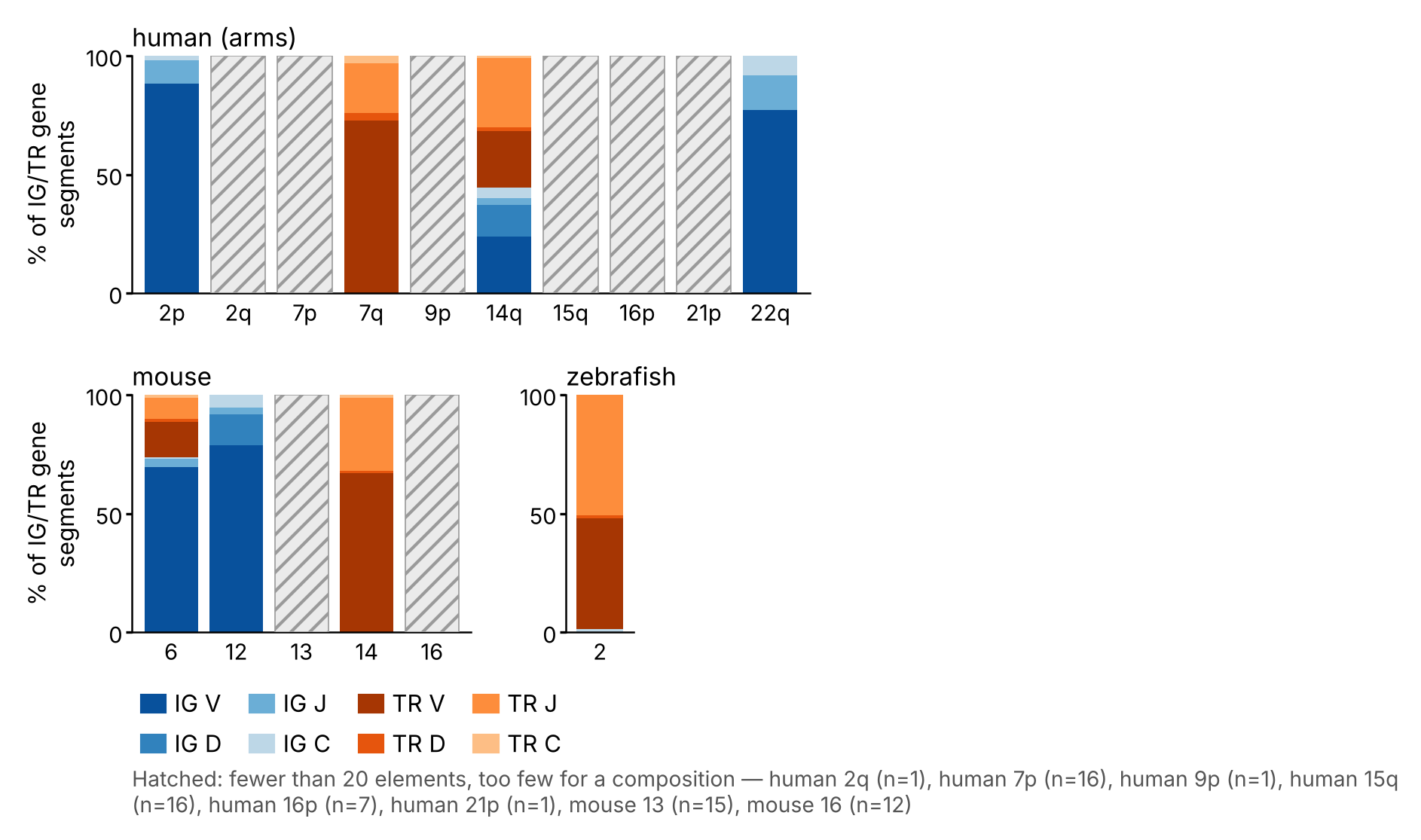}
\caption{Level-2-5 IG/TR segment per unit. Each locus is almost a single class (IG V, or
TR J and V), so the units that carry a locus are unlike one another and unlike the
genome.}
\label{fig:igunit}
\end{figure}

\paragraph{IG/TR segments.} The IG/TR segments are the textbook case of a clustered
class: each is a member of one of a handful of loci, the IG heavy, $\kappa$ and $\lambda$
loci and the TR $\alpha/\delta$, $\beta$ and $\gamma$ loci \cite{lefranc}. Only three human
and three mouse units carry at least 20 segments. None of them matches its genome (human
\texttt{2p} 0.418, \texttt{22q} 0.432, \texttt{7q} 0.478; mouse \texttt{14} 0.454,
\texttt{12} 0.541), excess is 39--44 points at every cut, and $\Uadj$ is 0.190 (human)
and 0.386 (mouse) (Fig.~\ref{fig:igunit}). This is heterogeneity by construction: the
class is defined by the loci. It serves as a positive control for the scan --- a class
known to be clustered is reported as maximally clustered --- rather than as a finding.
Zebrafish's 150 segments are mostly TR J and V segments on one chromosome and are not
scored.

\subsection{Levels 3-2 and 4-1: deep gene families}
\label{sec:het-deep}

\begin{figure}[!htbp]
\centering
\includegraphics[width=\textwidth]{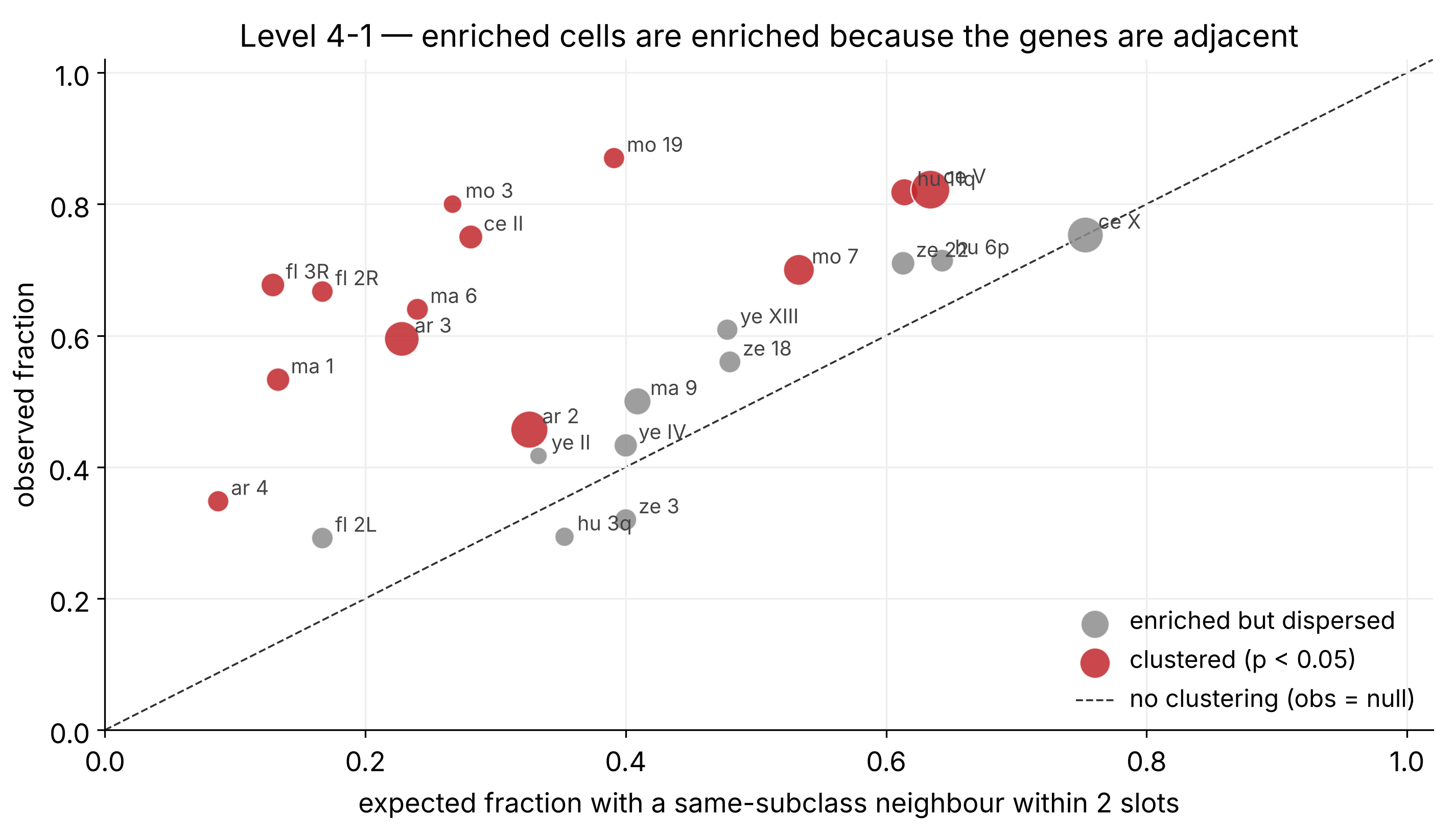}
\caption{EC subclass: the strongest enriched unit $\times$ subclass cells, split by
whether the excess genes are also adjacent in gene order. Eleven of 21 are tandem arrays.}
\label{fig:arrays}
\end{figure}

EC class showed no chromosome-scale organisation anywhere. One EC digit lower,
\textbf{six of eight genomes reject} a shared per-chromosome composition: \Ce, fly, human,
mouse and zebrafish at $p \leq 0.001$, and \At{} at $p = 0.002$. Maize ($p = 0.07$) and
yeast ($p = 0.17$) do not. \Ce{} is the strongest case ($\Uadj = 0.018$). The effect is
real but small (\S\ref{sec:homog-ecsub}), and partly tandem duplication. Of
the 21 strongest enriched cells, 11 are also clustered in gene order
(Fig.~\ref{fig:arrays}): fly's \textit{PH4alpha} prolyl-4-hydroxylases and
\textit{Mal-A} glycosylases, mouse \textit{Cyp2c} and \textit{Hsd3b}, human
\textit{PGA3/4/5} pepsinogens, and the \At{} ureide pathway. Yeast, with almost no tandem
enzyme arrays, is the clean negative.

\begin{figure}[!htbp]
\centering
\includegraphics[width=\textwidth,height=0.82\textheight,keepaspectratio]{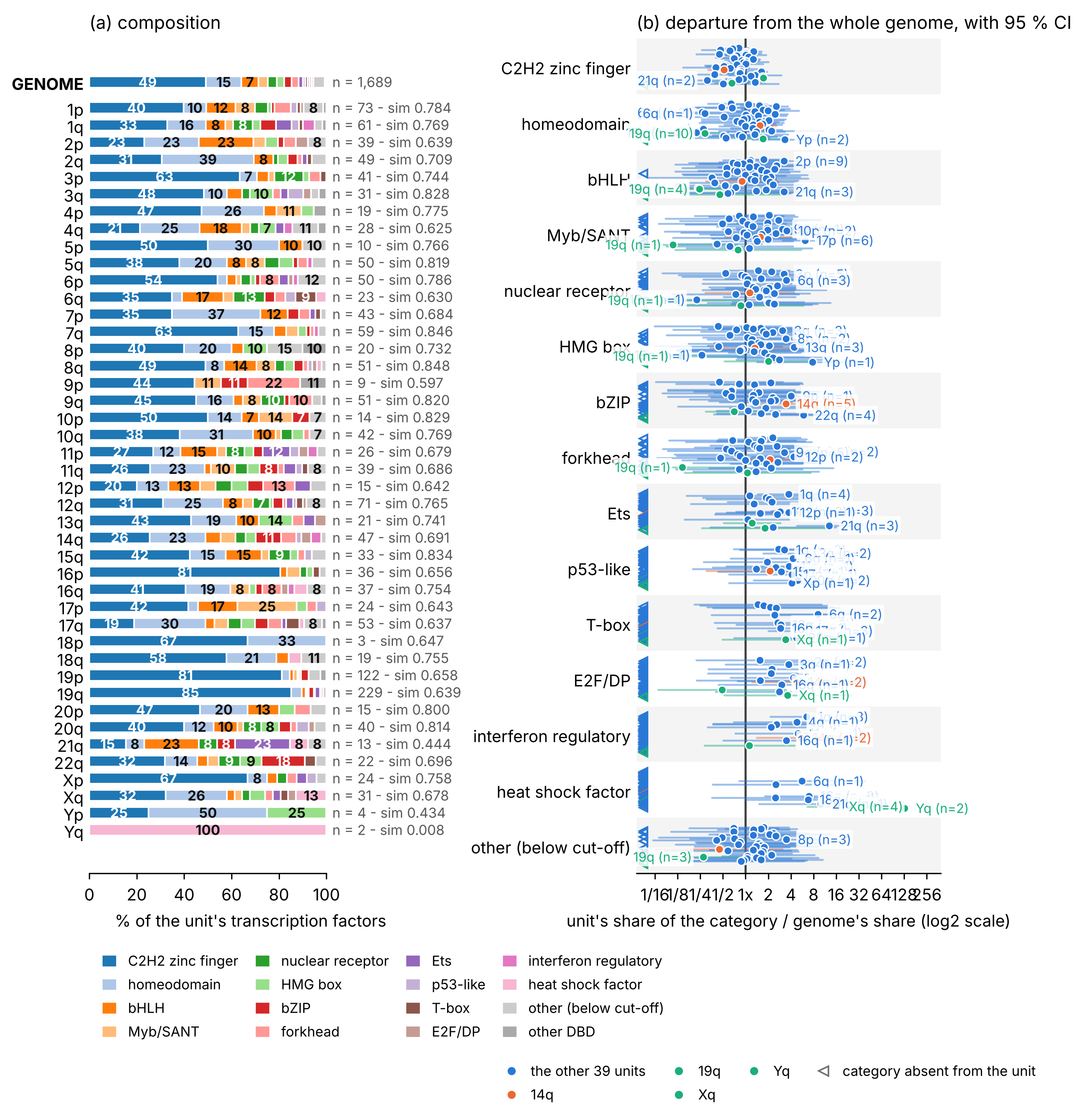}
\caption{Human TF DNA-binding domain, arm by arm. The \texttt{19p}/\texttt{19q} C2H2
zinc-finger excess dominates.}
\label{fig:hstf}
\end{figure}

The TF DNA-binding domain is the most heterogeneous protein level in the vertebrates and
\Ce{} ($\Uadj = 0.035$--$0.089$), and among the least in the plants (0.001--0.004). The
cause is the same paralogue clusters as at level~2-1, now resolved into families. Human
\texttt{19p} and \texttt{19q} are 0.637 and 0.644 (C2H2 zinc fingers;
Fig.~\ref{fig:hstf}), zebrafish \texttt{4} is 0.581, mouse \texttt{X} is 0.624, and
\Ce{} \texttt{V} is 0.652. On \Ce{} \texttt{V}, 70\% of classified TFs are nuclear
hormone receptors, against 35\% genome-wide --- the nematode-specific HNF4-derived
expansion \cite{robinsonrechavi}, one of the lineage-specific expansions that dominate
eukaryotic gene-family evolution \cite{lespinet}.

\subsection{Finer cuts: compartments inside homogeneous units}
\label{sec:het-cuts}

Cutting natural units into equal parts turns up heterogeneity that a whole-unit average
hides. The partition tests count gene classes only (\S\ref{sec:scores}), and in every
genome but yeast most units reject a single shared composition once cut. What
distinguishes the genomes is how large the departures are and what drives them
(Table~\ref{tab:partitions}), and there are only two drivers: a compartment, or a named
locus.

\begin{table}[!htbp]
\centering
\caption{The most heterogeneous unit of each genome under equal base-pair cuts, gene classes
only (protein-coding, ncRNA, pseudogene, IG/TR; TE copies not counted). Units with at least
200 genes are ranked by their median bias-corrected Cram\'er's $V$ over cuts into 2, 3, 4,
5 and 10 parts. ``$V$'' is the range over those cuts; ``after'' is the range once the named
driver is removed (its class from the unit, or its array window); ``all units'' is the
genome's median $V$ over every major unit and cut.}
\label{tab:partitions}
\small
\setlength{\tabcolsep}{4pt}
\begin{tabular}{llrlll>{\raggedright\arraybackslash}p{5.2cm}}
\toprule
species & unit & genes & $V$ & after & all units & driver \\
\midrule
fly         & \texttt{2R}  & 3{,}649  & 0.06--0.21 & 0.05--0.17 & 0.09 & a cluster of 126 lncRNA genes at 19.6~Mb (array window removed) \\
human       & \texttt{17p} & 1{,}098  & 0.25--0.34 & 0.24--0.29 & 0.13 & gradient: pseudogenes and ncRNA rise, protein-coding falls towards the centromere (pseudogenes removed) \\
mouse       & \texttt{14}  & 3{,}393  & 0.19--0.29 & 0.10--0.20 & 0.12 & TR-$\alpha/\delta$ locus at 54.3~Mb (IG/TR removed) \\
zebrafish   & \texttt{2}   & 1{,}621  & 0.25--0.35 & 0.01--0.07 & 0.07 & TR-$\alpha/\delta$ locus at 36.1~Mb (IG/TR removed) \\
yeast       & \texttt{III} & 200      & 0.06--0.21 & ---        & 0.06 & no single class; halves not distinguishable ($p = 0.30$) \\
\Ce         & \texttt{IV}  & 19{,}704 & 0.23--0.39 & 0.05--0.14 & 0.12 & piRNA loci in the two arm domains (ncRNA removed) \\
\At         & \texttt{2}   & 6{,}312  & 0.30--0.40 & 0.04--0.09 & 0.32 & pericentromeric pseudogenes (pseudogenes removed) \\
maize       & \texttt{6}   & 4{,}610  & 0.31--0.39 & 0.10--0.17 & 0.07 & rRNA genes of the nucleolus organiser at 16--23~Mb (ncRNA removed) \\
\bottomrule
\end{tabular}
\end{table}

\begin{figure}[!htbp]
\centering
\includegraphics[width=\textwidth]{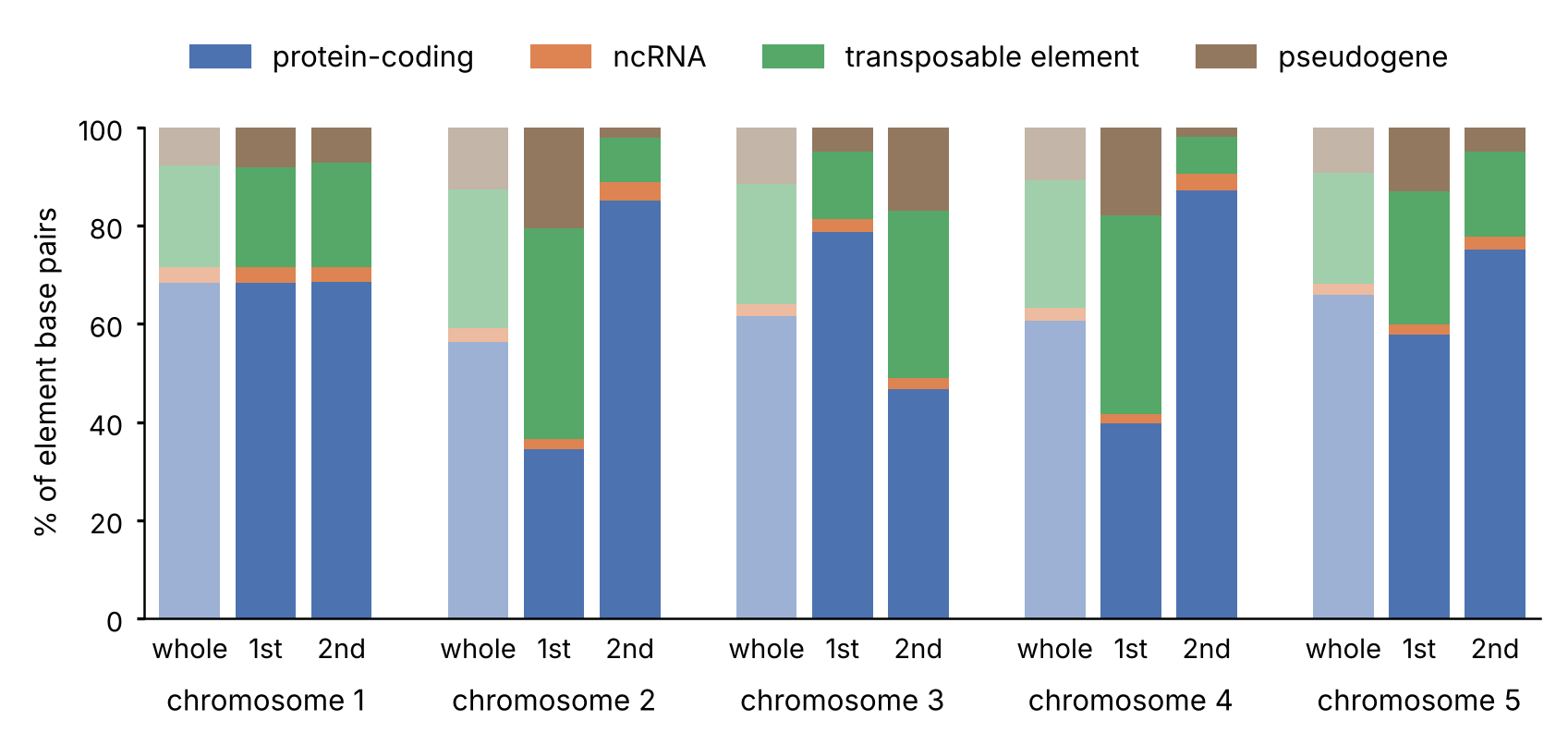}
\caption{\At{} level~1 by base pairs, each chromosome whole (pale) and as its first and
second halves. The half that holds the centromere carries the TEs and pseudogenes; on
chromosome~\texttt{1} the centromere lies near the middle, so both halves share it.}
\label{fig:athalves}
\end{figure}

\textbf{Compartments.} \At{} is the sharpest example of what a whole-unit panel hides,
and by this test the most heterogeneous genome of the eight (median $V = 0.32$ over all its
units and cuts, against 0.06--0.13 elsewhere). Each chromosome carries a block around its
centromere in which pseudogenes displace protein-coding genes: on chromosome \texttt{5} the
central decile is 80\% pseudogenes against 1--6\% in the distal deciles
(Fig.~\ref{fig:cuts}, bottom row), and on chromosome \texttt{2} pseudogenes fall from
31.5\% of genes in the first half to 3.0\% in the second. Removing pseudogenes takes
$V$ from 0.30--0.40 to 0.04--0.09 on chromosome \texttt{2} and from 0.17--0.45 to
0.01--0.06 on \texttt{5}. By base pairs the same halves differ further still, because the
pericentromeres are also where \At{} keeps its TEs \cite{agi,copenhaver}: on chromosome
\texttt{2}, TEs are 43\% of element base pairs in the half that holds the centromere and
9\% in the other, and protein-coding genes 34\% and 85\% (Fig.~\ref{fig:athalves}). That
is why \At's level-1 base-pair excess rises from 2.3 points for whole chromosomes to 16--26
for smaller pieces, and its match share from 60\% to 0--10\%, the largest collapse in the
study. Human \texttt{9q} (Fig.~\ref{fig:cuts}, third row) is the same shape as a continuous
gradient: from the pericentromeric to the distal decile, pseudogenes fall from 44\% of genes
to 7\% and protein-coding genes rise from 3\% to 39\%; human's most heterogeneous arm,
\texttt{17p}, is the same gradient on a p arm. Removing pseudogenes barely changes either
($V$ 0.19--0.21 on \texttt{9q}), because protein-coding and ncRNA genes shift too: these
are compartments of several classes at once, not single loci. \Ce{} \texttt{IV} is a
compartment of a different kind (Fig.~\ref{fig:cuts}, fifth row): ncRNA, nearly all
piRNA loci, makes up 64--87\% of genes in deciles 3--4 and 76--96\% in deciles 8--10,
against 35--39\% in the central deciles, in line with the arm--centre structure of the
\Ce{} chromosomes \cite{rockman,ruby}. Removing ncRNA takes its $V$ from 0.23--0.39 to
0.05--0.14. Fly, by contrast, is nearly flat once its TE insertions are left out: the median
$V$ over all fly units and cuts is 0.09, and fly's TE-rich pericentromeres appear only in
the base-pair analyses. The median largest half-to-half shift in a gene class tells the
same story: 16.7 percentage points in \At, 7.9--9.6 in the mammals, and 2.7--5.0 in
yeast, fly, maize, zebrafish and \Ce.

\textbf{Named loci.} Maize \texttt{6}, zebrafish \texttt{2} and \texttt{4}, mouse
\texttt{14} and yeast \texttt{XII} are the array cases. In maize \texttt{6} the rRNA genes
of the nucleolus organiser at 16--23~Mb \cite{mcclintock} make ncRNA 61--70\% of the
genes in the first two deciles (Fig.~\ref{fig:cuts}, fourth row); removing ncRNA takes
$V$ from 0.31--0.39 to 0.10--0.17, and maize's other nine chromosomes are among the most
homogeneous units in the study (median $V = 0.07$ for the genome). Zebrafish \texttt{2}
departs through a single TR locus, whose segments are 46\% of the genes in its seventh
decile; without IG/TR segments its $V$ falls from 0.25--0.35 to 0.01--0.07. Zebrafish
\texttt{4} carries its snRNA genes on the long arm (ncRNA 49--54\% of genes in deciles
4--8; $V$ 0.16--0.24, and 0.09--0.13 without ncRNA). Mouse \texttt{14} carries the
TR-$\alpha/\delta$ locus ($V$ 0.19--0.29, 0.10--0.20 without IG/TR), mouse \texttt{7} the
imprinted \textit{Snord115/116} cluster \cite{cavaille} (ncRNA 81\% of the genes in its
fifth decile), and cutting out
yeast \texttt{XII}'s snR/tRNA window around the rDNA at 0.35--0.49~Mb takes its $V$ from
0.07--0.21 to 0.07--0.11. Otherwise yeast is the most homogeneous genome by this test
(median $V = 0.06$; \S\ref{sec:homog-cuts}).

\subsection{Cytogenetic bands: heterogeneous, but not at the boundaries}
\label{sec:het-bands}

\begin{figure}[!htbp]
\centering
\includegraphics[width=\textwidth]{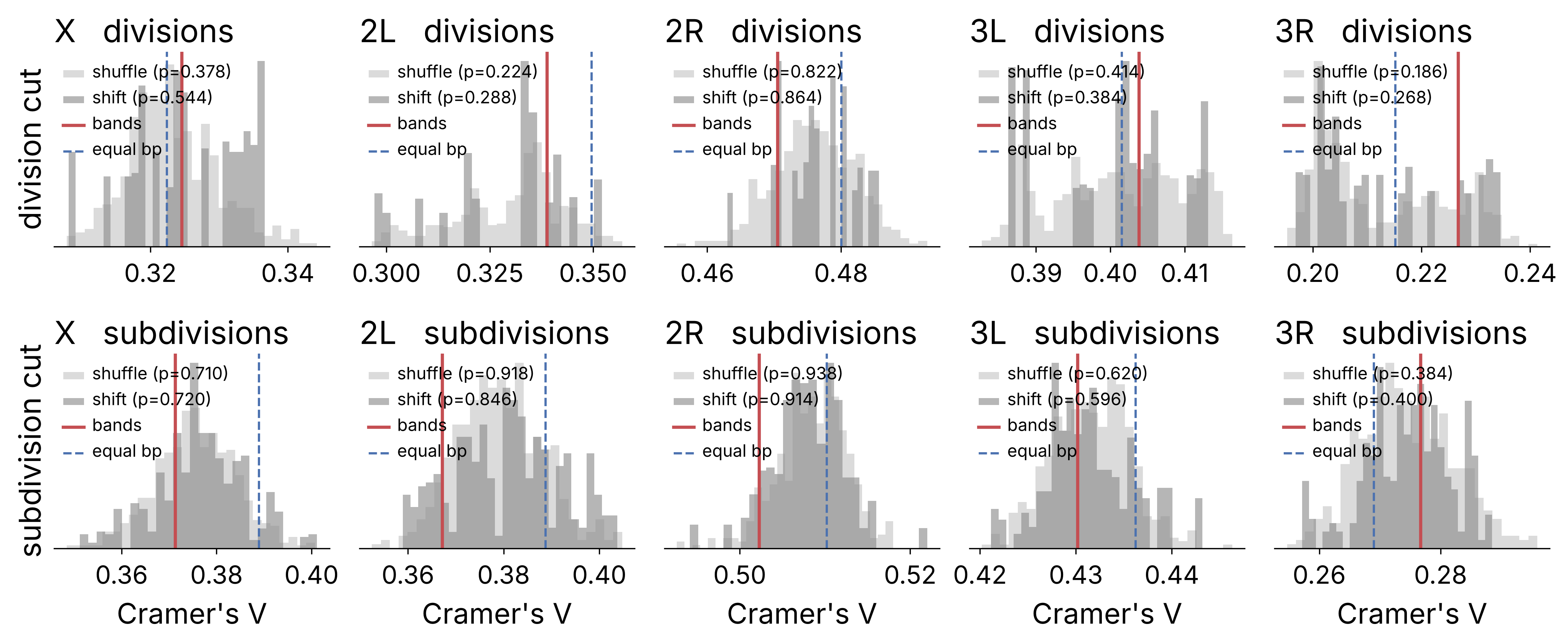}\\[0.5em]
\includegraphics[width=\textwidth]{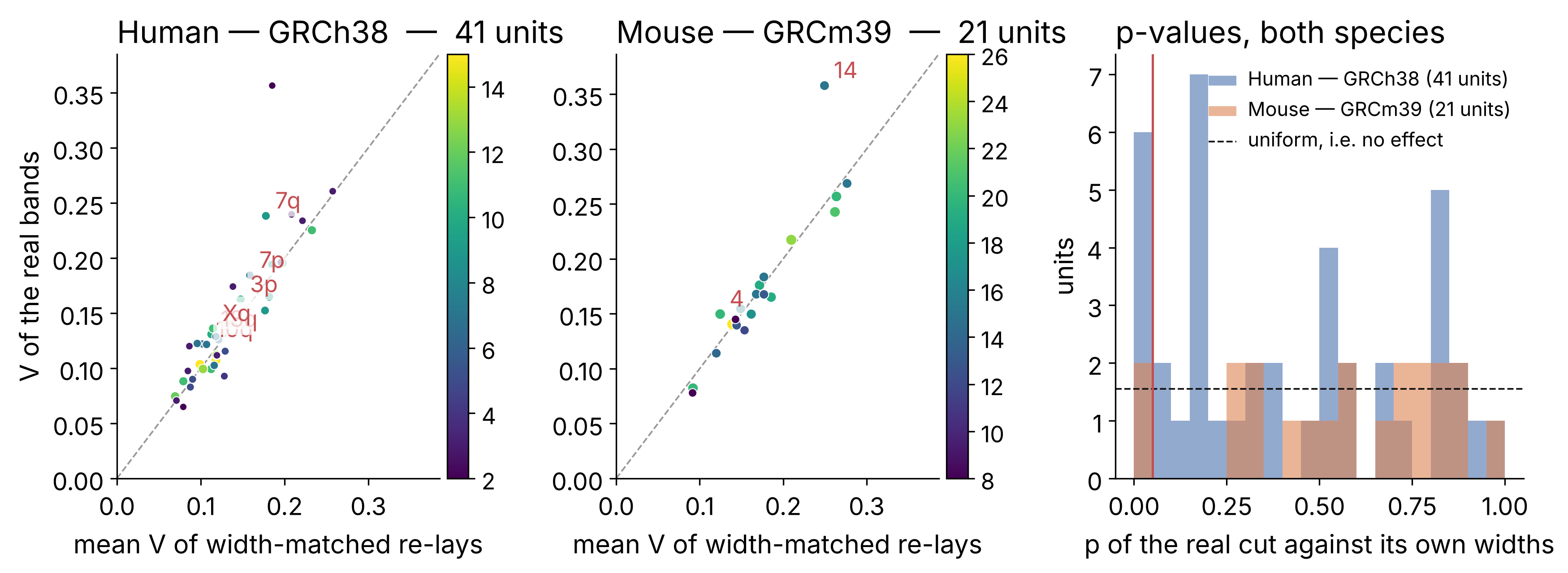}
\caption{\textbf{Band partitions against a width-matched null.} Top: fly Bridges
divisions and subdivisions, one panel per arm; histograms are re-lays of the same band widths
at shuffled or shifted positions, the red line is the real band cut and the dashed line the
equal-width cut. Bottom: human ISCN and mouse Giemsa bands, one dot per unit (colour: number
of bands on the unit), the real cut's $V$ against the mean of its width-matched re-lays, and
the distribution of the per-unit $p$-values. The observed $V$ sits inside the distribution
of re-lays of the same widths nearly everywhere.}
\label{fig:bands}
\end{figure}

Three genomes have a band map: the Bridges polytene map for fly \cite{bridges}, ISCN
850-band for human \cite{iscn,furey} and Giemsa bands for mouse. Cut at the bands, units
are strongly heterogeneous --- Cramér's $V$ is 0.23--0.47 across fly arms at divisions
and 0.13--0.16 at the median in human and mouse. But the boundaries carry no information
beyond the widths of the parts. Laying the same multiset of band widths at random positions
reproduces the observed $V$ (Fig.~\ref{fig:bands}): no fly arm reaches $p < 0.05$ at
divisions or subdivisions ($p = 0.19$--$0.94$); 6 of 41 human units under a shuffled re-lay
and none under a shifted one; and 2 of 21 mouse units. In fly, what does carry composition
is position: genes in the proximal two divisions of each arm are 72.3\% TE insertions
against 10.2\% in the distal 18 ($V = 0.59$), on every arm. The weak excess of one
lettered subdivision over another ($V = 0.07$, block null $p = 0.06$) is that same
proximal TE block seen through the letters, and an arm cut into six oriented equal parts
gives $V = 0.30$ against a block null of 0.07.

\begin{figure}[!htbp]
\centering
\includegraphics[width=\textwidth]{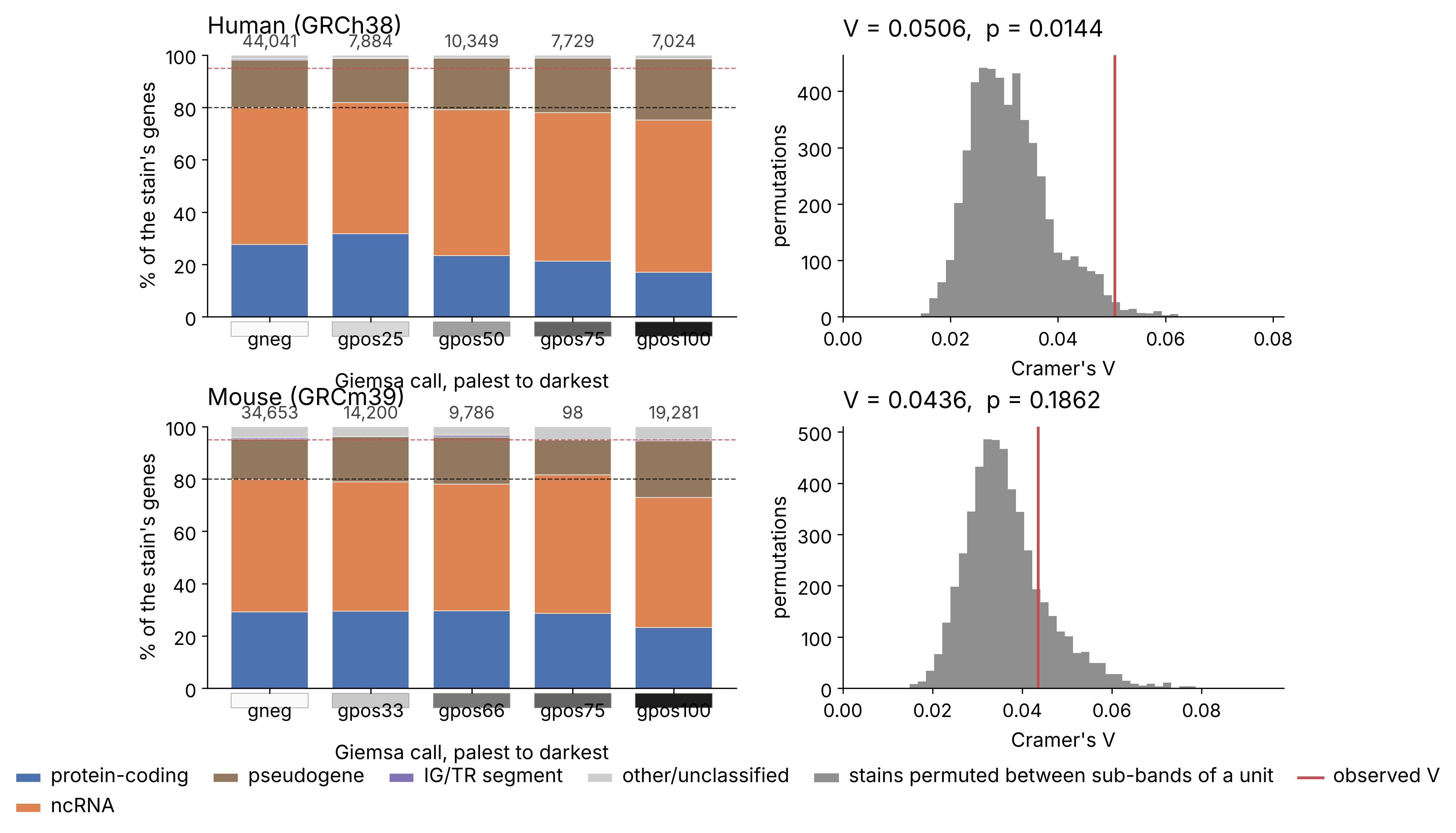}
\caption{Level-1 composition by Giemsa stain intensity --- the positive control for the
band machinery. Monotonic in human; not resolved on the coarser mouse scale.}
\label{fig:stain}
\end{figure}

The method is not blind. It recovers the textbook Giemsa gradient in human ---
protein-coding share 27.6\% of genes in \texttt{gneg} against 17.1\% in
\texttt{gpos100}, $V = 0.051$, $p = 0.014$ (Fig.~\ref{fig:stain}). The TE share by base
pairs, in contrast, is almost flat across stains (48--50\%). Mouse's coarser stain scale
does not resolve the gradient ($p = 0.19$).

\paragraph{Bands as units, at every level.} The band units can also be put through the
same similarity scan as the equal cuts, at every level of the hierarchy
(Table~\ref{tab:bands}). They behave like equal cuts of the same size. Match share falls
from the coarsest band unit to the finest, just as it does from halves to deciles: at
level~1 by base pairs from 40\% of fly divisions and 42\% of human regions to 10--14\% of
the finest units. The species order of Part~I survives: at protein function, fly's
divisions and subdivisions depart from the genome by 2.5--3.4 points beyond sampling,
human's band units by 4.7--6.5 and mouse's by 7.7--8.7. EC class is again close to
sampling everywhere ($\leq 3.1$ points), TF domain is the most heterogeneous protein level
in human and mouse (10--30 points), and IG/TR the most heterogeneous level overall
(42--44 points at regions). Fly's TE excess climbs to 16--22 points at subdivisions and
bands, and its ncRNA excess to 80 at bands, where pieces hold a handful of genes.

\begin{table}[!htbp]
\centering
\caption{Bands as units: excess dissimilarity (percentage points) and, in parentheses,
match share (\% of parts with $\sm \geq 0.95$), by band resolution and level. Fly units
are Bridges divisions (e.g.\ \texttt{21}), subdivisions (\texttt{21B}) and bands
(\texttt{21B1}); human units are ISCN regions (\texttt{1p3}), bands (\texttt{1p36}) and
sub-bands (\texttt{1p36.33}); mouse units are Giemsa regions (\texttt{1A}), bands
(\texttt{1A1}) and sub-bands, which split only some mouse bands. Dashes: fewer than ten
scorable parts, or a level the species cannot supply. Match share is omitted where it is
0.}
\label{tab:bands}
\small
\setlength{\tabcolsep}{3.5pt}
\begin{tabular}{llrrrrrrrr}
\toprule
species & unit & L1 (bp) & 2-1 fn. & 2-2 ncRNA & 2-3 TE & 2-4 pseudo. & 2-5 IG/TR & 3-1 EC & 3-2 TF \\
\midrule
fly   & division    &  3.2 (40) & 3.4 & 7.4  &  8.1 (5)  & ---       & ---  &  2.0     & --- \\
      & subdivision &  2.1 (22) & 2.5 & 9.5  & 15.6 (3)  & ---       & ---  &  3.1     & --- \\
      & band        & 16.1 (10) & 4.4 & 79.6 & 21.9 (2)  & ---       & ---  &  ---     & --- \\
\addlinespace
human & region      &  5.5 (42) & 5.1 & 1.1 (71) &  5.2 (44) &  6.5 (23) & 43.7 & $-0.3$ (6) & 10.3 \\
      & band        &  8.1 (20) & 4.7 & 1.3 (39) &  7.9 (27) & 10.0 (11) & ---  &  0.5     & 15.4 \\
      & sub-band    & 10.9 (12) & 6.5 & 1.1 (19) &  8.4 (19) & 12.0 (10) & ---  &  1.0     & 29.8 \\
\addlinespace
mouse & region      &  6.8 (30) & 7.7 & 1.1 (68) &  5.6 (40) & 11.0 (17) & 41.5 &  0.2     & 12.9 \\
      & band        &  9.6 (16) & 8.7 & 1.4 (35) & 10.0 (20) & 11.2 (8)  & ---  &  1.7     & 15.5 \\
      & sub-band    & 10.3 (14) & 8.4 & 1.3 (33) & 10.3 (19) & 11.2 (8)  & ---  &  2.3     & 12.0 \\
\bottomrule
\end{tabular}
\end{table}

\subsection{Random segments: how long before a piece looks like the whole?}
\label{sec:het-random}

\begin{figure}[!htbp]
\centering
\includegraphics[width=0.85\textwidth]{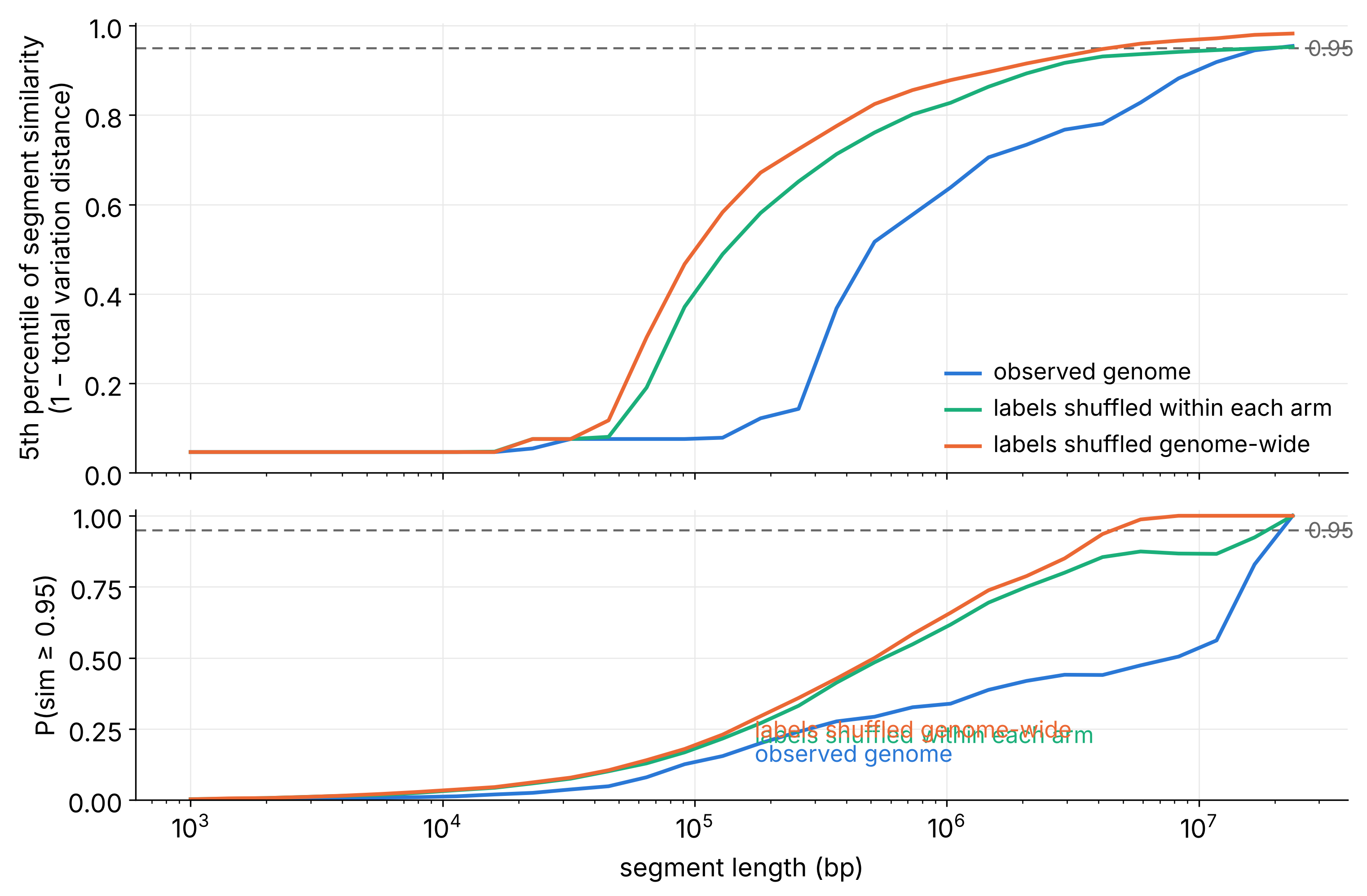}
\caption{Fly: similarity of random windows to the whole genome, by window length, with a
within-arm and a genome-wide label shuffle. The within-arm shuffle reaches 0.95 at the
median by 0.6~Mb, the observed genome only at 7.9~Mb: fly's level-1 heterogeneity lies
within arms, along them.}
\label{fig:flylen}
\end{figure}

The random-window sweep gives the heterogeneity a length scale (Table~\ref{tab:seglen}).
Raw lengths are not comparable across species, because they depend on genome size and on
how many elements a window holds. The ratio to a genome-wide label shuffle is comparable.
By that ratio the genomes fall into two groups. In the first, a window needs at most about
five times the shuffle length: maize, zebrafish and human (\S\ref{sec:homog-cuts}), and
yeast and mouse. In fly, \At{} and \Ce{} it needs 15--60 times as long: their heterogeneity
is concentrated in compartments --- pericentromeres in fly and \At, recombination arms in
\Ce{} --- that a window must be long enough to average out. Fly's windows, with TEs in
level~1, need 7.9~Mb, $15\times$ its shuffle (Fig.~\ref{fig:flylen}), because FBti
insertions crowd into the proximal fifth of every arm. The human and mouse windows reach 0.95 at the
median, but never at the 5th percentile: a window that falls in a TE-rich or a gene-dense
isochore stays unlike the genome at any length that fits inside one chromosome.

\begin{figure}[!htbp]
\centering
\includegraphics[width=0.49\textwidth]{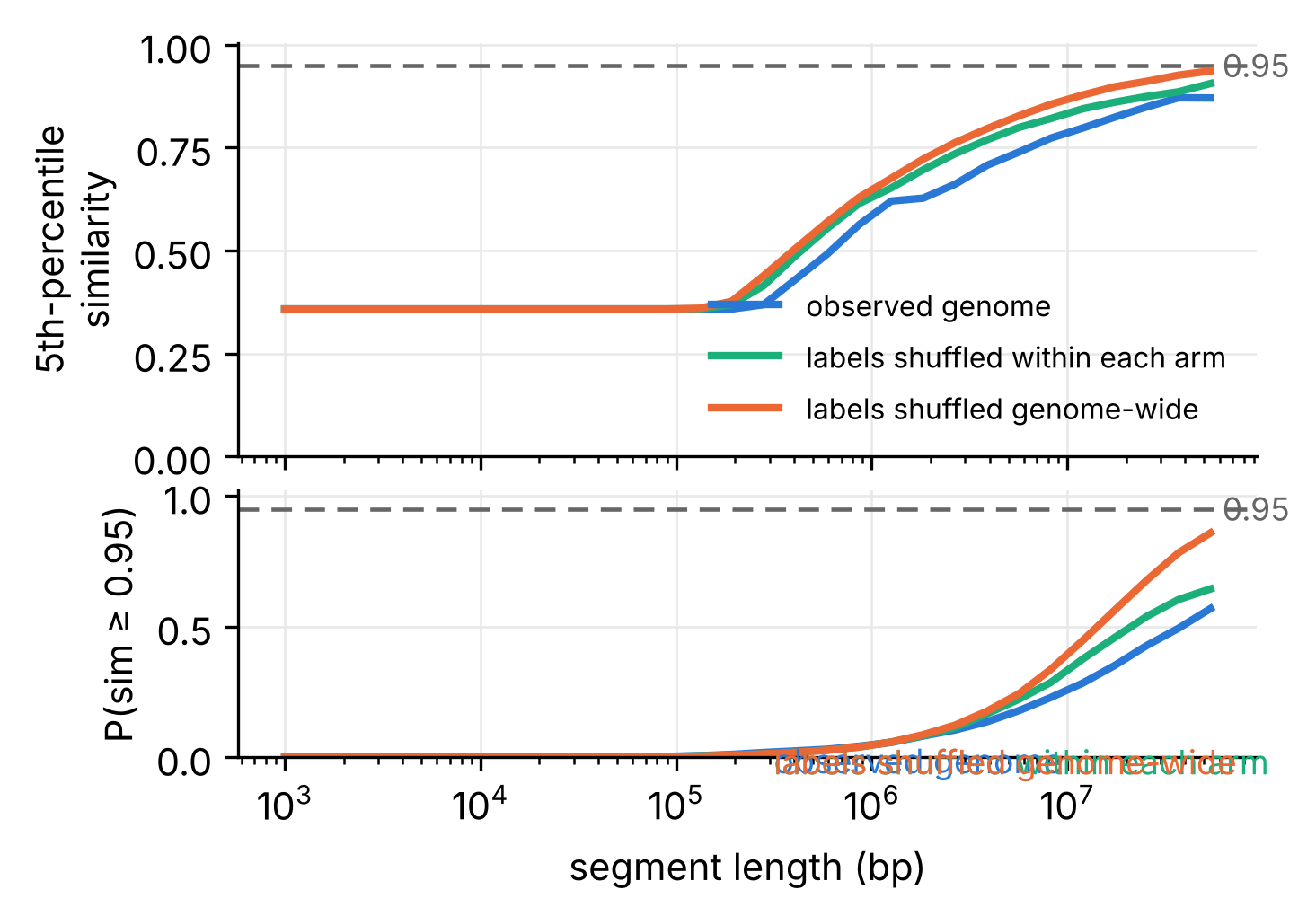}
\hfill
\includegraphics[width=0.49\textwidth]{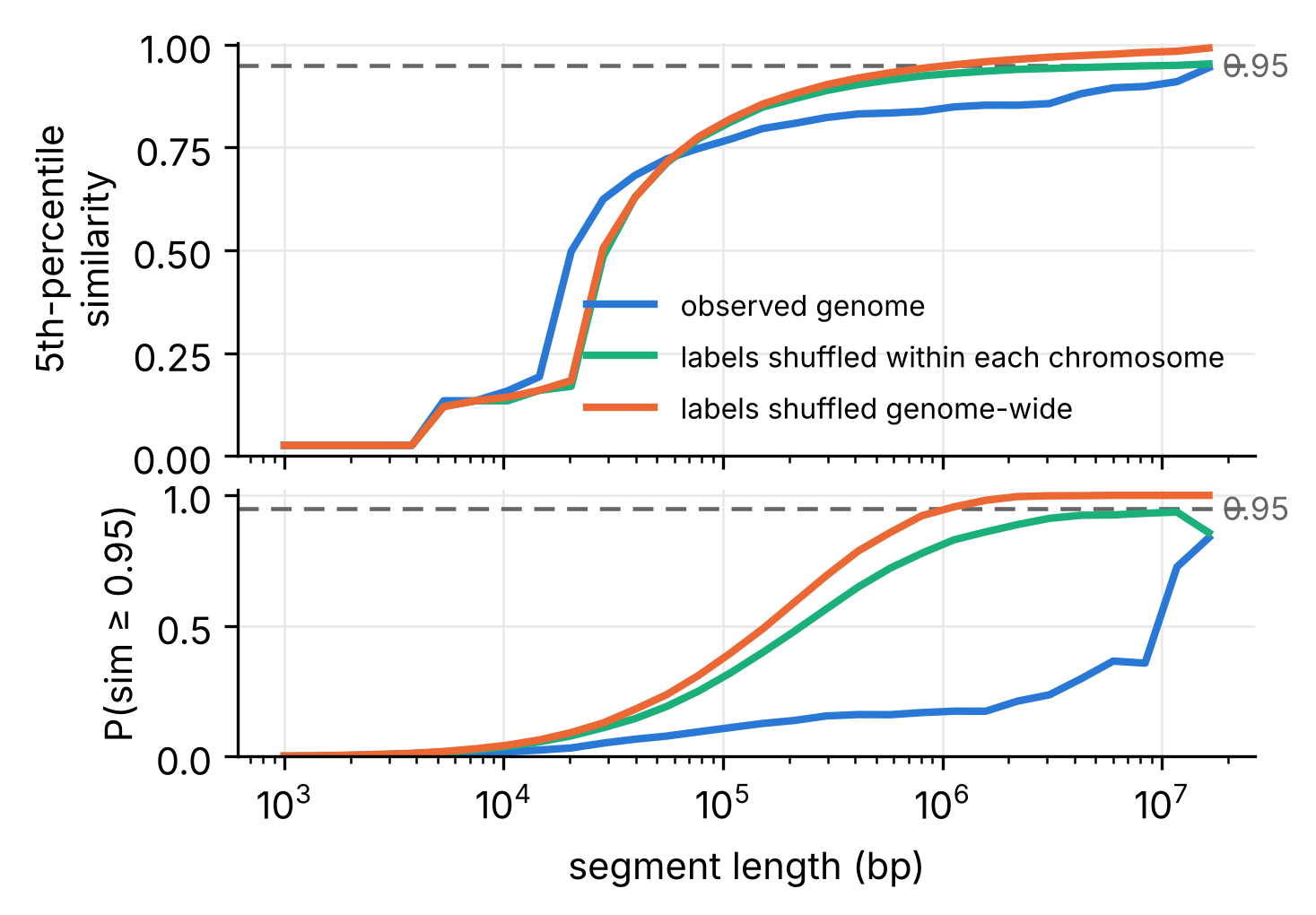}
\caption{Similarity of random windows to the whole genome by length, with within- and
between-unit shuffle nulls. Left: human. Right: \Ce.}
\label{fig:simlen}
\end{figure}

\begin{figure}[!htbp]
\centering
\includegraphics[width=\textwidth]{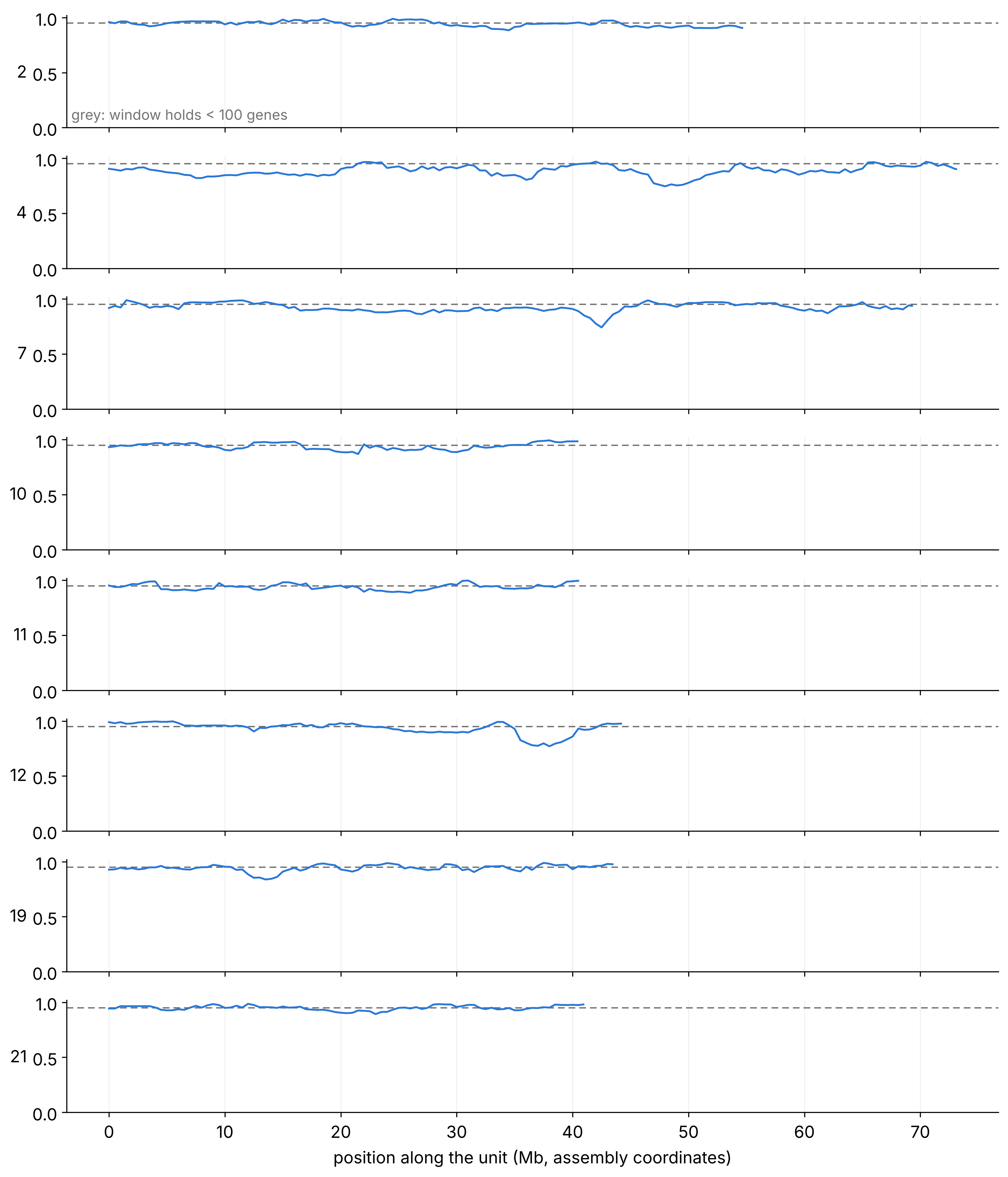}
\caption{Sliding-window similarity to the whole genome along eight zebrafish
chromosomes, level~1 by base pairs (5~Mb windows stepped by 500~kb; dashed line 0.95). Most
of every chromosome sits near the line; the deepest dips are on chromosome~\texttt{7} at
42.5--47.5~Mb, where ncRNA is 24\% of element base pairs, on the long arm of
\texttt{4}, and near the end of \texttt{12}.}
\label{fig:sliding}
\end{figure}

Decomposing the shortfall with a within-unit and a genome-wide shuffle divides the genomes
by mechanism (Fig.~\ref{fig:simlen}). In human the within-unit shuffle needs 21~Mb and the
genome-wide one 14~Mb, so most of the shortfall is local clustering within arms. In \Ce{}
the within-unit shuffle needs only 0.23~Mb, so the shortfall is position along each
chromosome. A sliding window then locates the least similar stretches in zebrafish
(Fig.~\ref{fig:sliding}): chromosome \texttt{7} at 42.5--47.5~Mb ($\sm = 0.742$; 24\%
ncRNA by base pairs against 4\% genome-wide) and the long arm of chromosome \texttt{4} from
47.5~Mb ($\sm = 0.745$--$0.763$).

\section{Synthesis: is composition more uniform in plants and flies?}
\label{sec:synthesis}

\begin{figure}[!htbp]
\centering
\includegraphics[width=\textwidth]{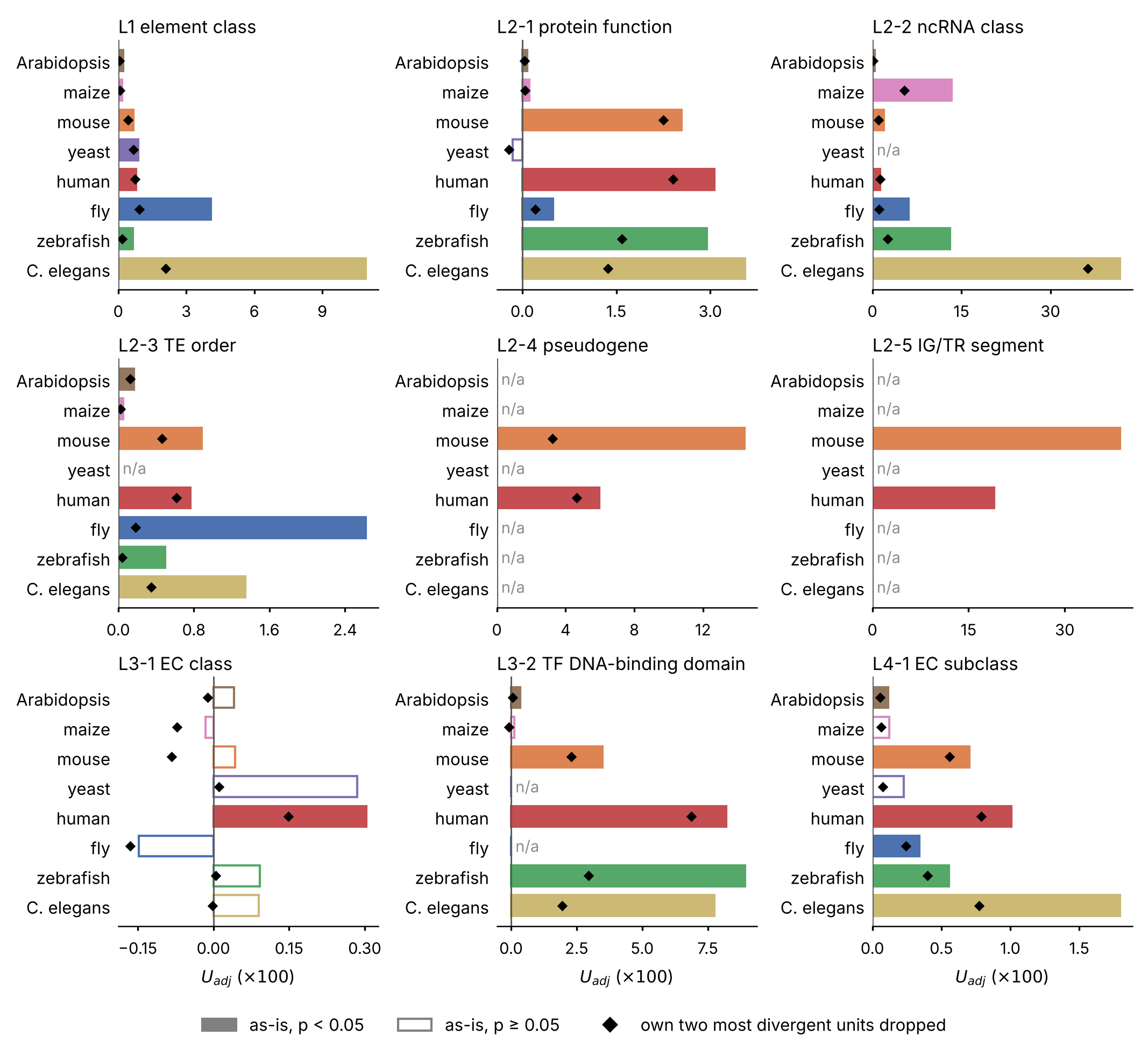}
\caption{\textbf{$\Uadj$ by level and genome} ($\times 100$; element counts). Bars are the
as-is values, filled where the permutation $p < 0.05$ and hollow otherwise; diamonds are
the same genome with its own two most divergent units dropped, the robustness check of
Table~\ref{tab:uadj}. Genomes are ordered by their mean rank across levels, most uniform at
the top; ``n/a'': a level the genome cannot supply.}
\label{fig:uadj}
\end{figure}

Parts~I and~II use the same scale and the same order, so we can test the hypothesis level
by level. Table~\ref{tab:uadj} adds an independent statistic, $\Uadj$, which uses no
threshold and no cut; Fig.~\ref{fig:uadj} plots it by level and species, with and
without each genome's two most divergent units. Table~\ref{tab:verdict} gathers the verdict, which the paragraphs
below justify level by level.

\begin{table}[!htbp]
\centering
\caption{The hypothesis --- composition is more uniform along plant and fly genomes than
along mammalian ones --- tested level by level. ``Evidence'' names the statistics that
agree; ``what decides uniformity'' names what sets each genome's score where the
hypothesis fails.}
\label{tab:verdict}
\small
\begin{tabular}{>{\raggedright\arraybackslash}p{3.0cm}>{\raggedright\arraybackslash}p{2.2cm}>{\raggedright\arraybackslash}p{4.3cm}>{\raggedright\arraybackslash}p{4.6cm}}
\toprule
level & verdict & evidence & what decides uniformity \\
\midrule
2-1 protein function & holds & match share, excess, $\Uadj$, at every cut &
cis-clustered paralogue families in vertebrates and \Ce \\
3-2 TF domain & holds & excess and $\Uadj$ & the same paralogue clusters, by family \\
4-1 EC subclass & holds, weakly & $\Uadj$; excess $\leq 3.7$ pp everywhere & tandem
enzyme arrays \\
1 element class & does not hold & mammals least uniform whole, \At{} least uniform cut &
where TEs and pseudogenes sit: along arms or in pericentromeres \\
2-3 TE order & reverses for fly and \At{} once cut & excess, $\Uadj$ & repeat biology of
each lineage \\
2-2, 2-4, 2-5 & not about lineage & $\Uadj$, named loci & single loci (piRNA, rDNA,
snRNA, IG/TR) \\
3-1 EC class & no contrast & every genome homogeneous & --- \\
\bottomrule
\end{tabular}
\end{table}

\begin{table}[!htbp]
\centering
\caption{$\Uadj$ ($\times 100$), the share of class entropy predictable from an element's
unit above a size-matched null, for eight genomes at nine levels (element counts). Larger
means more across-unit structure. Bold: permutation $p \leq 0.002$. Dashes are levels a
species cannot supply.}
\label{tab:uadj}
\small
\setlength{\tabcolsep}{4.5pt}
\begin{tabular}{lrrrrrrrr}
\toprule
level & \At & maize & fly & yeast & \Ce & zebrafish & mouse & human \\
\midrule
1 element class     & \textbf{0.23} & \textbf{0.17} & \textbf{4.09} & \textbf{0.88} & \textbf{10.92} & \textbf{0.65} & \textbf{0.68} & \textbf{0.79} \\
2-1 protein function & \textbf{0.08} & \textbf{0.11} & \textbf{0.49} & $-0.15$       & \textbf{3.55}  & \textbf{2.95} & \textbf{2.54} & \textbf{3.07} \\
2-2 ncRNA class      & \textbf{0.35} & \textbf{13.37} & \textbf{6.10} & ---          & \textbf{41.72} & \textbf{13.10} & \textbf{1.91} & \textbf{1.34} \\
2-3 TE order         & \textbf{0.17} & \textbf{0.05} & \textbf{2.62} & ---           & \textbf{1.34}  & \textbf{0.50} & \textbf{0.88} & \textbf{0.76} \\
2-4 pseudogene       & ---           & ---           & ---           & ---           & ---            & ---           & \textbf{14.42} & \textbf{5.97} \\
2-5 IG/TR segment    & ---           & ---           & ---           & ---           & ---            & ---           & \textbf{38.62} & \textbf{18.97} \\
3-1 EC class         & 0.04          & $-0.02$       & $-0.15$       & 0.29          & 0.09           & 0.09          & 0.04          & 0.30 \\
3-2 TF domain        & 0.35          & 0.12          & ---           & ---           & \textbf{7.75}  & \textbf{8.92} & \textbf{3.48} & \textbf{8.21} \\
4-1 EC subclass      & \textbf{0.11} & 0.12          & \textbf{0.34} & 0.23          & \textbf{1.80}  & \textbf{0.56} & \textbf{0.70} & \textbf{1.01} \\
\bottomrule
\end{tabular}
\end{table}

\paragraph{Where the hypothesis holds: protein function, TF domain, EC subclass.} At
level~2-1 all three instruments agree. The match share is 80--100\% for plant and fly units
against 0--4\% for vertebrate units. Excess dissimilarity is $\leq 0.6$ against 4.8--8.5
points. $\Uadj$ is 0.08--0.49 against 2.5--3.1. The order holds at every cut from whole
units to deciles. At TF domain the plants' $\Uadj$ is 0.1--0.4 against 3.5--8.9 for the
vertebrates. At EC subclass, which depends on EC assignment rather than on
function-annotation coverage, the plants (0.11--0.12) and fly (0.34) have the lowest
$\Uadj$, human (1.01) is highest among vertebrates, and \Ce{} (1.80) is highest overall.
The common mechanism is lineage-specific paralogue expansion --- KRAB zinc fingers,
olfactory and chemoreceptors, nuclear hormone receptors --- which in vertebrates and
nematodes has been clustered in cis, and in the plants and fly has not.

\paragraph{Where it does not hold: element class.} By base pairs at whole-unit scale,
the mammals are indeed the least homogeneous genomes (median 0.959--0.963, excess
3.6--3.9). But the most homogeneous are maize, zebrafish and yeast, not the plants and fly
as a group, and \At{} becomes the \emph{least} homogeneous genome as soon as its
chromosomes are cut. By count, $\Uadj$ is dominated by single dense classes: fly
(4.09; 0.94 without \texttt{4} and \texttt{Y}) and \Ce{} (10.92; 2.10 without its two
most divergent chromosomes) are high, the TE-swamped genomes low. The fair statement is
that level-1 heterogeneity is set by where each genome keeps its TEs and pseudogenes ---
spread along the arms (maize, zebrafish) or packed into pericentromeres (\At, fly) --- and
not by lineage.

\paragraph{Where it reverses: TE order.} Maize (0.05) and zebrafish (0.50) are the most
uniform repeat landscapes, and fly (2.62; 0.18 without \texttt{4} and \texttt{Y}) and
\At{} are the least uniform once their units are cut. Repeat composition follows repeat
biology, through the same placement that sets level~1: maize LTRs and zebrafish DNA
transposons are spread along the chromosome, fly and \At{} TEs packed into
pericentromeres --- not by lineage.

\paragraph{Where it is about loci: ncRNA, pseudogene and IG/TR.} These three classes are
the most clustered in the study, and in each case the clustering is one or a few named
loci: piRNA on \Ce{} \texttt{IV}, rDNA and snRNA arrays, unprocessed pseudogenes beside
their parents, and the IG/TR loci. \At{} is again the most uniform genome at ncRNA class,
but human and mouse come next; the high values in maize, zebrafish, fly and \Ce{} each
trace to one array.

\paragraph{Where there is nothing to separate: EC class.} Every genome is homogeneous, so
the hypothesis is neither supported nor contradicted.

\paragraph{The other three genomes.} Yeast is as homogeneous as the plants on every level
it can supply (\S\ref{sec:discussion} discusses what that implies). \Ce{} is the least homogeneous
genome at levels 2-1, 2-2 and 4-1, and at level~1 by count. Zebrafish is intermediate:
vertebrate at function and TF domain, but plant-like at level~1 and TE order.

\emph{The defensible statement:} along the genome, the protein-function and gene-family
composition of \At, maize and fly is markedly more uniform than that of human, mouse and
zebrafish --- and \Ce{} is the least uniform of all. The plant/fly advantage does not
extend to element class or TE order, where uniformity follows where each genome keeps its
repeats, nor to ncRNA, pseudogene and IG/TR classes, where it follows individual loci.

\section{Discussion}
\label{sec:discussion}

\paragraph{A well-mixed genome is a good first model.} The surprise of this study is how
much of a genome is a near-copy of the whole. A chromosome arm of fly, a chromosome of
zebrafish or yeast, or a 20~Mb decile of maize carries its genome's element-class
composition by base pairs, or its function or repeat composition, to within five percent.
At the level of reaction chemistry that is true of every chromosome in every genome
examined. The null model of a genome as a well-mixed pool, which one might expect to fail
everywhere given what is known about gene order \cite{hurst}, is a good first description
of most units at most levels. Gene-order studies and this one are not in conflict: the
clusters of co-expressed and housekeeping genes they describe are typically tens to
hundreds of kilobases long and are drawn from many functional classes
\cite{cohen,boutanaev,lercher}, so at the scale of a chromosome they average out.

\paragraph{Where it fails, it fails in three recognisable ways.} The three mechanisms of
Part~II (Fig.~\ref{fig:mechanisms}) recur across eight genomes and nine levels: named loci
(rDNA, snRNA, piRNA, IG/TR, pepsinogen, prolyl hydroxylase); megabase compartments
(pericentromeres, isochores, \Ce{} recombination arms)
\cite{bernardi,liebermanaiden,guelen,rockman}, which often shift several classes at once;
and clustered paralogue families (zinc fingers, chemoreceptors, nuclear hormone receptors),
the signature of lineage-specific expansion by tandem duplication \cite{lespinet}. The genomes differ in which mechanism dominates. That, more
than any overall degree of uniformity, is what separates the plants and fly from the
vertebrates and \Ce: the plants and fly carry few cis-clustered paralogue expansions, and
their heterogeneity is mostly compartmental.

\paragraph{What the hypothesis becomes.} The hypothesis survives only for protein function and
gene families (Table~\ref{tab:verdict}), and yeast's equal uniformity suggests that the underlying
contrast is less phylogenetic than structural: compact genomes, or genomes whose gene
families have not expanded in cis, look well mixed, whatever their lineage. For element
class and TEs the relevant variable is where a genome keeps its repeats. Maize spreads its
LTR retrotransposons between genes along the whole chromosome \cite{sanmiguel}, and its
chromosomes are near-copies of each other at every cut; \At{} and fly pack theirs into
pericentromeres \cite{copenhaver,hoskins}, and their chromosomes are near-copies only as
wholes.

\paragraph{Counting elements or base pairs.} Adding TEs to level~1 changes the coarse view
more than any other choice in the analysis. By count, a level that contains TE copies
measures TE copy number; by base pairs, it measures how the genome's sequence is divided
between genes and repeats, and where. The two disagree exactly where one would expect: for a
small, numerous class (\Ce{} piRNA loci, which dominate counts and vanish in base pairs) and
for short, numerous TE fragments (the mammalian SINEs). We therefore read level~1 by base
pairs and use counts only as a check.

\paragraph{Two methodological lessons.} First, raw similarity and $p$-values both report
the denominator: maize's TE copy counts make most cells ``significant'' at $\sm \geq
0.97$, while a few hundred TFs per unit make nothing significant at $\sm \approx 0.7$. Only
a count-matched floor reports the genome. Second, named cytogenetic boundaries are not
compositional boundaries: the position along an arm, not the band, is what carries
composition.

\paragraph{Outlook.} Two developments would sharpen the picture. Telomere-to-telomere
assemblies \cite{nurk,altemose} now resolve the heterochromatin, satellites and rDNA arrays
that every assembly used here collapses or omits, and so would let the special units and
the pericentromeric compartments be scored rather than set aside. And a single annotation
pipeline applied to all genomes would remove the softest comparison in the study, that
between RefSeq-annotated plants and Ensembl-annotated animals (\S\ref{sec:caveats}).

\section{Caveats}
\label{sec:caveats}

\begin{enumerate}
\item \textbf{Annotation, not biology.} Level-1 and level-2-2 composition depend on
curation depth: lncRNA spans are a quarter of human element base pairs, and piRNA loci are
62\% of \Ce's ncRNA genes. \At{} and maize genes come from RefSeq, fly from FlyBase, the
other five from Ensembl, which makes plant-versus-animal the softest comparison at these
levels.
\item \textbf{Repeat annotation differs between genomes.} TEs come from five kinds of
source: FlyBase FBti, UCSC RepeatMasker, SGD, TAIR and MaizeGDB. RepeatMasker reports
fragments, the others mostly whole insertions. Base-pair weighting reduces but does not
remove this difference. Fly TE is FBti-only (canonical insertions, 6.1\% of the assembly),
a floor in which heterochromatic TEs are under-represented.
\item \textbf{TEs in count-based tests.} The equal-cut partition tests count gene classes
only, so TE placement along a unit enters them not at all and is read from the base-pair
analyses instead. $\Uadj$ at level~1 is computed on element counts with TE copies
included (Table~\ref{tab:uadj}), so its level-1 values partly reflect how each repeat
annotation fragments its copies; the base-pair similarity scan is the main level-1 view.
\item \textbf{Overlapping elements.} Element base pairs are summed without resolving
overlaps, so intronic TEs count in both a gene and the TE class, and long lncRNA spans
weigh heavily in the mammals.
\item \textbf{The sampling floor is count-based} while the level-1 and TE similarities
are in base pairs. Their excess dissimilarity is therefore an upper bound.
\item \textbf{Function coverage.} Only 30\% of maize's and 46\% of \At's coding genes
carry a named function, against about 63\% in the mammals, and the named genes are the
conserved InterPro core. Part of the plants' function homogeneity may be that confound. It
does not apply to TF domain, where the order is at its clearest.
\item \textbf{Sampling floors at deep and small levels.} At EC subclass, TF domain,
pseudogene origin and IG/TR a unit may hold a few dozen elements, so a match share of 0
there often means ``too small to tell'', not ``different''. Excess dissimilarity and
$\Uadj$ are the statistics to read at these levels. Levels 2-4 and 2-5 are scored only in
human and mouse.
\item \textbf{Special units} are excused by a rule stated in advance, not by result.
Including them would lower fly's and human's match shares. In Table~\ref{tab:uadj} it
changes fly's level-1 and TE $\Uadj$ (0.94 and 0.18 without \texttt{4} and \texttt{Y}).
\item \textbf{Unit definition is uneven.} Human is split into arms and every other genome
except fly is not. Re-scoring human on whole chromosomes barely moves $\Uadj$ (level~1
$0.79 \to 0.68$).
\item \textbf{rRNA is under-counted} in every assembly because rDNA arrays collapse, so
yeast \texttt{XII} and maize \texttt{6} are floors.
\end{enumerate}

\section*{Data and code availability}
\label{sec:data}

All analyses use public annotation releases: FlyBase FB2026\_01 (dm6), genes and FBti
insertions \cite{flybase}; Ensembl and Ensembl Plants (GRCh38, GRCm39, GRCz11, R64,
WBcel235) \cite{ensembl}; NCBI RefSeq (TAIR10.1, Zm-B73-NAM-5.0) \cite{refseq}; WormBase
for \Ce{} gene classes \cite{wormbase}; UniProt EC and InterPro family assignments
\cite{interpro}; UCSC RepeatMasker tracks (hg38, mm39, danRer11, ce11); the SGD
chromosomal-feature file; the TAIR10 transposable-element table; the MaizeGDB structural TE
annotation; and the cytoband and Bridges maps distributed with \texttt{ideogram.js}
\cite{ideogram}. No new sequence data were generated. Per-species figures are produced by
notebooks 22--53 of the project repository, and the overview scan by notebook 52 and the
\texttt{fly\_genome.homogeneity} module. The gene-class partition tests
(Table~\ref{tab:partitions}, Fig.~\ref{fig:cuts}) are recomputed by
\texttt{analysis\_v6/partitions\_genes\_only\_v6.py} and \texttt{drivers\_v6.py} with the
same \texttt{fly\_genome.partitions} machinery; the figures are drawn at print size by the
scripts in \texttt{figs\_v6/}, from the same cached results. The per-chromosome level-1
figures of maize, zebrafish and yeast (Figs.~\ref{fig:mzl1}--\ref{fig:scl1}) are those of
notebooks 46, 43 and 44 (\texttt{figs\_v8/}).

\section*{Acknowledgements}

This work used no funding and no computing resources beyond a laptop. Analyses were
carried out in Python with \texttt{numpy}, \texttt{pandas}, \texttt{scipy},
\texttt{matplotlib} and \texttt{bioframe}.

The author would like to sincerely thank Anthropic's Claude Opus~4.8, 5.0, 5.1 and 5.5 for
their incredibly intelligent aid in the baking of this manuscript.


\begin{thebibliography}{99}

\bibitem{bernardi}
G.\ Bernardi, B.\ Olofsson, J.\ Filipski, M.\ Zerial, J.\ Salinas, G.\ Cuny,
M.\ Meunier-Rotival and F.\ Rodier.
\newblock The mosaic genome of warm-blooded vertebrates.
\newblock \textit{Science} 228:953--958, 1985.

\bibitem{eyrewalker}
A.\ Eyre-Walker and L.\ D.\ Hurst.
\newblock The evolution of isochores.
\newblock \textit{Nature Reviews Genetics} 2:549--555, 2001.

\bibitem{cohen}
B.\ A.\ Cohen, R.\ D.\ Mitra, J.\ D.\ Hughes and G.\ M.\ Church.
\newblock A computational analysis of whole-genome expression data reveals chromosomal
domains of gene expression.
\newblock \textit{Nature Genetics} 26:183--186, 2000.

\bibitem{caron}
H.\ Caron, B.\ van Schaik, M.\ van der Mee, et al.
\newblock The human transcriptome map: clustering of highly expressed genes in
chromosomal domains.
\newblock \textit{Science} 291:1289--1292, 2001.

\bibitem{spellman}
P.\ T.\ Spellman and G.\ M.\ Rubin.
\newblock Evidence for large domains of similarly expressed genes in the
\textit{Drosophila} genome.
\newblock \textit{Journal of Biology} 1:5, 2002.

\bibitem{boutanaev}
A.\ M.\ Boutanaev, A.\ I.\ Kalmykova, Y.\ Y.\ Shevelyov and D.\ I.\ Nurminsky.
\newblock Large clusters of co-expressed genes in the \textit{Drosophila} genome.
\newblock \textit{Nature} 420:666--669, 2002.

\bibitem{lercher}
M.\ J.\ Lercher, A.\ O.\ Urrutia and L.\ D.\ Hurst.
\newblock Clustering of housekeeping genes provides a unified model of gene order in
the human genome.
\newblock \textit{Nature Genetics} 31:180--183, 2002.

\bibitem{versteeg}
R.\ Versteeg, B.\ D.\ C.\ van Schaik, M.\ F.\ van Batenburg, et al.
\newblock The human transcriptome map reveals extremes in gene density, intron length,
GC content, and repeat pattern for domains of highly and weakly expressed genes.
\newblock \textit{Genome Research} 13:1998--2004, 2003.

\bibitem{hurst}
L.\ D.\ Hurst, C.\ P\'al and M.\ J.\ Lercher.
\newblock The evolutionary dynamics of eukaryotic gene order.
\newblock \textit{Nature Reviews Genetics} 5:299--310, 2004.

\bibitem{liebermanaiden}
E.\ Lieberman-Aiden, N.\ L.\ van Berkum, L.\ Williams, et al.
\newblock Comprehensive mapping of long-range interactions reveals folding principles
of the human genome.
\newblock \textit{Science} 326:289--293, 2009.

\bibitem{filion}
G.\ J.\ Filion, J.\ G.\ van Bemmel, U.\ Braunschweig, et al.
\newblock Systematic protein location mapping reveals five principal chromatin types in
\textit{Drosophila} cells.
\newblock \textit{Cell} 143:212--224, 2010.

\bibitem{kharchenko}
P.\ V.\ Kharchenko, A.\ A.\ Alekseyenko, Y.\ B.\ Schwartz, et al.
\newblock Comprehensive analysis of the chromatin landscape in \textit{Drosophila
melanogaster}.
\newblock \textit{Nature} 471:480--485, 2011.

\bibitem{guelen}
L.\ Guelen, L.\ Pagie, E.\ Brasset, et al.
\newblock Domain organization of human chromosomes revealed by mapping of nuclear lamina
interactions.
\newblock \textit{Nature} 453:948--951, 2008.

\bibitem{copenhaver}
G.\ P.\ Copenhaver, K.\ Nickel, T.\ Kuromori, et al.
\newblock Genetic definition and sequence analysis of \textit{Arabidopsis} centromeres.
\newblock \textit{Science} 286:2468--2474, 1999.

\bibitem{hoskins}
R.\ A.\ Hoskins, J.\ W.\ Carlson, K.\ H.\ Wan, et al.
\newblock The Release 6 reference genome sequence of \textit{Drosophila melanogaster}.
\newblock \textit{Genome Research} 25:445--458, 2015.

\bibitem{altemose}
N.\ Altemose, G.\ A.\ Logsdon, A.\ V.\ Bzikadze, et al.
\newblock Complete genomic and epigenetic maps of human centromeres.
\newblock \textit{Science} 376:eabl4178, 2022.

\bibitem{rockman}
M.\ V.\ Rockman and L.\ Kruglyak.
\newblock Recombinational landscape and population genomics of
\textit{Caenorhabditis elegans}.
\newblock \textit{PLoS Genetics} 5:e1000419, 2009.

\bibitem{comeron}
J.\ M.\ Comeron, R.\ Ratnappan and S.\ Bailin.
\newblock The many landscapes of recombination in \textit{Drosophila melanogaster}.
\newblock \textit{PLoS Genetics} 8:e1002905, 2012.

\bibitem{lespinet}
O.\ Lespinet, Y.\ I.\ Wolf, E.\ V.\ Koonin and L.\ Aravind.
\newblock The role of lineage-specific gene family expansion in the evolution of
eukaryotes.
\newblock \textit{Genome Research} 12:1048--1059, 2002.

\bibitem{muller}
H.\ J.\ Muller.
\newblock Bearings of the `\textit{Drosophila}' work on systematics.
\newblock In J.\ Huxley (ed.), \textit{The New Systematics}, pp.\ 185--268. Clarendon
Press, Oxford, 1940.

\bibitem{bhutkar}
A.\ Bhutkar, S.\ W.\ Schaeffer, S.\ M.\ Russo, et al.
\newblock Chromosomal rearrangement inferred from comparisons of 12 \textit{Drosophila}
genomes.
\newblock \textit{Genetics} 179:1657--1680, 2008.

\bibitem{passarge}
E.\ Passarge, B.\ Horsthemke and R.\ A.\ Farber.
\newblock Incorrect use of the term synteny.
\newblock \textit{Nature Genetics} 23:387, 1999.

\bibitem{levin}
D.\ A.\ Levin and Y.\ Peres.
\newblock \textit{Markov Chains and Mixing Times}, 2nd edn.
\newblock American Mathematical Society, Providence, RI, 2017.

\bibitem{koonin}
E.\ V.\ Koonin.
\newblock Orthologs, paralogs, and evolutionary genomics.
\newblock \textit{Annual Review of Genetics} 39:309--338, 2005.

\bibitem{adams}
M.\ D.\ Adams, S.\ E.\ Celniker, R.\ A.\ Holt, et al.
\newblock The genome sequence of \textit{Drosophila melanogaster}.
\newblock \textit{Science} 287:2185--2195, 2000.

\bibitem{flybase}
A.\ \"Ozt\"urk-\c{C}olak, S.\ J.\ Marygold, G.\ Antonazzo, et al.
\newblock FlyBase: updates to the \textit{Drosophila} genes and genomes database.
\newblock \textit{Genetics} 227:iyad211, 2024. Release FB2026\_01 used here.

\bibitem{lander}
International Human Genome Sequencing Consortium.
\newblock Initial sequencing and analysis of the human genome.
\newblock \textit{Nature} 409:860--921, 2001.

\bibitem{schneider}
V.\ A.\ Schneider, T.\ Graves-Lindsay, K.\ Howe, et al.
\newblock Evaluation of GRCh38 and \textit{de novo} haploid genome assemblies
demonstrates the enduring quality of the reference assembly.
\newblock \textit{Genome Research} 27:849--864, 2017.

\bibitem{mouse}
Mouse Genome Sequencing Consortium.
\newblock Initial sequencing and comparative analysis of the mouse genome.
\newblock \textit{Nature} 420:520--562, 2002.

\bibitem{howe}
K.\ Howe, M.\ D.\ Clark, C.\ F.\ Torroja, et al.
\newblock The zebrafish reference genome sequence and its relationship to the human
genome.
\newblock \textit{Nature} 496:498--503, 2013.

\bibitem{goffeau}
A.\ Goffeau, B.\ G.\ Barrell, H.\ Bussey, et al.
\newblock Life with 6000 genes.
\newblock \textit{Science} 274:546, 563--567, 1996.

\bibitem{celegans}
\textit{C.~elegans} Sequencing Consortium.
\newblock Genome sequence of the nematode \textit{C.~elegans}: a platform for
investigating biology.
\newblock \textit{Science} 282:2012--2018, 1998.

\bibitem{wormbase}
P.\ W.\ Sternberg, K.\ Van Auken, Q.\ Wang, et al.
\newblock WormBase 2024: status and transitioning to Alliance infrastructure.
\newblock \textit{Genetics} 227:iyae050, 2024.

\bibitem{agi}
Arabidopsis Genome Initiative.
\newblock Analysis of the genome sequence of the flowering plant \textit{Arabidopsis
thaliana}.
\newblock \textit{Nature} 408:796--815, 2000.

\bibitem{tair}
P.\ Lamesch, T.\ Z.\ Berardini, D.\ Li, et al.
\newblock The Arabidopsis Information Resource (TAIR): improved gene annotation and new
tools.
\newblock \textit{Nucleic Acids Research} 40:D1202--D1210, 2012.

\bibitem{schnable}
P.\ S.\ Schnable, D.\ Ware, R.\ S.\ Fulton, et al.
\newblock The B73 maize genome: complexity, diversity, and dynamics.
\newblock \textit{Science} 326:1112--1115, 2009.

\bibitem{hufford}
M.\ B.\ Hufford, A.\ S.\ Seetharam, M.\ R.\ Woodhouse, et al.
\newblock \textit{De novo} assembly, annotation, and comparative analysis of 26 diverse
maize genomes.
\newblock \textit{Science} 373:655--662, 2021.

\bibitem{ensembl}
P.\ W.\ Harrison, M.\ R.\ Amode, O.\ Austine-Orimoloye, et al.
\newblock Ensembl 2024.
\newblock \textit{Nucleic Acids Research} 52:D891--D899, 2024.

\bibitem{frankish}
A.\ Frankish, M.\ Diekhans, A.-M.\ Ferreira, et al.
\newblock GENCODE reference annotation for the human and mouse genomes.
\newblock \textit{Nucleic Acids Research} 47:D766--D773, 2019.

\bibitem{refseq}
N.\ A.\ O'Leary, M.\ W.\ Wright, J.\ R.\ Brister, et al.
\newblock Reference sequence (RefSeq) database at NCBI: current status, taxonomic
expansion, and functional annotation.
\newblock \textit{Nucleic Acids Research} 44:D733--D745, 2016.

\bibitem{ensemblgenomes}
A.\ D.\ Yates, J.\ Allen, R.\ M.\ Amode, et al.
\newblock Ensembl Genomes 2022: an expanding genome resource for non-vertebrates.
\newblock \textit{Nucleic Acids Research} 50:D996--D1003, 2022.

\bibitem{uniprot}
UniProt Consortium.
\newblock UniProt: the Universal Protein Knowledgebase in 2023.
\newblock \textit{Nucleic Acids Research} 51:D523--D531, 2023.

\bibitem{interpro}
M.\ Blum, A.\ Andreeva, L.\ C.\ Florentino, et al.
\newblock InterPro: the protein sequence classification resource in 2025.
\newblock \textit{Nucleic Acids Research} 53:D444--D456, 2025.

\bibitem{kaminker}
J.\ S.\ Kaminker, C.\ M.\ Bergman, B.\ Kronmiller, et al.
\newblock The transposable elements of the \textit{Drosophila melanogaster}
euchromatin: a genomics perspective.
\newblock \textit{Genome Biology} 3:research0084, 2002.

\bibitem{ucsc}
W.\ J.\ Kent, C.\ W.\ Sugnet, T.\ S.\ Furey, et al.
\newblock The human genome browser at UCSC.
\newblock \textit{Genome Research} 12:996--1006, 2002.

\bibitem{repeatmasker}
A.\ F.\ A.\ Smit, R.\ Hubley and P.\ Green.
\newblock \textit{RepeatMasker Open-4.0}, 2013--2015.
\newblock \url{http://www.repeatmasker.org}

\bibitem{maddox}
P.\ S.\ Maddox, K.\ Oegema, A.\ Desai and I.\ M.\ Cheeseman.
\newblock ``Holo''er than thou: chromosome segregation and kinetochore function in
\textit{C.~elegans}.
\newblock \textit{Chromosome Research} 12:641--653, 2004.

\bibitem{bridges}
C.\ B.\ Bridges.
\newblock Salivary chromosome maps, with a key to the banding of the chromosomes of
\textit{Drosophila melanogaster}.
\newblock \textit{Journal of Heredity} 26:60--64, 1935.

\bibitem{iscn}
J.\ McGowan-Jordan, R.\ J.\ Hastings and S.\ Moore (eds.).
\newblock \textit{ISCN 2020: An International System for Human Cytogenomic
Nomenclature}.
\newblock Karger, Basel, 2020.

\bibitem{furey}
T.\ S.\ Furey and D.\ Haussler.
\newblock Integration of the cytogenetic map with the draft human genome sequence.
\newblock \textit{Human Molecular Genetics} 12:1037--1044, 2003.

\bibitem{riddle}
N.\ C.\ Riddle and S.\ C.\ R.\ Elgin.
\newblock The \textit{Drosophila} dot chromosome: where genes flourish amidst repeats.
\newblock \textit{Genetics} 210:757--772, 2018.

\bibitem{carvalho}
A.\ B.\ Carvalho, L.\ B.\ Koerich and A.\ G.\ Clark.
\newblock Origin and evolution of Y chromosomes: \textit{Drosophila} tales.
\newblock \textit{Trends in Genetics} 25:270--277, 2009.

\bibitem{mcstay}
B.\ McStay.
\newblock Nucleolar organizer regions: genomic `dark matter' requiring illumination.
\newblock \textit{Genes \& Development} 30:1598--1610, 2016.

\bibitem{nurk}
S.\ Nurk, S.\ Koren, A.\ Rhie, et al.
\newblock The complete sequence of a human genome.
\newblock \textit{Science} 376:44--53, 2022.

\bibitem{wicker}
T.\ Wicker, F.\ Sabot, A.\ Hua-Van, et al.
\newblock A unified classification system for eukaryotic transposable elements.
\newblock \textit{Nature Reviews Genetics} 8:973--982, 2007.

\bibitem{webb}
E.\ C.\ Webb (ed.).
\newblock \textit{Enzyme Nomenclature 1992: Recommendations of the Nomenclature
Committee of the International Union of Biochemistry and Molecular Biology}.
\newblock Academic Press, San Diego, 1992.

\bibitem{ec}
A.\ G.\ McDonald, S.\ Boyce and K.\ F.\ Tipton.
\newblock ExplorEnz: the primary source of the IUBMB enzyme list.
\newblock \textit{Nucleic Acids Research} 37:D593--D597, 2009.

\bibitem{wingender}
E.\ Wingender, T.\ Schoeps and J.\ D\"onitz.
\newblock TFClass: an expandable hierarchical classification of human transcription
factors.
\newblock \textit{Nucleic Acids Research} 41:D165--D170, 2013.

\bibitem{lambert}
S.\ A.\ Lambert, A.\ Jolma, L.\ F.\ Campitelli, et al.
\newblock The human transcription factors.
\newblock \textit{Cell} 172:650--665, 2018.

\bibitem{go}
M.\ Ashburner, C.\ A.\ Ball, J.\ A.\ Blake, et al.
\newblock Gene Ontology: tool for the unification of biology.
\newblock \textit{Nature Genetics} 25:25--29, 2000.

\bibitem{pei}
B.\ Pei, C.\ Sisu, A.\ Frankish, et al.
\newblock The GENCODE pseudogene resource.
\newblock \textit{Genome Biology} 13:R51, 2012.

\bibitem{lefranc}
M.-P.\ Lefranc, V.\ Giudicelli, C.\ Ginestoux, et al.
\newblock IMGT, the international ImMunoGeneTics information system.
\newblock \textit{Nucleic Acids Research} 37:D1006--D1012, 2009.

\bibitem{mattick}
J.\ S.\ Mattick, P.\ P.\ Amaral, P.\ Carninci, et al.
\newblock Long non-coding RNAs: definitions, functions, challenges and recommendations.
\newblock \textit{Nature Reviews Molecular Cell Biology} 24:430--447, 2023.

\bibitem{bartel}
D.\ P.\ Bartel.
\newblock Metazoan microRNAs.
\newblock \textit{Cell} 173:20--51, 2018.

\bibitem{gtrnadb}
P.\ P.\ Chan and T.\ M.\ Lowe.
\newblock GtRNAdb 2.0: an expanded database of transfer RNA genes identified in
complete and draft genomes.
\newblock \textit{Nucleic Acids Research} 44:D184--D189, 2016.

\bibitem{will}
C.\ L.\ Will and R.\ L\"uhrmann.
\newblock Spliceosome structure and function.
\newblock \textit{Cold Spring Harbor Perspectives in Biology} 3:a003707, 2011.

\bibitem{kiss}
T.\ Kiss.
\newblock Small nucleolar RNAs: an abundant group of noncoding RNAs with diverse
cellular functions.
\newblock \textit{Cell} 109:145--148, 2002.

\bibitem{darzacq}
X.\ Darzacq, B.\ E.\ J\'ady, C.\ Verheggen, et al.
\newblock Cajal body-specific small nuclear RNAs: a novel class of 2$'$-O-methylation
and pseudouridylation guide RNAs.
\newblock \textit{EMBO Journal} 21:2746--2756, 2002.

\bibitem{ozata}
D.\ M.\ Ozata, I.\ Gainetdinov, A.\ Zoch, D.\ O'Carroll and P.\ D.\ Zamore.
\newblock PIWI-interacting RNAs: small RNAs with big functions.
\newblock \textit{Nature Reviews Genetics} 20:89--108, 2019.

\bibitem{ruby}
J.\ G.\ Ruby, C.\ Jan, C.\ Player, et al.
\newblock Large-scale sequencing reveals 21U-RNAs and additional microRNAs and
endogenous siRNAs in \textit{C.~elegans}.
\newblock \textit{Cell} 127:1193--1207, 2006.

\bibitem{cover}
T.\ M.\ Cover and J.\ A.\ Thomas.
\newblock \textit{Elements of Information Theory}, 2nd edition.
\newblock Wiley, Hoboken, NJ, 2006.

\bibitem{phipson}
B.\ Phipson and G.\ K.\ Smyth.
\newblock Permutation $p$-values should never be zero: calculating exact $p$-values when
permutations are randomly drawn.
\newblock \textit{Statistical Applications in Genetics and Molecular Biology} 9:39, 2010.

\bibitem{bergsma}
W.\ Bergsma.
\newblock A bias-correction for Cram\'er's $V$ and Tschuprow's $T$.
\newblock \textit{Journal of the Korean Statistical Society} 42:323--328, 2013.

\bibitem{timetree}
S.\ Kumar, M.\ Suleski, J.\ M.\ Craig, et al.
\newblock TimeTree 5: an expanded resource for species divergence times.
\newblock \textit{Molecular Biology and Evolution} 39:msac174, 2022.

\bibitem{aguinaldo}
A.\ M.\ A.\ Aguinaldo, J.\ M.\ Turbeville, L.\ S.\ Linford, et al.
\newblock Evidence for a clade of nematodes, arthropods and other moulting animals.
\newblock \textit{Nature} 387:489--493, 1997.

\bibitem{sanmiguel}
P.\ SanMiguel, A.\ Tikhonov, Y.-K.\ Jin, et al.
\newblock Nested retrotransposons in the intergenic regions of the maize genome.
\newblock \textit{Science} 274:765--768, 1996.

\bibitem{bailey}
J.\ A.\ Bailey, L.\ Carrel, A.\ Chakravarti and E.\ E.\ Eichler.
\newblock Molecular evidence for a relationship between LINE-1 elements and X chromosome
inactivation: the Lyon repeat hypothesis.
\newblock \textit{Proceedings of the National Academy of Sciences USA} 97:6634--6639, 2000.

\bibitem{huntley}
S.\ Huntley, D.\ M.\ Baggott, A.\ T.\ Hamilton, et al.
\newblock A comprehensive catalog of human KRAB-associated zinc finger genes: insights
into the evolutionary history of a large family of transcriptional repressors.
\newblock \textit{Genome Research} 16:669--677, 2006.

\bibitem{grimwood}
J.\ Grimwood, L.\ A.\ Gordon, A.\ Olsen, et al.
\newblock The DNA sequence and biology of human chromosome 19.
\newblock \textit{Nature} 428:529--535, 2004.

\bibitem{imbeault}
M.\ Imbeault, P.-Y.\ Helleboid and D.\ Trono.
\newblock KRAB zinc-finger proteins contribute to the evolution of gene regulatory
networks.
\newblock \textit{Nature} 543:550--554, 2017.

\bibitem{robertson}
H.\ M.\ Robertson and J.\ H.\ Thomas.
\newblock The putative chemoreceptor families of \textit{C.~elegans}.
\newblock In \textit{WormBook}, ed.\ The \textit{C.~elegans} Research Community, 2006.
\newblock doi:10.1895/wormbook.1.66.1

\bibitem{zhangfirestein}
X.\ Zhang and S.\ Firestein.
\newblock The olfactory receptor gene superfamily of the mouse.
\newblock \textit{Nature Neuroscience} 5:124--133, 2002.

\bibitem{mcclintock}
B.\ McClintock.
\newblock The relation of a particular chromosomal element to the development of the
nucleoli in \textit{Zea mays}.
\newblock \textit{Zeitschrift f\"ur Zellforschung und Mikroskopische Anatomie}
21:294--326, 1934.

\bibitem{robinsonrechavi}
M.\ Robinson-Rechavi, C.\ V.\ Maina, C.\ R.\ Gissendanner, V.\ Laudet and A.\ Sluder.
\newblock Explosive lineage-specific expansion of the orphan nuclear receptor HNF4 in
nematodes.
\newblock \textit{Journal of Molecular Evolution} 60:577--586, 2005.

\bibitem{cavaille}
J.\ Cavaill\'e, K.\ Buiting, M.\ Kiefmann, et al.
\newblock Identification of brain-specific and imprinted small nucleolar RNA genes
exhibiting an unusual genomic organization.
\newblock \textit{Proceedings of the National Academy of Sciences USA}
97:14311--14316, 2000.

\bibitem{ideogram}
E.\ Weitz.
\newblock \textit{ideogram: chromosome visualization for the web}.
\newblock \url{https://github.com/eweitz/ideogram}

\end{thebibliography}
\end{document}